\documentclass[10pt,draft,twoside,fleqn]{book}
\usepackage{epsf,latexsym,graphics}
\usepackage{cite}

\newcommand{\Ef}{\ensuremath{\epsilon_{\mathrm{F}}}}
\newcommand{\kf}{\ensuremath{k_{\mathrm{F}}}}
\newcommand{\etal}{\textit{et al.}}

\newtheorem{exercise}{Exercise}[chapter]

\newcounter{saveeqn}
\newcommand{\alpheqn}{\setcounter{saveeqn}{\value{equation}}%
\stepcounter{saveeqn}\setcounter{equation}{0}%
\renewcommand{\theequation}{\mbox{\arabic{chapter}.\arabic{saveeqn}\alph{equation}}}}
\newcommand{\reseteqn}{\setcounter{equation}{\value{saveeqn}}%
\renewcommand{\theequation}{\arabic{chapter}.\arabic{equation}}}

\newcommand{\boxedtext}[1]{\medskip\noindent\fbox{\begin{minipage}[h]{4.65in}
#1\end{minipage}}\medskip}

\newcommand{\dedication}[1]{\medskip\noindent\fbox{\fbox{\begin{minipage}[h]{3.65in}
\large #1\end{minipage}}}\medskip}

\frontmatter
\title{The Quantum Hall Effect:\\
Novel Excitations and Broken Symmetries}
\author{Steven M. Girvin\thanks{\copyright S.M. Girvin, 1998\hfill\break
Lectures delivered at Ecole d'Et\'{e} Les Houches, July 1998\hfill\break
To be published by Springer Verlag and Les Editions de Physique in 1999.}\\ 
Indiana University\\
Department of Physics\\
Bloomington, IN  47405\\
United States of America}
\date{\copyright 1998}
\vfill
\begin{document}
\maketitle
\newpage
\null\vfill
\begin{center}
\dedication{These lectures are dedicated to the memory of Heinz Schulz, a great
friend and a wonderful physicist.}
\end{center}
\vfill
\tableofcontents

\mainmatter
\chapter{The Quantum Hall Effect}
\label{chap:quantum_hall}

\section{Introduction}
\label{sec:introduction}

The quantum Hall effect (QHE) is one of the most remarkable condensed-matter
phenomena discovered in the second half of the 20th century. It rivals
superconductivity in its fundamental significance as a manifestation of quantum
mechanics on macroscopic scales. The basic experimental observation is the
nearly vanishing dissipation
\begin{equation}
\sigma_{xx} \rightarrow 0
\end{equation}
and the quantization of the Hall conductance
\begin{equation}
\sigma_{xy} = \nu \frac{e^{2}}{h} \label{eq:9812-01}
\end{equation}
of a real (as opposed to some theorist's fantasy) transistor-like device
(similar in some cases to the transistors in computer chips) containing a
two-dimensional electron gas subjected to a strong magnetic field. This
quantization is universal and independent of all microscopic details such as the
type of semiconductor material, the purity of the sample, the precise value of
the magnetic field, and so forth. As a result, the effect is now used to
maintain\footnote{Maintain does \textit{not} mean \textit{define}. The SI ohm is
defined in terms of the kilogram, the second and the speed of light (formerly
the meter). It is best realized using the reactive impedance of a capacitor
whose capacitance is computed from first principles. This is an extremely
tedious procedure and the QHE is a very convenient method for realizing a fixed,
reproducible impedance to check for drifts of resistance standards. It does not
however \textit{define} the ohm. Eq.~(\ref{eq:9812-01}) is given in cgs units.
When converted to SI units the quantum of resistance is $h/e^{2}(\mathrm{cgs})
\rightarrow \frac{Z}{2\alpha} \approx 25,812.80~\Omega~(\mathrm{SI})$ where
$\alpha$ is the fine structure constant and $Z \equiv
\sqrt{\mu_{0}/\epsilon_{0}}$ is the impedance of free space.} the standard of
electrical resistance by metrology laboratories around the world. In addition,
since the speed of light is now defined, a measurement of $e^{2}/h$ is
equivalent to a measurement of the fine structure constant of fundamental
importance in quantum electrodynamics.

In the so-called integer quantum Hall effect (IQHE) discovered by von Klitzing
in 1980, the quantum number $\nu$ is a simple integer with a precision of about
$10^{-10}$ and an absolute accuracy of about $10^{-8}$ (both being limited by
our ability to do resistance metrology).

In 1982, Tsui, St\"{o}rmer and Gossard discovered that in certain devices with
reduced (but still non-zero) disorder, the quantum number $\nu$ could take on
rational fractional values. This so-called fractional quantum Hall effect (FQHE)
is the result of quite different underlying physics involving strong Coulomb
interactions and correlations among the electrons. The particles condense into
special quantum states whose excitations have the bizarre property of being
described by fractional quantum numbers, including fractional charge and
fractional statistics that are intermediate between ordinary Bose and Fermi
statistics. The FQHE has proven to be a rich and surprising arena for the
testing of our understanding of strongly correlated quantum systems. With a
simple twist of a dial on her apparatus, the quantum Hall experimentalist can
cause the electrons to condense into a bewildering array of new `vacua', each of
which is described by a different quantum field theory. The novel order
parameters describing each of these phases are completely unprecedented.

We begin with a brief description of why two-dimensionality is important to the
universality of the result and how modern semiconductor processing techniques
can be used to generate a nearly ideal two-dimensional electron gas (2DEG). We
then give a review of the classical and semi-classical theories of the motion of
charged particles in a magnetic field. Next we consider the limit of low
temperatures and strong fields where a full quantum treatment of the dynamics is
required. After that we will be in a position to understand the localization
phase transition in the IQHE. We will then study the origins of the FQHE and the
physics described by the novel wave function invented by Robert Laughlin to
describe the special condensed state of the electrons. Finally we will discuss
topological excitations and broken symmetries in quantum Hall ferromagnets.

The review presented here is by no means complete. It is primarily an
introduction to the basics followed by a more advanced discussion of recent
developments in quantum Hall ferromagnetism. Among the many topics which receive
little or no discussion are the FQHE hierarchical states, interlayer drag
effects, FQHE edge state tunneling and the composite boson \cite{compositeboson}
and fermion \cite{compositefermion} pictures of the FQHE. A number of general
reviews exist which the reader may be interested in consulting
\cite{SMGBOOK,TAPASHbook,macdbook,DasSarmabook,Hajdu,stonebook,sciam,sczhang,macdleshouches}

\subsection{Why 2D Is Important}

As one learns in the study of scaling in the localization transition,
resistivity (which is what theorists calculate) and resistance (which is what
experimentalists measure) for classical systems (in the shape of a hypercube) of
size $L$ are related by \cite{LeeRamakrishnan,sondhiRMP97}
\begin{equation}
R = \rho L^{(2-d)}.
\end{equation}
Two dimensions is therefore special since in this case the resistance of the
sample is scale invariant and $(e^{2}/h)R$ is dimensionless. This turns out to
be crucial to the universality of the result. In particular it means that one
does not have to measure the physical dimensions of the sample to one part in
$10^{10}$ in order to obtain the resistivity to that precision. Since the
locations of the edges of the sample are not well-defined enough to even
contemplate such a measurement, this is a very fortunate feature of having
available a 2DEG. It further turns out that, since the dissipation is nearly
zero in the QHE states, even the shape of the sample and the precise location of
the Hall voltage probes are almost completely irrelevant.

\subsection{Constructing the 2DEG}

\begin{figure}
\centerline{\epsfxsize=10cm
 \epsffile{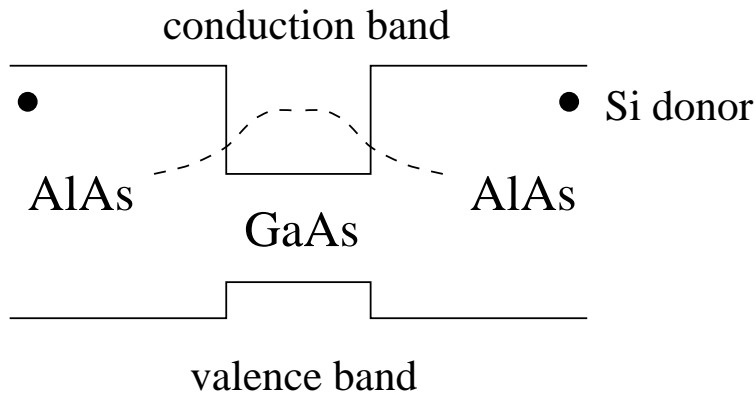}}
\caption[]{Schematic illustration of a GaAs/AlAs heterostructure quantum well.
The vertical axis is band energy and the horizontal axis is position in the MBE
growth direction. The dark circles indicate the Si$^{+}$ ions which have donated
electrons into the quantum well. The lowest electric subband wave function of the
quantum well is illustrated by the dashed line. It is common to use an alloy of
GaAs and AlAs rather than pure AlAs for the barrier region as illustrated here.}
\label{fig:2DEG}
\end{figure}

There are a variety of techniques to construct two-dimensional electron gases.
Fig.~(\ref{fig:2DEG})
shows one example in which the energy bands in a GaAs/AlAs heterostructure are
used to create a `quantum well'. Electrons from a Si donor layer fall into the
quantum well to create the 2DEG. The energy level (`electric subband') spacing
for the `particle in a box' states of the well can be of order $10^{3}~\mbox{K}$
which is much larger than the cryogenic temperatures at which QHE experiments
are performed. Hence all the electrons are frozen into the lowest electric
subband (if this is consistent with the Pauli principle) but remain free to move
in the plane of the GaAs layer forming the well. The dynamics of the electrons
is therefore effectively two-dimensional even though the quantum well is not
literally two-dimensional.

Heterostructures that are grown one atomic layer at a time by Molecular Beam
Epitaxy (MBE) are nearly perfectly ordered on the atomic scale. In addition the
Si donor layer can be set back a considerable distance ($\sim 0.5\mu\mathrm{m}$)
to minimize the random scattering from the ionized Si donors. Using these
techniques, electron mobilities of $10^{7}~\mathrm{cm^{2}/Vs}$ can be achieved
at low temperatures corresponding to incredibly long mean free paths of $\sim
0.1~\mbox{mm}$. As a result of the extremely low disorder in these systems,
subtle electronic correlation energies come to the fore and yield a remarkable
variety of quantum ground states, some of which we shall explore here.

The same MBE and remote doping technology is used to make GaAs quantum well High
Electron Mobility Transistors (HEMTs) which are used in all cellular telephones
and in radio telescope receivers where they are prized for their low noise and
ability to amplify extremely weak signals. The same technology is widely
utilized to produce the quantum well lasers used in compact disk players.

\subsection{Why is Disorder and Localization Important?}

Paradoxically, the extreme universality of the transport properties in the
quantum Hall regime occurs because of, rather than in spite of, the random
disorder and uncontrolled imperfections which the devices contain. Anderson
localization in the presence of disorder plays an essential role in the
quantization, but this localization is strongly modified by the strong magnetic
field.

In two dimensions (for zero magnetic field and non-interacting electrons) all
states are localized even for arbitrarily weak disorder. The essence of this
weak localization effect is the current `echo' associated with the quantum
interference corrections to classical transport \cite{Bergmann}. These quantum
interference effects rely crucially on the existence of time-reversal symmetry.
In the presence of a strong quantizing magnetic field, time-reversal symmetry is
destroyed and the localization properties of the disordered 2D electron gas are
radically altered. We will shortly see that there exists a novel phase
transition, not between a metal and insulator, but rather between two distinctly
different insulating states.

In the absence of any impurities the 2DEG is translationally invariant and there
is no preferred frame of reference.\footnote{This assumes that we can ignore the
periodic potential of the crystal which is of course fixed in the lab frame.
Within the effective mass approximation this potential modifies the mass but
does not destroy the Galilean invariance since the energy is still quadratic in
the momentum.} As a result we can transform to a frame of reference moving with
velocity $-\vec{v}$ relative to the lab frame. In this frame the electrons
appear to be moving at velocity $+\vec{v}$ and carrying current density
\begin{equation}
\vec{J} = -ne\vec{v},
\end{equation}
where $n$ is the areal density and we use the convention that the electron
charge is $-e$. In the lab frame, the electromagnetic fields are
\begin{eqnarray}
\vec{E} &=& \vec{0}\\
\vec{B} &=& B \hat{z}.
\end{eqnarray}
In the moving frame they are (to lowest order in $v/c$)
\begin{eqnarray}
\vec{E} &=& -\frac{1}{c}\vec{v}\times \vec{B}\\
\vec{B} &=& B \hat{z}.
\end{eqnarray}
This Lorentz transformation picture is precisely equivalent to the usual
statement that an electric field must exist which just cancels the Lorentz force
$\frac{-e}{c} \vec{v}\times \vec{B}$ in order for the device to carry the current
straight through without deflection. Thus we have
\begin{equation}
\vec{E} = \frac{B}{nec} \vec{J}\times \hat{B}.
\end{equation}
The resistivity tensor is defined by
\begin{equation}
E^{\mu} = \rho_{\mu\nu} J^{\nu}.
\end{equation}
Hence we can make the identification
\begin{equation}
\underline{\underline{\rho}} = \frac{B}{nec}
\left(\begin{array}{cc}
0&+1\\
-1&0
\end{array}\right)
\end{equation}

The conductivity tensor is the matrix inverse of this so that
\begin{equation}
J^{\mu} = \sigma_{\mu\nu} E^{\nu},
\end{equation}
and
\begin{equation}
\underline{\underline{\sigma}} = \frac{nec}{B}
\left(\begin{array}{cc}
0&-1\\
+1&0
\end{array}\right)
\end{equation}
Notice that, paradoxically, the system looks insulating since $\sigma_{xx}=0$
and yet it looks like a perfect conductor since $\rho_{xx} = 0$. In an ordinary
insulator $\sigma_{xy}=0$ and so $\rho_{xx}=\infty$. Here $\sigma_{xy} =
\frac{nec}{B} \ne 0$ and so the inverse exists.

The argument given above relies only on Lorentz covariance. The only property of
the 2DEG that entered was the density. The argument works equally well whether
the system is classical or quantum, whether the electron state is liquid, vapor,
or solid. It simply does not matter. Thus, in the absence of disorder, the Hall
effect teaches us nothing about the system other than its density. The Hall
resistivity is simply a linear function of magnetic field whose slope tells us
about the density. In the quantum Hall regime we would therefore see none of the
novel physics in the absence of disorder since disorder is needed to destroy
translation invariance. Once the translation invariance is destroyed there is a
preferred frame of reference and the Lorentz covariance argument given above
fails.

\begin{figure}
\centerline{\epsfxsize=12cm
 \epsffile{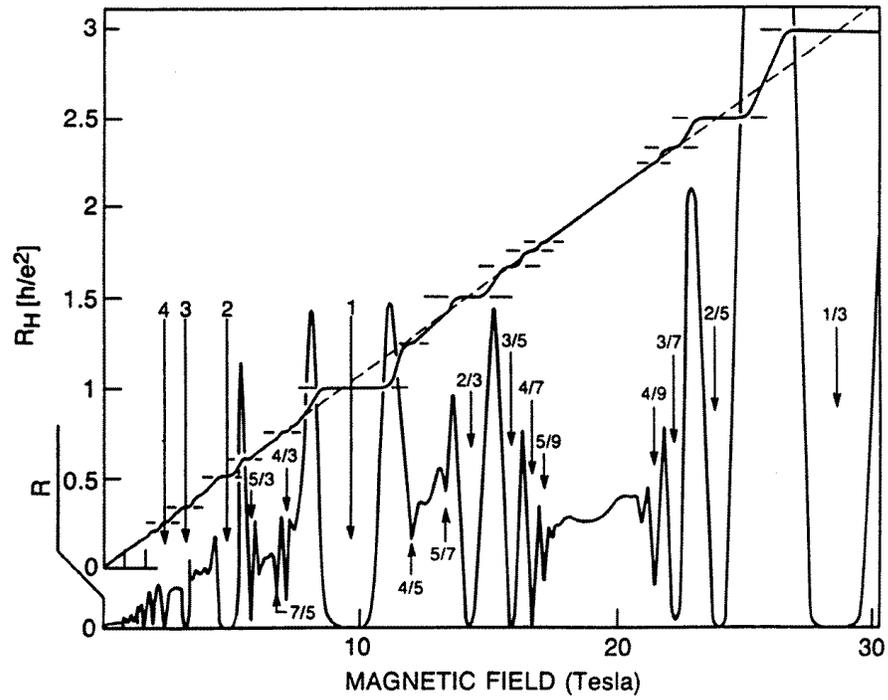}}
\caption[]{Integer and fractional quantum Hall transport data showing the
plateau regions in the Hall resistance $R_{\rm H}$ and associated dips in the
dissipative resistance $R$. The numbers indicate the Landau level filling
factors at which various features occur. After ref.~\cite{transport-data}.}
\label{fig:qhedata}
\end{figure}

\begin{figure}
\centerline{\epsfxsize=6cm
 \epsffile{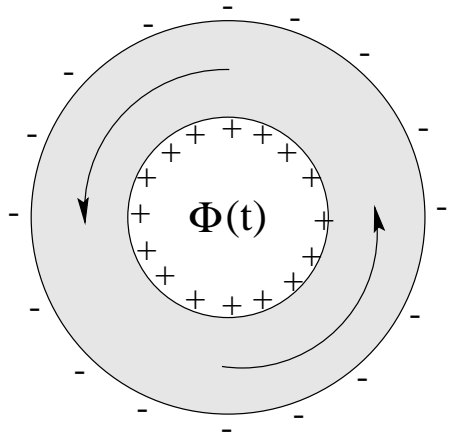}}
\caption[]{Persistent current circulating in a quantum Hall device having the
Corbino geometry. The radial electric field is maintained by the charges which
can not flow back together because $\sigma_{xx}$ is nearly zero. These charges
result from the radial current pulse associated with the azimuthal electric
field pulse produced by the applied flux $\Phi(t)$.}
\label{fig:corbino}
\end{figure}

Figure~(\ref{fig:qhedata})
shows the remarkable transport data for a real device in the quantum Hall
regime. Instead of a Hall resistivity which is simply a linear function of
magnetic field, we see a series of so-called \textit{Hall plateaus} in which
$\rho_{xy}$ is a universal constant
\begin{equation}
\rho_{xy} = -\frac{1}{\nu}\frac{h}{e^{2}}
\end{equation}
independent of all microscopic details (including the precise value of the
magnetic field). Associated with each of these plateaus is a dramatic decrease
in the dissipative resistivity $\rho_{xx}\longrightarrow 0$ which drops as much
as 13 orders of magnitude in the plateau regions. Clearly the system is
undergoing some sort of sequence of phase transitions into highly idealized
dissipationless states. Just as in a superconductor, the dissipationless state
supports persistent currents. These can be produced in devices having the
Corbino ring geometry shown in fig.~(\ref{fig:corbino}).
Applying additional flux through the ring produces a temporary azimuthal
electric field by Faraday induction. A current pulse is induced at right angles
to the $E$ field and produces a radial charge polarization as shown. This
polarization induces a (quasi-) permanent radial electric field which in turn
causes persistent azimuthal currents. Torque magnetometer measurements
\cite{torque} have shown that the currents can persist $\sim 10^{3}~\mbox{secs}$
at very low temperatures. After this time the tiny $\sigma_{xx}$ gradually
allows the radial charge polarization to dissipate. We can think of the
azimuthal currents as gradually spiraling outwards due to the Hall angle
(between current and electric field) being very slightly less than $90^{\circ}$
(by $\sim 10^{-13}$).

We have shown that the random impurity potential (and by implication Anderson
localization) is a necessary condition for Hall plateaus to occur, but we have
not yet understood precisely how this novel behavior comes about. That is our
next task.

\section{Classical and Semi-Classical Dynamics}

\subsection{Classical Approximation}

The classical equations of motion for an electron of charge $-e$ moving in two
dimensions under the influence of the Lorentz force $\frac{-e}{c}\vec{v}\times
\vec{B}$ caused by a magnetic field $\vec{B} = B\hat{z}$ are
\begin{eqnarray}
m \ddot{x} &=& -\frac{eB}{c} \dot{y} \label{eq:9812-02}\\
m \ddot{y} &=& +\frac{eB}{c} \dot{x}.
\label{eq:lorentz}
\end{eqnarray}
The general solution of these equations corresponds to motion in a circle of
arbitrary radius $R$
\begin{equation}
\vec{r} = R\left(\cos(\omega_{c} t+\delta),\sin(\omega_{c} t+\delta)\right).
\end{equation}
Here $\delta$ is an arbitrary phase for the motion and
\begin{equation}
\omega_{c}\equiv \frac{eB}{mc}
\end{equation}
is known as the classical cyclotron frequency. Notice that the period of the
orbit is independent of the radius and that the tangential speed
\begin{equation}
v = R\omega_{c}
\end{equation}
controls the radius. A fast particle travels in a large circle but returns to
the starting point in the same length of time as a slow particle which
(necessarily) travels in a small circle. The motion is thus \textit{isochronous}
much like that of a harmonic oscillator whose period is independent of the
amplitude of the motion. This apparent analogy is not an accident as we shall
see when we study the Hamiltonian (which we will need for the full quantum
solution).

Because of some subtleties involving distinctions between canonical and
mechanical momentum in the presence of a magnetic field, it is worth reviewing
the formal Lagrangian and Hamiltonian approaches to this problem. The above
classical equations of motion follow from the Lagrangian
\begin{equation}
\mathcal{L} = \frac{1}{2}m\dot{x}^{\mu}\dot{x}^{\mu} - \frac{e}{c}\dot{x}^{\mu}
A^{\mu},
\label{eq:classicalLagrangian}
\end{equation}
where $\mu=1,2$ refers to $x$ and $y$ respectively and $\vec{A}$ is the vector
potential evaluated at the position of the particle. (We use the Einstein
summation convention throughout this discussion.) Using
\begin{equation}
\frac{\delta \mathcal{L}}{\delta x^{\nu}} = -\frac{e}{c} \dot{x}^{\mu}\,
\partial_{\nu} A^{\mu}
\end{equation}
and
\begin{equation}
\frac{\delta \mathcal{L}}{\delta \dot{x}^{\nu}} = m\dot{x}^{\nu} -\frac{e}{c}
A^{\nu}
\end{equation}
the Euler-Lagrange equation of motion becomes
\begin{equation}
m \ddot{x}^{\nu} = -\frac{e}{c}\left[\partial_{\nu} A^{\mu} - \partial_{\mu}
A^{\nu}\right]\dot{x}^{\mu}.
\label{eq:euler-lagrange}
\end{equation}
Using
\begin{eqnarray}
\vec{B} &=& \vec\nabla\times\vec{A}\\
B^{\alpha} &=& \epsilon^{\alpha\beta\gamma}\partial_{\beta} A^{\gamma}
\end{eqnarray}
shows that this is equivalent to eqs.~(\ref{eq:9812-02}--\ref{eq:lorentz}).

Once we have the Lagrangian we can deduce the canonical momentum
\begin{eqnarray}
p^{\mu} &\equiv& \frac{\delta \mathcal{L}}{\delta \dot{x}^{\mu}}\nonumber\\
&=& m\dot{x}^{\mu} -\frac{e}{c}A^{\mu},
\end{eqnarray}
and the Hamiltonian
\begin{eqnarray}
H[\vec{p},\vec{x}] &\equiv& \dot{x}^{\mu} p^{\mu} - \mathcal{L}(\dot{\vec{x}},
\vec{x})\nonumber\\
&=& \frac{1}{2m} \left(p^{\mu} + \frac{e}{c}A^{\mu}\right) \left(p^{\mu} +
\frac{e}{c}A^{\mu}\right).
\end{eqnarray}
(Recall that the Lagrangian is canonically a function of the positions and
velocities while the Hamiltonian is canonically a function of the positions and
momenta). The quantity
\begin{equation}
p_{\mathrm{mech}}^{\mu} \equiv p^{\mu} + \frac{e}{c}A^{\mu}
\end{equation}
is known as the \textit{mechanical} momentum.  Hamilton's equations of motion
\begin{eqnarray}
\dot{x}^{\mu} &=& \frac{\partial H}{\partial p^{\mu}} =
\frac{1}{m}p_{\mathrm{mech}}^{\mu} \label{eq:9812-03}\\
\dot{p}^{\mu} &=& -\frac{\partial H}{\partial x^{\mu}} = -\frac{e}{mc}
\left(p^{\nu} + \frac{e}{c}A^{\nu}\right)\partial_{\mu} A^{\nu}
\label{eq:hamilton}
\end{eqnarray}
show that it is the mechanical momentum, not the canonical momentum, which is
equal to the usual expression related to the velocity
\begin{equation}
p_{\mathrm{mech}}^{\mu} = m \dot{x}^{\mu}.
\label{eq:mechanical}
\end{equation}
Using Hamilton's equations of motion we can recover Newton's law for the Lorentz
force given in eq.~(\ref{eq:euler-lagrange}) by simply taking a time derivative
of $\dot{x}^{\mu}$ in eq.~(\ref{eq:9812-03}) and then using
eq.~(\ref{eq:hamilton}).

The distinction between canonical and mechanical momentum can lead to confusion.
For example it is possible for the particle to have a finite velocity while
having zero (canonical) momentum! Furthermore the canonical momentum is
dependent (as we will see later) on the choice of gauge for the vector potential
and \textit{hence is not a physical observable}. The mechanical momentum, being
simply related to the velocity (and hence the current) is physically observable
and gauge invariant. The classical equations of motion only involve the curl of
the vector potential and so the particular gauge choice is not very important at
the classical level. We will therefore delay discussion of gauge choices until
we study the full quantum solution, where the issue is unavoidable.

\subsection{Semi-classical Approximation}

Recall that in the semi-classical approximation used in transport theory we
consider wave packets $\Psi_{\vec{R}(t),\vec{K}(t)}(\vec{r},t)$ made up of a
linear superposition of Bloch waves. These packets are large on the scale of the
de Broglie wavelength so that they have a well-defined central wave vector
$\vec{K}(t)$, but they are small on the scale of everything else (external
potentials, etc.) so that they simultaneously can be considered to have
well-defined mean position $R(t)$. (Note that $\vec{K}$ and $\vec R$ are
\textit{parameters} labeling the wave packet not arguments.) We then argue (and
will discuss further below) that the solution of the Schr\"{o}dinger equation in
this semiclassical limit gives a wave packet whose parameters $\vec{K}(t)$ and
$\vec{R}(t)$ obey the appropriate analog of the classical Hamilton equations of
motion
\begin{eqnarray}
\dot{R}^{\mu} &=& \frac{\partial \langle
\Psi_{\vec{R},\vec{K}}|H|\Psi_{\vec{R},\vec{K}}\rangle} {\partial \hbar K^{\mu}}\\
\hbar\dot{K}^{\mu} &=& -\frac{\partial \langle
\Psi_{\vec{R},\vec{K}}|H|\Psi_{\vec{R},\vec{K}}\rangle} {\partial R^{\mu}}.
\label{eq:semiclassical}
\end{eqnarray}
Naturally this leads to the same circular motion of the wave packet at the
classical cyclotron frequency discussed above. For weak fields and fast
electrons the radius of these circular orbits will be large compared to the size
of the wave packets and the semi-classical approximation will be valid. However
at strong fields, the approximation begins to break down because the orbits are
too small and because $\hbar\omega_{c}$ becomes a significant (large) energy.
Thus we anticipate that the semi-classical regime requires $\hbar\omega_{c} \ll
\Ef$, where $\Ef$ is the Fermi energy.

We have already seen hints that the problem we are studying is really a harmonic
oscillator problem. For the harmonic oscillator there is a characteristic energy
scale $\hbar\omega$ (in this case $\hbar\omega_{c}$) and a characteristic length
scale $\ell$ for the zero-point fluctuations of the position in the ground
state. The analog quantity in this problem is the so-called magnetic length
\begin{equation}
\ell \equiv \sqrt{\frac{\hbar c}{eB}} = \frac{257\hbox{\AA}}{\sqrt{\frac{B}{1
\mathrm{tesla}}}}.
\end{equation}
The physical interpretation of this length is that the area $2\pi\ell^{2}$
contains one quantum of magnetic flux $\Phi_{0}$ where\footnote{Note that in the
study of superconductors the flux quantum is defined with a factor of $2e$
rather than $e$ to account for the pairing of the electrons in the condensate.}
\begin{equation}
\Phi_{0} = \frac{hc}{e}.
\end{equation}
That is to say, the density of magnetic flux is
\begin{equation}
B = \frac{\Phi_{0}}{2\pi\ell^{2}}.
\end{equation}

To be in the semiclassical limit then requires that the Fermi wavelength be
small on the scale of the magnetic length so that $\kf\ell \gg 1$. This
condition turns out to be equivalent to $\hbar\omega_{c} \ll \Ef$ so they are
not separate constraints.

\boxedtext{\begin{exercise}
Use the Bohr-Sommerfeld quantization condition that the orbit have a
circumference containing an integral number of de Broglie wavelengths to find the
allowed orbits of a 2D electron moving in a uniform magnetic field. Show that
each successive orbit encloses precisely one additional quantum of flux in its
interior. Hint: It is important to make the distinction between the canonical
momentum (which controls the de Broglie wavelength) and the mechanical momentum
(which controls the velocity). The calculation is simplified if one uses the
symmetric gauge $\vec{A} = -\frac{1}{2}\vec{r} \times \vec{B}$ in which the
vector potential is purely azimuthal and independent of the azimuthal angle.
\label{ex:stateperflux}
\end{exercise}}

\section{Quantum Dynamics in Strong B Fields}

\begin{figure}
\centerline{\epsfysize=6cm
 \epsffile{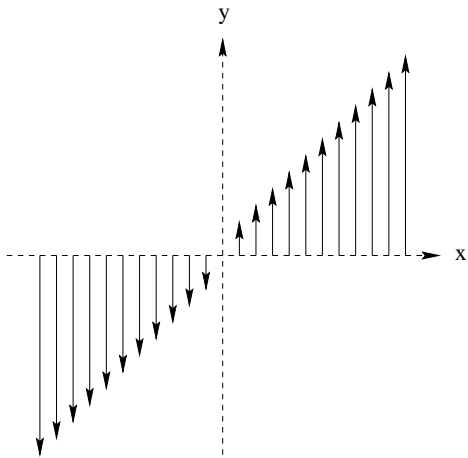}}
\caption[]{Illustration of the Landau gauge vector potential $\vec{A} =
xB\hat{y}$. The magnetic field is perfectly uniform, but the vector potential
has a preferred origin and orientation corresponding to the particular gauge
choice.}
\label{fig:gauge}
\end{figure}

Since we will be dealing with the Hamiltonian and the Schr\"{o}dinger equation,
our first order of business is to choose a gauge for the vector potential. One
convenient choice is the so-called Landau gauge:
\begin{equation}
\vec{A}(\vec{r}\,) = xB\hat{y}
\end{equation}
which obeys $\vec{\nabla} \times \vec{A} = B\hat{z}$. In this gauge the vector
potential points in the $y$ direction but varies only with the $x$ position, as
illustrated in fig.~(\ref{fig:gauge}).
Hence the system still has translation invariance in the $y$ direction. Notice
that the magnetic field (and hence all the physics) is translationally
invariant, but the Hamiltonian is not! (See exercise~\ref{ex:9801}.) This is one
of many peculiarities of dealing with vector potentials.

\boxedtext{\begin{exercise}
Show for the Landau gauge that even though the Hamiltonian is not invariant for
translations in the $x$ direction, the physics is still invariant since the
change in the Hamiltonian that occurs under translation is simply equivalent to
a gauge change. Prove this for any arbitrary gauge, assuming only that the
magnetic field is uniform.
\label{ex:9801}
\end{exercise}}

The Hamiltonian can be written in the Landau gauge as
\begin{equation}
H = \frac{1}{2m}\left( p_{x}^{2} + (p_{y} + \frac{eB}{c}x)^{2} \right)
\end{equation}
Taking advantage of the translation symmetry in the $y$ direction, let us attempt
a separation of variables by writing the wave function in the form
\begin{equation}
\psi_{k}(x,y) = e^{i ky} f_{k}(x).
\end{equation}
This has the advantage that it is an eigenstate of $p_{y}$ and hence we can make
the replacement $p_{y} \longrightarrow \hbar k$ in the Hamiltonian. After
separating variables we have the effective one-dimensional Schr\"{o}dinger
equation
\begin{equation}
h_{k} f_{k}(x) = \epsilon_{k} f_{k}(x),
\end{equation}
where
\begin{equation}
h_{k} \equiv \frac{1}{2m} p_{x}^{2}+\frac{1}{2m}\left(\hbar k +
\frac{eB}{c}x\right)^{2}.
\end{equation}
This is simply a one-dimensional displaced harmonic oscillator\footnote{Thus we
have arrived at the harmonic oscillator hinted at semiclassically, but
paradoxically it is only one-dimensional, not two. The other degree of freedom
appears (in this gauge) in the $y$ momentum.}
\begin{equation}
h_{k} = \frac{1}{2m} p_{x}^{2} + \frac{1}{2}m\omega_{c}^{2} \left(x +
k\ell^{2}\right)^{2}
\label{eq:1d-displaced}
\end{equation}
whose frequency is the classical cyclotron frequency and whose central position
$X_{k} = -k\ell^{2}$ is (somewhat paradoxically) determined by the $y$ momentum
quantum number. Thus for each plane wave chosen for the $y$ direction there will
be an entire family of energy eigenvalues
\begin{equation}
\epsilon_{kn} = (n+\frac{1}{2})\hbar\omega_{c}
\end{equation}
which depend only on $n$ are completely independent of the $y$ momentum
$\hbar k$. The corresponding (unnormalized) eigenfunctions are
\begin{equation}
\psi_{nk}(\vec{r}\,) = \frac{1}{\sqrt{L}} e^{iky}
H_{n}(x+k\ell^{2})e^{-\frac{1}{2\ell^{2}}(x+k\ell^{2})^{2}},
\label{eq:landaupsi}
\end{equation}
where $H_{n}$ is (as usual for harmonic oscillators) the $n$th Hermite
polynomial (in this case displaced to the new central position $X_{k}$).

\boxedtext{\begin{exercise}
Verify that eq.~(\ref{eq:landaupsi}) is in fact a solution of the
Schr\"{o}dinger equation as claimed.
\label{ex:9802}
\end{exercise}}

These harmonic oscillator levels are called Landau levels. Due to the lack of
dependence of the energy on $k$, the degeneracy of each level is enormous, as we
will now show. We assume periodic boundary conditions in the $y$ direction.
Because of the vector potential, it is \textit{impossible} to simultaneously
have periodic boundary conditions in the $x$ direction. However since the basis
wave functions are harmonic oscillator polynomials multiplied by strongly
converging gaussians, they rapidly vanish for positions away from the center
position $X_{0} = -k\ell^{2}$. Let us suppose that the sample is rectangular
with dimensions $L_{x},L_{y}$ and that the left hand edge is at $x=-L_{x}$ and
the right hand edge is at $x=0$. Then the values of the wavevector $k$ for which
the basis state is substantially inside the sample run from $k=0$ to
$k=L_{x}/\ell^{2}$. It is clear that the states at the left edge and the right
edge differ strongly in their $k$ values and hence periodic boundary conditions
are impossible.\footnote{The best one can achieve is so-called quasi-periodic
boundary conditions in which the phase difference between the left and right
edges is zero at the bottom and rises linearly with height, reaching $2\pi
N_{\Phi} \equiv L_{x}L_{y}/\ell^{2}$ at the top. The eigenfunctions with these
boundary conditions are elliptic theta functions which are linear combinations
of the gaussians discussed here. See the discussion by Haldane in
Ref.~\cite{SMGBOOK}.}

The total number of states in \textit{each} Landau level is then
\begin{equation}
N = \frac{L_{y}}{2\pi}\int_{0}^{L_{x}/\ell^{2}} dk =
\frac{L_{x}L_{y}}{2\pi\ell^{2}} = N_{\Phi}
\end{equation}
where
\begin{equation}
N_{\Phi}\equiv\frac{BL_{x}L_{y}}{\Phi_{0}}
\end{equation}
is the number of flux quanta penetrating the sample. Thus there is one state per
Landau level per flux quantum which is consistent with the semiclassical result
from Exercise~(\ref{ex:stateperflux}). Notice that even though the family of
allowed wavevectors is only one-dimensional, we find that the degeneracy of each
Landau level is extensive in the two-dimensional area. The reason for this is
that the spacing between wave vectors allowed by the periodic boundary
conditions $\Delta_{k} = \frac{2\pi}{L_{y}}$ \textit{decreases} while the
range of allowed wave vectors $[0,L_{x}/\ell^{2}]$ \textit{increases} with
increasing $L$. The reader may also worry that for very large samples, the
range of allowed values of $k$ will be so large that it will fall outside the
first Brillouin zone forcing us to include band mixing and the periodic lattice
potential beyond the effective mass approximation. This is not true however,
since the canonical momentum is a gauge dependent quantity. The value of $k$ in
any particular region of the sample can be made small by shifting the origin of
the coordinate system to that region (thereby making a gauge transformation).

The width of the harmonic oscillator wave functions in the $n$th Landau level is
of order $\sqrt{n}\ell$. This is microscopic compared to the system size, but
note that the spacing between the centers
\begin{equation}
\Delta = \Delta_{k}\ell^{2} = \frac{2\pi\ell^{2}}{L_{y}}
\end{equation}
is vastly smaller (assuming $L_{y} >> \ell$). Thus the supports of the different
basis states are strongly overlapping (but they are still orthogonal).

\boxedtext{\begin{exercise} 
Using the fact that the energy for the $n$th harmonic oscillator state is
$(n+\frac{1}{2})\hbar\omega_{c}$, present a semi-classical argument explaining
the result claimed above that the width of the support of the wave function
scales as $\sqrt{n}\ell$.
\label{ex:9803}
\end{exercise}

\begin{exercise}
Using the Landau gauge, construct a gaussian wave packet in the lowest Landau
level of the form
\[
\Psi(x,y) = \int_{-\infty}^{+\infty} a_{k}
e^{iky}e^{-\frac{1}{2\ell^{2}}(x+k\ell^{2})^{2}},
\]
choosing $a_{k}$ in such a way that the wave packet is localized as closely as
possible around some point $\vec{R}$. What is the smallest size wave packet that
can be constructed without mixing in higher Landau levels?
\label{ex:landaupacket}
\end{exercise}}

Having now found the eigenfunctions for an electron in a strong magnetic field
we can relate them back to the semi-classical picture of wave packets undergoing
circular cyclotron motion. Consider an initial semiclassical wave packet located
at some position and having some specified momentum. In the semiclassical limit
the mean energy of this packet will greatly exceed the cyclotron energy
$\frac{\hbar^{2}K^{2}}{2m}\gg\hbar\omega_{c}$ and hence it will be made up of a
linear combination of a large number of different Landau level states centered
around $\bar n = \frac{\hbar^{2}K^{2}}{2m\hbar\omega_{c}}$
\begin{equation}
\Psi(\vec{r},t) = \sum_{n}\int L_{y}\frac{dk}{2\pi}\, a_{n}(\vec{k})
\psi_{nk}(\vec{r}\,) e^{-i(n+\frac{1}{2})\omega_{c} t}.
\end{equation}
Notice that in an ordinary 2D problem at zero field, the complete set of plane
wave states would be labeled by a 2D continuous momentum label. Here we have
one discrete label (the Landau level index) and a 1D continuous labels (the $y$
wave vector). Thus the `sum' over the complete set of states is actually a
combination of a summation and an integration.

The details of the initial position and momentum are controlled by the
amplitudes $a_{n}(\vec{k})$. We can immediately see however, that since the
energy levels are exactly evenly spaced that the motion is exactly periodic:
\begin{equation}
\Psi(\vec{r},t+\frac{2\pi}{\omega_{c}}) = \Psi(\vec{r},t).
\end{equation}
If one works through the details, one finds that the motion is indeed circular
and corresponds to the expected semi-classical cyclotron orbit.

For simplicity we will restrict the remainder of our discussion to the lowest
Landau level where the (correctly normalized) eigenfunctions in the Landau gauge
are (dropping the index $n=0$ from now on):
\begin{equation}
\psi_{k}(\vec{r}\,) = \frac{1}{\sqrt{\pi^{1/2}L\ell}} e^{iky}
e^{-\frac{1}{2\ell^{2}}(x+k\ell^{2})^{2}}
\label{eq:lowlandaupsi}
\end{equation}
and every state has the same energy eigenvalue $\epsilon_{k}=
\frac{1}{2}\hbar\omega_{c}$.

We imagine that the magnetic field (and hence the Landau level splitting) is
very large so that we can ignore higher Landau levels. (There are some
subtleties here to which we will return.) Because the states are all degenerate,
any wave packet made up of any combination of the basis states will be a
stationary state. The total current will therefore be zero. We anticipate
however from semiclassical considerations that there should be some remnant of
the classical circular motion visible in the local current density. To see this
note that the expectation value of the current in the $k$th basis state is
\begin{equation}
\langle\vec{J}\,\rangle = -e\frac{1}{m} \left\langle\Psi_{k}\left|\left(\vec{p}
+ \frac{e}{c}\vec{A}\,\right)\right|\Psi_{k}\right\rangle.
\end{equation}
The $y$ component of the current is
\begin{eqnarray}
\langle J_{y}\rangle &=& -\frac{e}{m\pi^{1/2}\ell} \int
dx\,e^{-\frac{1}{2\ell^{2}}(x+k\ell^{2})^{2}} \left(\hbar k +
\frac{eB}{c}x\right) e^{-\frac{1}{2\ell^{2}}(x+k\ell^{2})^{2}}\nonumber\\
&=& -\frac{e\omega_{c}}{\pi^{1/2}\ell} \int
dx\,e^{-\frac{1}{\ell^{2}}(x+k\ell^{2})^{2}} \left(x+k\ell^{2}\right)
\end{eqnarray}
We see from the integrand that the current density is antisymmetric about the
peak of the gaussian and hence the total current vanishes. This antisymmetry
(positive vertical current on the left, negative vertical current on the right)
is the remnant of the semiclassical circular motion.

Let us now consider the case of a uniform electric field pointing in the $x$
direction and giving rise to the potential energy
\begin{equation}
V(\vec{r}\,) = +eEx.
\end{equation}
This still has translation symmetry in the $y$ direction and so our Landau gauge
choice is still the most convenient. Again separating variables we see that the
solution is nearly the same as before, except that the displacement of the
harmonic oscillator is slightly different. The Hamiltonian in
eq.~(\ref{eq:1d-displacedE}) becomes
\begin{equation}
h_{k} = \frac{1}{2m} p_{x}^{2} + \frac{1}{2}m\omega_{c}^{2} \left(x +
k\ell^{2}\right)^{2} +eEx.
\label{eq:1d-displacedE}
\end{equation}
Completing the square we see that the oscillator is now centered at the new
position
\begin{equation}
X_{k} = -k\ell^{2} - \frac{eE}{m\omega_{c}^{2}}
\end{equation}
and the energy eigenvalue is now linearly dependent on the particle's peak
position $X_{k}$ (and therefore linear in the $y$ momentum)
\begin{equation}
\epsilon_{k} = \frac{1}{2}\hbar\omega_{c} +eEX_{k} + \frac{1}{2}m\bar{v}^{2},
\label{eq:driftepsilon}
\end{equation}
where
\begin{equation}
\bar{v} \equiv -c\frac{E}{B}.
\end{equation}
Because of the shift in the peak position of the wavefunction, the perfect
antisymmetry of the current distribution is destroyed and there is a net current
\begin{equation}
\langle J_{y}\rangle = -e\bar{v}
\end{equation}
showing that $\bar{v}\hat{y}$ is simply the usual $c\vec{E} \times
\vec{B}/B^{2}$ drift velocity. This result can be derived either by explicitly
doing the integral for the current or by noting that the wave packet group
velocity is 
\begin{equation}
\frac{1}{\hbar}\frac{\partial \epsilon_{k}}{\partial k} =
\frac{eE}{\hbar}\frac{\partial X_{k}}{\partial k} = \bar{v}
\end{equation}
independent of the value of $k$ (since the electric field is a constant in this
case, giving rise to a strictly linear potential). Thus we have recovered the
correct kinematics from our quantum solution.

\begin{figure}
\centerline{\epsfxsize=10cm
 \epsffile{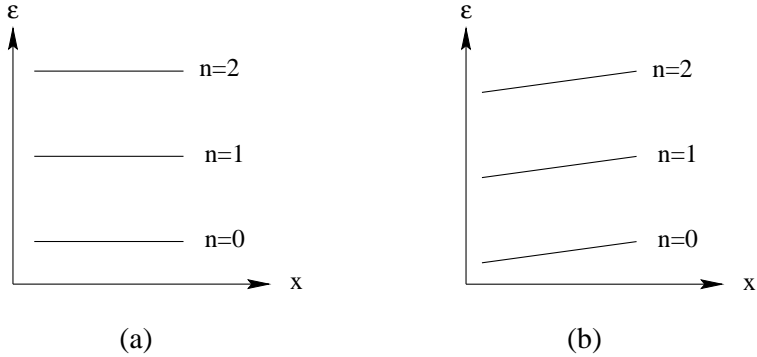}}
\caption[]{Illustration of electron Landau energy levels
$\left(n + \frac{1}{2}\right)\hbar\omega_{\mathrm{c}}$ vs.\ position $x_{k} =
-k\ell^{2}$. (a) Zero electric field case. (b) Case with finite electric field
pointing in the $+\hat{x}$ direction.}
\label{fig:LLtilt}
\end{figure}

It should be noted that the applied electric field `tilts' the Landau levels in
the sense that their energy is now linear in position as illustrated in
fig.(\ref{fig:LLtilt}).
This means that there are degeneracies between different Landau level states
because different kinetic energy can compensate different potential energy in
the electric field. Nevertheless, we have found the exact eigenstates (i.e., the
stationary states). It is not possible for an electron to decay into one of the
other degenerate states because they have different canonical momenta. If however
disorder or phonons are available to break translation symmetry, then these
decays become allowed and dissipation can appear. The matrix elements for such
processes are small if the electric field is weak because the degenerate states
are widely separated spatially due to the small tilt of the Landau levels.

\boxedtext{\begin{exercise}
It is interesting to note that the exact eigenstates in the presence of the
electric field can be viewed as displaced oscillator states in the original
(zero $E$ field) basis. In this basis the displaced states are linear
combinations of all the Landau level excited states of the same $k$. Use
first-order perturbation theory to find the amount by which the $n=1$ Landau
level is mixed into the $n=0$ state. Compare this with the exact amount of
mixing computed using the exact displaced oscillator state. Show that the two
results agree to first order in $E$. Because the displaced state is a linear
combination of more than one Landau level, it can carry a finite current. Give
an argument, based on perturbation theory why the amount of this current is
inversely proportional to the $B$ field, but is independent of the mass of the
particle. Hint: how does the mass affect the Landau level energy spacing and the
current operator?
\label{ex:9804}
\end{exercise}}

\section{IQHE Edge States}

\begin{figure}
\centerline{\epsfxsize=10cm
 \epsffile{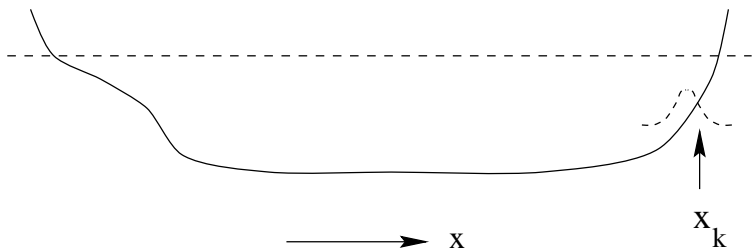}}
\caption[]{Illustration of a smooth confining potential which varies only in the
$x$ direction. The horizontal dashed line indicates the equilibrium fermi level.
The dashed curve indicates the wave packet envelope $f_{k}$ which is displaced
from its nominal position $x_{k} \equiv -k\ell^{2}$ by the slope of the
potential.}
\label{fig:hallbarLL}
\end{figure}

\begin{figure}
\centerline{\epsfysize=6cm
 \epsffile{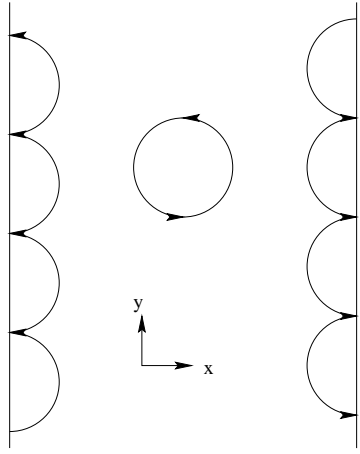}}
\caption[]{Semi-classical view of skipping orbits at the fermi level at the two
edges of the sample where the confining electric field causes $\vec{E} \times
\vec{B}$ drift. The circular orbit illustrated in the center of the sample
carries no net drift current if the local electric field is zero.}
\label{fig:skipping}
\end{figure}

Now that we understand drift in a uniform electric field, we can consider the
problem of electrons confined in a Hall bar of finite width by a non-uniform
electric field. For simplicity, we will consider the situation where the
potential $V(x)$ is smooth on the scale of the magnetic length, but this is not
central to the discussion. If we assume that the system still has translation
symmetry in the $y$ direction, the solution to the Schr\"{o}dinger equation must
still be of the form
\begin{equation}
\psi(x,y) = \frac{1}{\sqrt{L_{y}}}e^{iky}f_{k}(x).
\label{eq:psiHallbar}
\end{equation}
The function $f_{k}$ will no longer be a simple harmonic wave function as we
found in the case of the uniform electric field. However we can anticipate that
$f_{k}$ will still be peaked near (but in general not precisely at) the point
$X_{k}\equiv -k\ell^{2}$. The eigenvalues $\epsilon_{k}$ will no longer be
precisely linear in $k$ but will still reflect the kinetic energy of the
cyclotron motion plus the local potential energy $V(X_{k})$ (plus small
corrections analogous to the one in eq.~(\ref{eq:driftepsilon})). This is
illustrated in fig.~(\ref{fig:hallbarLL}).
We see that the group velocity 
\begin{equation}
\vec{v}_{k} = \frac{1}{\hbar}\frac{\partial \epsilon_{k}}{\partial k} \hat{y}
\label{eq:groupv}
\end{equation}
has the opposite sign on the two edges of the sample. This means that in the
ground state there are edge currents of opposite sign flowing in the sample. The
semi-classical interpretation of these currents is that they represent `skipping
orbits' in which the circular cyclotron motion is interrupted by collisions with
the walls at the edges as illustrated in fig.~(\ref{fig:skipping}).

One way to analyze the Hall effect in this system is quite analogous to the
Landauer picture of transport in narrow wires \cite{KaneFisher,Buttiker}. The
edge states play the role of the left and right moving states at the two fermi
points. Because (as we saw earlier) momentum in a magnetic field corresponds to
position, the edge states are essentially real space realizations of the fermi
surface. A Hall voltage drop across the sample in the $x$ direction corresponds
to a difference in electrochemical potential between the two edges. Borrowing
from the Landauer formulation of transport, we will choose to apply this in the
form of a chemical potential difference and ignore any changes in electrostatic
potential.\footnote{This has led to various confusions in the literature. If
there is an electrostatic potential gradient then some of the net Hall current
may be carried in the bulk rather than at the edges, but the final answer is the
same. In any case, the essential part of the physics is that the only place
where there are low lying excitations is at the edges.} What this does is
increase the number of electrons in skipping orbits on one edge of the sample
and/or decrease the number on the other edge. Previously the net current due to
the two edges was zero, but now there is a net Hall current. To calculate this
current we have to add up the group velocities of all the occupied states 
\begin{equation}
I = -\frac{e}{L_{y}}\int_{-\infty}^{+\infty} dk\frac{L_{y}}{2\pi}\,
\frac{1}{\hbar}\frac{\partial\epsilon_{k}}{\partial k} n_{k},
\end{equation}
where for the moment we assume that in the bulk, only a single Landau level is
occupied and $n_{k}$ is the probability that state $k$ in that Landau level is
occupied. Assuming zero temperature and noting that the integrand is a perfect
derivative, we have
\begin{equation}
I = -\frac{e}{h}\int_{\mu_{R}}^{\mu_{L}} d\epsilon = -\frac{e}{h}\left[\mu_{L} -
\mu_{R}\right].
\end{equation}
(To understand the order of limits of integration, recall that as $k$ increases,
$X_{k}$ decreases.) The definition of the Hall voltage drop is\footnote{To get
the signs straight here, note that an increase in chemical potential brings in
more electrons. This is equivalent to a more positive voltage and hence a more
negative potential energy $-eV$. Since $H-\mu N$ enters the thermodynamics,
electrostatic potential energy and chemical potential move the electron density
oppositely. $V$ and $\mu$ thus have the same sign of effect because electrons
are negatively charged.}
\begin{equation}
(+e)V_{H} \equiv (+e)\left[V_{R} - V_{L}\right] = \left[\mu_{R} -
\mu_{L}\right].
\end{equation}
Hence 
\begin{equation}
I = -\nu \frac{e^{2}}{h}V_{H},
\end{equation}
where we have now allowed for the possibility that $\nu$ different Landau levels
are occupied in the bulk and hence there are $\nu$ separate edge channels
contributing to the current. This is the analog of having $\nu$ `open' channels
in the Landauer transport picture. In the Landauer picture for an ordinary wire,
we are considering the longitudinal voltage drop (and computing $\sigma_{xx}$),
while here we have the Hall voltage drop (and are computing $\sigma_{xy}$). The
analogy is quite precise however because we view the right and left movers as
having distributions controlled by separate chemical potentials. It just happens
in the QHE case that the right and left movers are physically separated in such
a way that the voltage drop is transverse to the current. Using the above result
and the fact that the current flows at right angles to the voltage drop we have
the desired results
\begin{eqnarray}
\sigma_{xx} &=& 0\\
\sigma_{xy} &=& -\nu\frac{e^{2}}{h},
\end{eqnarray}
with the quantum number $\nu$ being an integer.

So far we have been ignoring the possible effects of disorder. Recall that for a
single-channel one-dimensional wire in the Landauer picture, a disordered region
in the middle of the wire will reduce the conductivity to
\begin{equation}
I = \frac{e^{2}}{h} |T|^{2},
\end{equation}
where $|T|^{2}$ is the probability for an electron to be transmitted through the
disordered region. The reduction in transmitted current is due to \textit{back
scattering}. Remarkably, in the QHE case, the back scattering is essentially
zero in very wide samples. To see this note that in the case of the Hall bar,
scattering into a backward moving state would require transfer of the electron
from one edge of the sample to the other since the edge states are spatially
separated. For samples which are very wide compared to the magnetic length (more
precisely, to the Anderson localization length) the matrix element for this is
exponentially small. In short, there can be nothing but forward scattering. An
incoming wave given by eq.~(\ref{eq:psiHallbar}) can only be transmitted in the
forward direction, at most suffering a simple phase shift $\delta_{k}$
\begin{equation}
\psi_{\mathrm{out}}(x,y) = \frac{1}{\sqrt{L_{y}}}e^{i\delta_{k}}e^{iky}f_{k}(x).
\end{equation}
This is because no other states of the same energy are available. If the
disorder causes Landau level mixing at the edges to occur (because the confining
potential is relatively steep) then it is possible for an electron in one edge
channel to scatter into another, but the current is still going in the same
direction so that there is no reduction in overall transmission probability. It
is this \textit{chiral} (unidirectional) nature of the edge states which is
responsible for the fact that the Hall conductance is correctly quantized
independent of the disorder.

Disorder will broaden the Landau levels in the bulk and provide a reservoir of
(localized) states which will allow the chemical potential to vary smoothly with
density. These localized states will not contribute to the transport and so the
Hall conductance will be quantized over a plateau of finite width in $B$ (or
density) as seen in the data. Thus obtaining the universal value of quantized
Hall conductance to a precision of $10^{-10}$ does not require fine tuning the
applied $B$ field to a similar precision.

The localization of states in the bulk by disorder is an essential part of the
physics of the quantum Hall effect as we saw when we studied the role of
translation invariance. We learned previously that in zero magnetic field all
states are (weakly) localized in two dimensions. In the presence of a quantizing
magnetic field, most states are strongly localized as discussed above. However
if all states were localized then it would be impossible to have a quantum phase
transition from one QHE plateau to the next. To understand how this works it is
convenient to work in a semiclassical percolation picture to be described below.

\boxedtext{\begin{exercise}
Show that the number of edge channels whose energies lie in the gap between two
Landau levels scales with the length $L$ of the sample, while the number of bulk
states scales with the area. Use these facts to show that the range of magnetic
field in which the chemical potential lies in between two Landau levels scales
to zero in the thermodynamic limit. Hence finite width quantized Hall plateaus
can not occur in the absence of disorder that produces a reservoir of localized
states in the bulk whose number is proportional to the area.
\label{ex:edgecount}
\end{exercise}}

\section{Semiclassical Percolation Picture}

Let us consider a smooth random potential caused, say, by ionized silicon donors
remotely located away from the 2DEG in the GaAs semiconductor host. We take the
magnetic field to be very large so that the magnetic length is small on the
scale over which the potential varies. In addition, we ignore the Coulomb
interactions among the electrons. 

What is the nature of the eigenfunctions in this random potential? We have
learned how to solve the problem exactly for the case of a constant electric
field and know the general form of the solution when there is translation
invariance in one direction. We found that the wave functions were plane waves
running along lines of constant potential energy and having a width
perpendicular to this which is very small and on the order of the magnetic
length. The reason for this is the discreteness of the kinetic energy in a
strong magnetic field. It is impossible for an electron stuck in a given Landau
level to continuously vary its kinetic energy. Hence energy conservation
restricts its motion to regions of constant potential energy. In the limit of
infinite magnetic field where Landau level mixing is completely negligible, this
confinement to lines of constant potential becomes exact (as the magnetic length
goes to zero).

We are led to the following somewhat paradoxical picture. The strong magnetic
field should be viewed as putting the system in the quantum limit in the sense
that $\hbar\omega_{c}$ is a very large energy (comparable to $\Ef$). At the same
time (if one assumes the potential is smooth) one can argue that since the
magnetic length is small compared to the scale over which the random potential
varies, the system is in a semi-classical limit where small wave packets (on the
scale of $\ell$) follow classical $\vec{E}\times \vec{B}$ drift trajectories.

{}From this discussion it then seems very reasonable that in the presence of a
smooth random potential, with no particular translation symmetry, the
eigenfunctions will live on contour lines of constant energy on the random
energy surface. Thus low energy states will be found lying along contours in
deep valleys in the potential landscape while high energy states will be found
encircling `mountain tops' in the landscape. Naturally these extreme states will
be strongly localized about these extrema in the potential.

\boxedtext{\begin{exercise}
Using the Lagrangian for a charged particle in a magnetic field with a scalar potential
$V(\vec{r}\,)$, consider the high field limit by setting the mass to zero (thereby sending
the quantum cyclotron energy to infinity).  
\begin{enumerate}
\item Derive the classical equations of motion from the Lagrangian and show that they yield
simple $\vec{E} \times \vec{B}$ drift along isopotential contours.
\item Find the momentum conjugate to the coordinate $x$ and show that (with an appropriate
gauge choice) it is the coordinate $y$: 
\begin{equation}
p_{x} = -\frac{\hbar}{\ell^{2}}y
\end{equation}
 so that we have the strange
commutation relation
\begin{equation}
[x,y]=-i\ell^{2}.
\end{equation}
\end{enumerate}
In the infinite field limit where $\ell\rightarrow 0$ the coordinates commute and we recover
the semi-classical result in which effectively point particles drift along isopotentials.  
\label{ex:semiclassical}
\end{exercise}}

\begin{figure}
\centerline{\epsfysize=6cm
 \epsffile{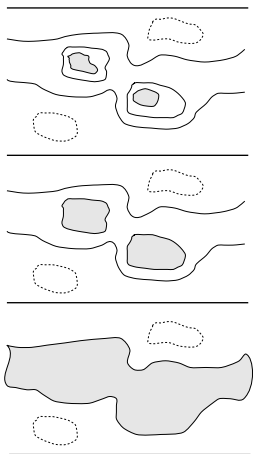}}
\caption[]{Contour map of a smooth random landscape. Closed dashed lines
indicate local mountain peaks. Closed solid lines indicate valleys. From top to
bottom, the gray filled areas indicate the increasing `sea level' whose
shoreline finally percolates from one edge of the sample to the other (bottom
panel). The particle-hole excitations live along the shoreline and become
gapless when the shoreline becomes infinite in extent.}
\label{fig:sealevel}
\end{figure}

To understand the nature of states at intermediate energies, it is useful to
imagine gradually filling a random landscape with water as illustrated in
fig.~(\ref{fig:sealevel}).
In this analogy, sea level represents the chemical potential for the electrons.
When only a small amount of water has been added, the water will fill the
deepest valleys and form small lakes. As the sea level is increased the lakes
will grow larger and their shorelines will begin to take on more complex shapes.
At a certain critical value of sea level a phase transition will occur in which
the shoreline percolates from one side of the system to the other. As the sea
level is raised still further, the ocean will cover the majority of the land and
only a few mountain tops will stick out above the water. The shore line will no
longer percolate but only surround the mountain tops. 

As the sea level is raised still higher additional percolation transitions will
occur successively as each successive Landau level passes under water. If Landau
level mixing is small and the disorder potential is symmetrically distributed
about zero, then the critical value of the chemical potential for the $n$th
percolation transition will occur near the center of the $n$th Landau level
\begin{equation}
\mu^{*}_{n} = (n+\frac{1}{2})\hbar\omega_{c}.
\end{equation}

This percolation transition corresponds to the transition between quantized Hall
plateaus. To see why, note that when the sea level is below the percolation
point, most of the sample is dry land. The electron gas is therefore insulating.
When sea level is above the percolation point, most of the sample is covered
with water. The electron gas is therefore connected throughout the majority of
the sample and a quantized Hall current can be carried. Another way to see this
is to note that when the sea level is above the percolation point, the confining
potential will make a shoreline along the full length of each edge of the
sample. The edge states will then carry current from one end of the sample to
the other.

We can also understand from this picture why the dissipative conductivity
$\sigma_{xx}$ has a sharp peak just as the plateau transition occurs. (Recall
the data in fig.~(\ref{fig:qhedata}).) Away from the critical point the
circumference of any particular patch of shoreline is finite. The period of the
semiclassical orbit around this is finite and hence so is the quantum level
spacing. Thus there are small energy gaps for excitation of states across these
real-space fermi levels. Adding an infinitesimal electric field will only weakly
perturb these states due to the gap and the finiteness of the perturbing matrix
element which will be limited to values on the order of $\sim eED$ where $D$ is
the diameter of the orbit. If however the shoreline percolates from one end of
the sample to the other then the orbital period diverges and the gap vanishes.
An infinitesimal electric field can then cause dissipation of energy. 

\begin{figure}
\centerline{\epsfysize=6cm
 \epsffile{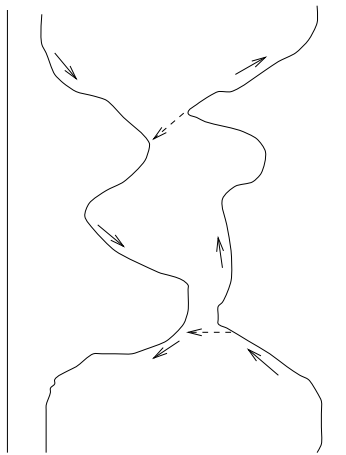}}
\caption[]{Illustration of edge states that wander deep into the bulk as the
quantum Hall localization transition is approached from the conducting side.
Solid arrows indicate the direction of drift along the isopotential lines.
Dashed arrows indicate quantum tunneling from one semi-classical orbit (edge
state) to the other. This backscattering localizes the eigenstates and prevents
transmission through the sample using the `edge' states (which become part of
the bulk localized states).}
\label{fig:zigzag}
\end{figure}

Another way to see this is that as the percolation level is approached from
above, the edge states on the two sides will begin taking detours deeper and
deeper into the bulk and begin communicating with each other as the localization
length diverges and the shoreline zig zags throughout the bulk of the sample.
Thus electrons in one edge state can be back scattered into the other edge
states and ultimately reflected from the sample as illustrated in
fig.~(\ref{fig:zigzag}).

Because the random potential broadens out the Landau level density of states,
the quantized Hall plateaus will have finite width. As the chemical potential is
varied in the regime of localized states in between the Landau level peaks, only
the occupancy of localized states is changing. Hence the transport properties
remain constant until the next percolation transition occurs. It is important to
have the disorder present to produce this finite density of states and to
localize those states.

It is known that as the (classical) percolation point is approached in two
dimensions, the characteristic size (diameter) of the shoreline orbits diverges
like
\begin{equation}
\xi \sim |\delta|^{-4/3},
\end{equation}
where $\delta$ measures the deviation of the sea level from its critical value.
The shoreline structure is not smooth and in fact its circumference diverges
with a larger exponent $7/3$ showing that these are highly ramified fractal
objects whose circumference scales as the $7/4$th power of the diameter.

So far we have assumed that the magnetic length is essentially zero. That is, we
have ignored the fact that the wave function support extends a small distance
transverse to the isopotential lines. If two different orbits with the same
energy pass near each other but are classically disconnected, the particle can
still tunnel between them if the magnetic length is finite. This quantum
tunneling causes the localization length to diverge faster than the classical
percolation model predicts. Numerical simulations find that the localization
length diverges like \cite{huckestein,chalker,HuoBhatt,DasSarmalocalizationbook}
\begin{equation}
\xi \sim |\delta|^{-\nu}
\end{equation}
where the exponent $\nu$ (not to be confused with the Landau level filling
factor!) has a value close (but probably not exactly equal to) $7/3$ rather than
the $4/3$ found in classical percolation. It is believed that this exponent is
universal and independent of Landau level index.

Experiments on the quantum critical behavior are quite difficult but there is
evidence \cite{Wei}, at least in selected samples which show good scaling, that
$\nu$ is indeed close to $7/3$ (although there is some recent controversy on this point.
\cite{shahar}) and that the conductivity tensor is universal at the critical point.
\cite{HuoBhatt,Yanguniversality}
Why Coulomb interactions that are present in
real samples do not spoil agreement with the numerical simulations is something
of a mystery at the time of this writing. For a discussion of some of these issues see
\cite{sondhiRMP97}.

\section{Fractional QHE}

Under some circumstances of weak (but non-zero) disorder, quantized Hall
plateaus appear which are characterized by simple rational fractional quantum
numbers. For example, at magnetic fields three times larger than those at which
the $\nu=1$ integer filling factor plateau occurs, the lowest Landau level is
only 1/3 occupied. The system ought to be below the percolation threshold and
hence be insulating. Instead a robust quantized Hall plateau is observed
indicating that electrons can travel through the sample and that (since
$\sigma_{xx}\longrightarrow 0$) there is an excitation gap. This novel and quite
unexpected physics is controlled by Coulomb repulsion between the electrons. It
is best understood by first ignoring the disorder and trying to discover the
nature of the special correlated many-body ground state into which the electrons
condense when the filling factor is a rational fraction.

For reasons that will become clear later, it is convenient to analyze the
problem in a new gauge
\begin{equation}
\vec{A} = -\frac{1}{2} \vec{r} \times \vec{B}
\end{equation} 
known as the symmetric gauge. Unlike the Landau gauge which preserves
translation symmetry in one direction, the symmetric gauge preserves rotational
symmetry about the origin. Hence we anticipate that angular momentum (rather
than $y$ linear momentum) will be a good quantum number in this gauge.

For simplicity we will restrict our attention to the lowest Landau level only
and (simply to avoid some awkward minus signs) change the sign of the $B$ field:
$\vec{B} = -B\hat{z}$. With these restrictions, it is not hard to show that the
solutions of the free-particle Schr\"{o}dinger equation having definite angular
momentum are
\begin{equation}
\varphi_{m} = \frac{1}{\sqrt{2\pi\ell^{2} 2^{m} m!} } z^{m}
e^{-\frac{1}{4}|z|^{2}}
\label{eq:symmgauge}
\end{equation}
where $z=(x+iy)/\ell$ is a dimensionless complex number representing the
position vector $\vec{r} \equiv (x,y)$ and $m\ge 0$ is an integer. 

\boxedtext{\begin{exercise}
Verify that the basis functions in eq.~(\ref{eq:symmgauge}) do solve the
Schr\"{o}dinger equation in the absence of a potential and do lie in the lowest
Landau level. Hint: Rewrite the kinetic energy in such a way that $\vec{p}\cdot
\vec{A}$ becomes $\vec{B}\cdot \vec{L}$.
\label{ex:ssymmgauge}
\end{exercise}}                     

The angular momentum of these basis states is of course $\hbar m$. If we
restrict our attention to the lowest Landau level, then there exists only one
state with any given angular momentum and only non-negative values of $m$ are
allowed. This `handedness' is a result of the chirality built into the problem
by the magnetic field.

It seems rather peculiar that in the Landau gauge we had a continuous
one-dimensional family of basis states for this two-dimensional problem. Now we
find that in a different gauge, we have a discrete one dimensional label for the
basis states! Nevertheless, we still end up with the correct density of states
per unit area. To see this note that the peak value of $|\varphi_{m}|^{2}$
occurs at a radius of $R_{\mathrm{peak}}=\sqrt{2m\ell^{2}}$. The area
$2\pi\ell^{2} m$ of a circle of this radius contains $m$ flux quanta. Hence we
obtain the standard result of one state per Landau level per quantum of flux
penetrating the sample.

Because all the basis states are degenerate, any linear combination of them is
also an allowed solution of the Schr\"{o}dinger equation. Hence any function of
the form \cite{girvinjach}
\begin{equation}
\Psi(x,y) = f(z) e^{-\frac{1}{4}|z|^{2}}
\end{equation}
is allowed so long as $f$ is \textit{analytic} in its argument. In particular,
arbitrary polynomials of any degree $N$ 
\begin{equation}
f(z) = \prod_{j=1}^{N} (z-Z_{j})
\end{equation}
are allowed (at least in the thermodynamic limit) and are conveniently defined
by the locations of their $N$ zeros $\{Z_{j}; j=1,2,\dots,N\}$. 

Another useful solution is the so-called coherent state which is a particular
infinite order polynomial
\begin{equation}
f_{\lambda}(z) \equiv \frac{1}{\sqrt{2\pi\ell^{2}}} e^{\frac{1}{2}\lambda^{*}
z}e^{-\frac{1}{4}\lambda^{*}\lambda} .
\end{equation}
The wave function using this polynomial has the property that it is a narrow
gaussian wave packet centered at the position defined by the complex number
$\lambda$. Completing the square shows that the probability density is given by
\begin{equation}
|\Psi_{\lambda}|^{2} = |f_{\lambda}|^{2} e^{-\frac{1}{2} |z|^{2}} =
\frac{1}{2\pi\ell^{2}}e^{-\frac{1}{2}|z-\lambda|^{2}}
\end{equation}
This is the smallest wave packet that can be constructed from states within the
lowest Landau level. The reader will find it instructive to compare this
gaussian packet to the one constructed in the Landau gauge in
exercise~(\ref{ex:landaupacket}).

Because the kinetic energy is completely degenerate, the effect of Coulomb
interactions among the particles is nontrivial. To develop a feel for the
problem, let us begin by solving the two-body problem. Recall that the standard
procedure is to take advantage of the rotational symmetry to write down a
solution with the relative angular momentum of the particles being a good
quantum number and then solve the Schr\"{o}dinger equation for the radial part
of the wave function. Here we find that the analyticity properties of the wave
functions in the lowest Landau level greatly simplifies the situation. If we
know the angular behavior of a wave function, analyticity uniquely defines the
radial behavior. Thus for example for a single particle, knowing that the
angular part of the wave function is $e^{im\theta}$, we know that the full wave
function is guaranteed to uniquely be $r^{m}e^{im\theta}e^{-\frac{1}{4}|z|^{2}}
= z^{m}e^{-\frac{1}{4}|z|^{2}}$. 

Consider now the two body problem for particles with relative angular momentum
$m$ and center of mass angular momentum $M$. The \textit{unique} analytic wave
function is (ignoring normalization factors)
\begin{equation}
\Psi_{mM}(z_{1},z_{2}) = (z_{1} - z_{2})^{m} (z_{1}+z_{2})^{M}
e^{-\frac{1}{4}(|z_{1}|^{2} + |z_{2}|^{2})}.
\end{equation}
If $m$ and $M$ are non-negative integers, then the prefactor of the exponential
is simply a polynomial in the two arguments and so is a state made up of linear
combinations of the degenerate one-body basis states $\varphi_{m}$ given in
eq.~(\ref{eq:symmgauge}) and therefore lies in the lowest Landau level. Note that
if the particles are spinless fermions then $m$ must be odd to give the correct
exchange symmetry. Remarkably, this is the exact (neglecting Landau level
mixing) solution for the Schr\"{o}dinger equation for \textit{any} central
potential $V(|z_{1}-z_{2}|)$ acting between the two particles.\footnote{Note
that neglecting Landau level mixing is a poor approximation for strong
potentials $V \gg \hbar\omega_{c}$ unless they are very smooth on the scale of
the magnetic length.} We do not need to solve any radial equation because of the
powerful restrictions due to analyticity. There is only one state in the (lowest
Landau level) Hilbert space with relative angular momentum $m$ and center of
mass angular momentum $M$. Hence (neglecting Landau level mixing) it is an exact
eigenstate of \textit{any} central potential. $\Psi_{mM}$ is the exact answer
independent of the Hamiltonian!

\begin{figure}
\centerline{\epsfxsize=10cm
 \epsffile{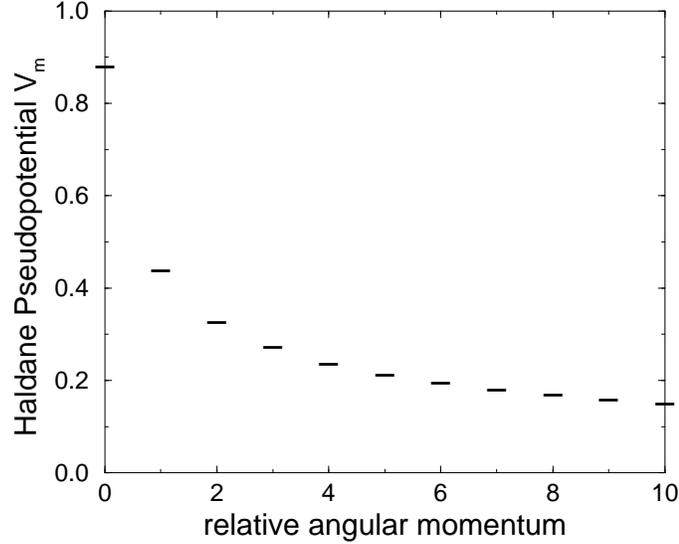}}
\caption[]{The Haldane pseudopotential $V_{m}$ vs. relative angular momentum $m$
for two particles interacting via the Coulomb interaction. Units are
${e^{2}}/{\epsilon\ell}$, where $\epsilon$ is the dielectric constant of the
host semiconductor and the finite thickness of the quantum well has been
neglected.}
\label{fig:pseudopots}
\end{figure}

The corresponding energy eigenvalue $v_{m}$ is independent of $M$ and is
referred to as the $m$th Haldane pseudopotential
\begin{equation}
v_{m} = \frac{\left\langle mM|V|mM\right\rangle}{\left\langle
mM|mM\right\rangle}.
\end{equation}
The Haldane pseudopotentials for the repulsive Coulomb potential are shown in
fig.~(\ref{fig:pseudopots}).
These discrete energy eigenstates represent bound states of the repulsive
potential. If there were no magnetic field present, a repulsive potential would
of course have only a continuous spectrum with no discrete bound states. However
in the presence of the magnetic field, there are effectively bound states
because the kinetic energy has been quenched. Ordinarily two particles that have
a lot of potential energy because of their repulsive interaction can fly apart
converting that potential energy into kinetic energy. Here however (neglecting
Landau level mixing) the particles all have fixed kinetic energy. Hence
particles that are repelling each other are stuck and can not escape from each
other. One can view this semi-classically as the two particles orbiting each
other under the influence of $\vec{E}\times\vec{B}$ drift with the Lorentz force
preventing them from flying apart. In the presence of an attractive potential
the eigenvalues change sign, but of course the eigenfunctions remain exactly the
same (since they are unique)!

The fact that a repulsive potential has a discrete spectrum for a pair of
particles is (as we will shortly see) the central feature of the physics
underlying the existence of an excitation gap in the fractional quantum Hall
effect. One might hope that since we have found analyticity to uniquely
determine the two-body eigenstates, we might be able to determine many-particle
eigenstates exactly. The situation is complicated however by the fact that for
three or more particles, the various relative angular momenta
$L_{12},L_{13},L_{23}$, etc.\ do not all commute. Thus we can not write down
general exact eigenstates. We will however be able to use the analyticity to
great advantage and make exact statements for certain special cases.

\boxedtext{\begin{exercise}
Express the exact lowest Landau level two-body eigenstate
\[
\Psi(z_{1},z_{2}) = (z_{1} - z_{2})^{3}\; 
e^{-\frac{1}{4}\left\{|z_{1}|^{2}+|z_{2}|^{2}\right\}}
\]
in terms of the basis of all possible two-body Slater determinants.
\label{ex:9805}
\end{exercise}

\begin{exercise}
Verify the claim that the Haldane pseudopotential $v_{m}$ is independent of the
center of mass angular momentum $M$.
\label{ex:pseudopot1}
\end{exercise}

\begin{exercise}
Evaluate the Haldane pseudopotentials for the Coulomb potential
$\frac{e^{2}}{\epsilon r}$. Express your answer in units of
$\frac{e^{2}}{\epsilon\ell}$. For the specific case of $\epsilon=10$ and
$B=10$T, express your answer in Kelvin.
\label{ex:pseudopot2}
\end{exercise}

\begin{exercise}
Take into account the finite thickness of the quantum well by assuming that the
one-particle basis states have the form
\begin{displaymath}
\psi_{m}(z,s) = \varphi_{m}(z)\Phi(s),
\end{displaymath}
where $s$ is the coordinate in the direction normal to the quantum well. Write
down (but do not evaluate) the formal expression for the Haldane
pseudo-potentials in this case. Qualitatively describe the effect of finite
thickness on the values of the different pseudopotentials for the case where the
well thickness is approximately equal to the magnetic length.
\label{ex:pseudopot3}
\end{exercise}}                     

\subsection{The $\nu=1$ many-body state}
\label{subsec:nuequalsone}

\begin{figure}
\centerline{\epsfxsize=6cm
 \epsffile{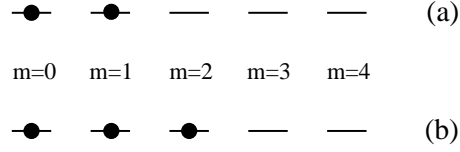}}
\caption[]{Orbital occupancies for the maximal density filled Landau level state
with (a) two particles and (b) three particles. There are no particle labels
here. In the Slater determinant wave function, the particles are labeled but a
sum is taken over all possible permutations of the labels in order to
antisymmetrize the wave function.}
\label{fig:slater2}
\end{figure}

So far we have found the one- and two-body states. Our next task is to write
down the wave function for a fully filled Landau level. We need to find
\begin{equation}
\psi[z] = f[z]\;
e^{-\frac{1}{4}\sum_{j} |z_{j}|^{2}}
\end{equation}
where $[z]$ stands for $(z_{1},z_{2},\ldots ,z_{N})$ and
$f$ is a polynomial representing the Slater determinant with all states
occupied. Consider the simple example of two particles. We want one particle in
the orbital $\varphi_{0}$ and one in $\varphi_1$, as illustrated schematically in
fig.~(\ref{fig:slater2}a).
Thus (again ignoring normalization)
\begin{eqnarray}
f[z] &=& \left|\begin{array}{cc}
(z_{1})^{0} & (z_{2})^{0}\\
(z_{1})^{1} & (z_{2})^{1}\end{array}\right| = (z_{1})^{0} (z_{2})^{1} -
(z_{2})^{0} (z_{1})^{1}\nonumber\\
 &=& (z_{2} - z_{1})
\end{eqnarray}
This is the lowest possible order polynomial that is antisymmetric. For the case
of three particles we have (see fig.~(\ref{fig:slater2}b))
\begin{eqnarray}
f[z] &=& \left|\begin{array}{ccc}
(z_{1})^{0} & (z_{2})^{0} & (z_{3})^{0}\\
(z_{1})^{1} & (z_{2})^{1} & (z_{3})^{1}\\
(z_{1})^{2} & (z_{2})^{2} & (z_{3})^{2}\end{array}\right| = z_{2}z_{3}^{2} -
z_{3}z_{2}^{2} - z_{1}^{1}z_{3}^{2} + z_{3}^{1}z_{1}^{2} + z_{1}z_{2}^{2} -
z_{2}^{1}z_{1}^{2}\nonumber\\
&=& -(z_{1} - z_{2}) (z_{1} - z_{3}) (z_{2} - z_{3})\nonumber\\
&=& -\prod_{i<j}^{3} (z_{i} - z_{j}) \label{eq:1282}
\end{eqnarray}
This form for the Slater determinant is known as the Vandermonde polynomial. The
overall minus sign is unimportant and we will drop it.

The single Slater determinant to fill the first $N$ angular momentum states is
a simple generalization of eq.~(\ref{eq:1282})
\begin{equation}
f_{N}[z] = \prod_{i<j}^{N} (z_{i} - z_{j}).
\end{equation}
To prove that this is true for general $N$, note that the polynomial is fully
antisymmetric and the highest power of any $z$ that appears is $z^{N-1}$. Thus
the highest angular momentum state that is occupied is $m = N - 1$. But since
the antisymmetry guarantees that no two particles can be in the same state, all
$N$ states from $m = 0$ to $m = N - 1$ must be occupied. This proves that we
have the correct Slater determinant.

\boxedtext{\begin{exercise}
Show carefully that the Vandermonde polynomial for $N$ particles is in fact
totally antisymmetric.
\label{ex:9806}
\end{exercise}}

One can also use induction to show that the Vandermonde polynomial is the correct
Slater determinant by writing
\begin{equation}
f_{N+1}(z) = f_{N}(z)\; \prod_{i=1}^{N} (z_{i} - z_{N+1})
\end{equation}
which can be shown to agree with the result of expanding the determinant of the
$(N + 1) \times (N + 1)$ matrix in terms of the minors associated with the
$(N+1)$st row or column. 

Note that since the Vandermonde polynomial corresponds to the filled Landau
level it is the unique state having the maximum density and hence is an exact
eigenstate for any form of interaction among the particles (neglecting Landau
level mixing and ignoring the degeneracy in the center of mass angular
momentum).

The (unnormalized) probability distribution for particles in the filled Landau
level state is
\begin{equation}
\left|\Psi[z]\right|^{2} = \prod_{i<j}^{N} |z_{i} - z_{j}|^{2}\; e^{-\frac{1}{2}
\sum_{j=1}^{N} |z_{j}|^{2}}.
\end{equation}
 This seems like
a rather complicated object about which it is hard to make any useful
statements. It is clear that the polynomial term tries to keep the particles
away from each other and gets larger as the particles spread out. It is also
clear that the exponential term is small if the particles spread out too much.
Such simple questions as, `Is the density uniform?', seem hard to answer
however.

It turns out that there is a beautiful analogy to plasma physics developed by
R.\ B.\ Laughlin which sheds a great deal of light on the nature of this many
particle probability distribution. To see how this works, let us pretend that
the norm of the wave function
\begin{equation}
Z \equiv \int d^{2}z_{1} \ldots \int d^{2}z_{N}\; |\psi_{[z]}|^{2}
\end{equation}
is the partition function of a classical statistical mechanics problem with
Boltzmann weight
\begin{equation}
\left|\Psi[z]\right|^{2} = e^{-\beta U_{\mathrm{class}}}
\end{equation}
where $\beta \equiv \frac{2}{m}$ and
\begin{equation}
U_{\mathrm{class}} \equiv m^{2} \sum_{i<j} \left(-\ln{|z_{i} - z_{j}|}\right) +
\frac{m}{4} \sum_{k} |z_{k}|^{2}.
\label{eq:Uclass}
\end{equation}
(The parameter $m=1$ in the present case but we introduce it for later
convenience.) It is perhaps not obvious at first glance that we have made
tremendous progress, but we have. This is because $U_{\mathrm{class}}$ turns out
to be the potential energy of a fake classical one-component plasma of particles
of charge $m$ in a uniform (`jellium') neutralizing background. Hence we can
bring to bear well-developed intuition about classical plasma physics to study
the properties of $|\Psi|^{2}$. Please remember however that all the statements
we make here are about a particular wave function. There are no actual
long-range logarithmic interactions in the quantum Hamiltonian for which this
wave function is the approximate groundstate.

To understand this, let us first review the electrostatics of charges in three
dimensions. For a charge $Q$ particle in 3D, the surface integral of the electric
field on a sphere of radius $R$ surrounding the charge obeys
\begin{equation}
\int d\vec{A} \cdot \vec{E} = 4\pi Q.
\end{equation}
Since the area of the sphere is $4\pi R^{2}$ we deduce
\begin{eqnarray}
\vec{E}(\vec{r}\,) &=& Q \frac{\hat{r}}{r^{2}}\\
\varphi(\vec{r}\,) &=& \frac{Q}{r}
\end{eqnarray}
and 
\begin{equation}
\vec{\nabla} \cdot \vec{E} = -\nabla^{2}\varphi = 4\pi Q\; \delta^{3}(\vec{r}\,)
\end{equation}
where $\varphi$ is the electrostatic potential. Now consider a two-dimensional
world where all the field lines are confined to a plane (or equivalently
consider the electrostatics of infinitely long charged rods in 3D). The
analogous equation for the line integral of the normal electric field on a
\textit{circle} of radius $R$ is
\begin{equation}
\int d\vec{s} \cdot \vec{E} = 2\pi Q
\end{equation}
where the $2\pi$ (instead of $4\pi$) appears because the circumference of a
circle is $2\pi R$ (and is analogous to $4\pi R^{2}$). Thus we find
\begin{eqnarray}
\vec{E}(\vec{r}\,) &=& \frac{Q\hat{r}}{r}\\
\varphi(\vec{r}\,) &=& Q \left(-\ln{\frac{r}{r_{0}}}\right)
\end{eqnarray}
and the 2D version of Poisson's equation is
\begin{equation}
\vec{\nabla} \cdot \vec{E} = -\nabla^{2}\varphi = 2\pi Q\; \delta^{2}(\vec{r}\,).
\end{equation}
Here $r_{0}$ is an arbitrary scale factor whose value is immaterial since it
only shifts $\varphi$ by a constant.

We now see why the potential energy of interaction among a group of objects with
charge $m$ is 
\begin{equation}
U_{0} = m^{2} \sum_{i<j} \left(-\ln{|z_{i} - z_{j}|}\right).
\end{equation}
(Since $z = (x + iy)/\ell$ we are using $r_{0} = \ell$.) This explains the first
term in eq.~(\ref{eq:Uclass}).

To understand the second term notice that
\begin{equation}
-\nabla^{2}\; \frac{1}{4}|z|^{2} = -\frac{1}{\ell^{2}} = 2\pi\rho_{\mathrm{B}}
\label{eq:9812-04}
\end{equation}
where
\begin{equation}
\rho_{\mathrm{B}} \equiv -\frac{1}{2\pi\ell^{2}}.
\end{equation}
Eq.~(\ref{eq:9812-04}) can be interpreted as Poisson's equation and tells us
that $\frac{1}{4}|z|^{2}$ represents the electrostatic potential of a constant
charge density $\rho_{\mathrm{B}}$. Thus the second term in
eq.~(\ref{eq:Uclass}) is the energy of charge $m$ objects interacting with this
negative background.

Notice that $2\pi\ell^{2}$ is precisely the area containing one quantum of flux.
Thus the background charge density is precisely $B/\Phi_{0}$, the density of
flux in units of the flux quantum.

The very long range forces in this fake plasma cost huge (fake) `energy' unless
the plasma is everywhere locally neutral (on length scales larger than the Debye
screening length which in this case is comparable to the particle spacing). In
order to be neutral, the density $n$ of particles must obey
\begin{eqnarray}
nm + \rho_{\mathrm{B}} &=& 0\\
\Rightarrow\qquad n &=& \frac{1}{m}\; \frac{1}{2\pi\ell^{2}}
\end{eqnarray}
since each particle carries (fake) charge $m$. For our filled Landau level with
$m=1$, this is of course the correct answer for the density since every
single-particle state is occupied and there is one state per quantum of flux.

\begin{figure}
\centerline{\epsfysize=5cm
 \epsffile{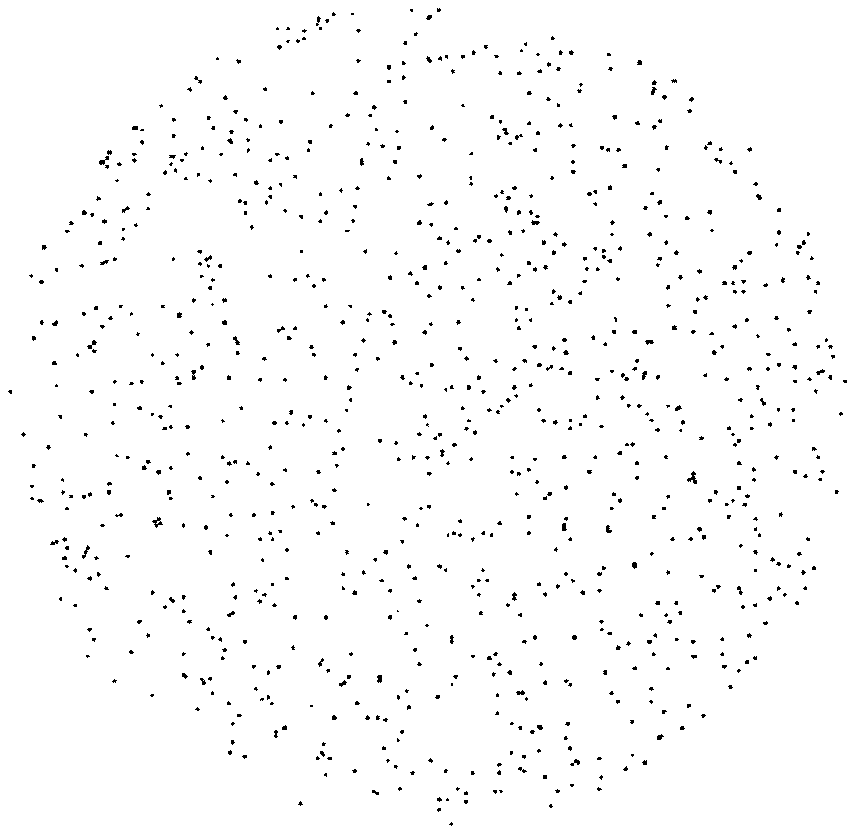}\hfill \epsfysize=5cm\epsffile{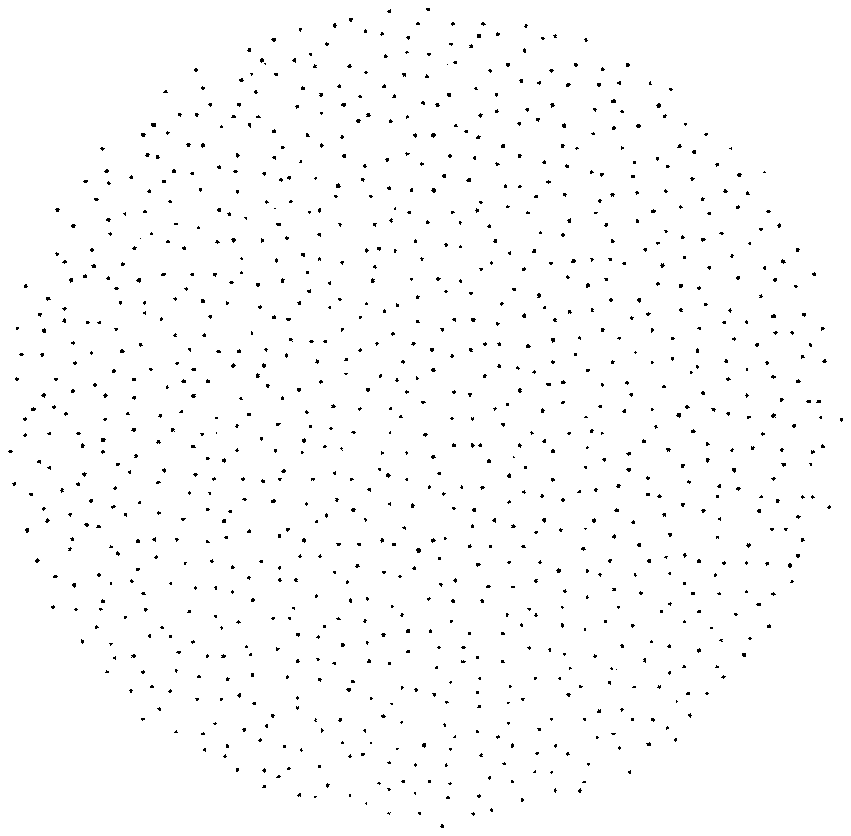}}
\caption[]{Comparison of typical configurations for a completely uncorrelated
(Poisson) distribution of 1000 particles (left panel) to the distribution given
by the Laughlin wave function for $m=3$ (right panel). The latter is a snapshot
taken during a Monte Carlo simulation of the distribution. The Monte Carlo
procedure consists of proposing a random trial move of one of the particles to a
new position. If this move increases the value of $|\Psi|^2$ it is always
accepted. If the move decreases the value of $|\Psi|^2$ by a factor $p$, then
the move is accepted with probability $p$. After equilibration of the plasma by
a large number of such moves one finds that the configurations generated are
distributed according to $|\Psi|^2$. (After R. B. Laughlin, Chap. 7 in
\cite{SMGBOOK}.)}
\label{fig:snapshot}
\end{figure}

We again emphasize that the energy of the fake plasma has nothing to do with the
quantum Hamiltonian and the true energy. The plasma analogy is merely a
statement about this particular choice of wave function. It says that the square
of the wave function is very small (because $U_\mathrm{class}$ is large) for
configurations in which the density deviates even a small amount from
$1/(2\pi\ell^2)$. The electrons can in principle be found anywhere, but the
overwhelming probability is that they are found in a configuration which is
locally random (liquid-like) but with negligible density fluctuations on long
length scales. We will discuss the nature of the typical configurations again
further below in connection with fig.~(\ref{fig:snapshot}).

When the fractional quantum Hall effect was discovered, Robert Laughlin realized
that one could write down a many-body variational wave function at filling
factor $\nu = 1/m$ by simply taking the $m$th power of the polynomial that
describes the filled Landau level
\begin{equation}
f_{N}^{m}[z] = \prod_{i<j}^{N} (z_{i} - z_{j})^{m}.
\end{equation}
In order for this to remain analytic, $m$ must be an integer. To preserve the
antisymmetry $m$ must be restricted to the odd integers. In the plasma analogy
the particles now have fake charge $m$ (rather than unity) and the density of
electrons is $n = \frac{1}{m}\; \frac{1}{2\pi\ell^{2}}$ so the Landau level
filling factor $\nu = \frac{1}{m} = \frac{1}{3}, \frac{1}{5}, \frac{1}{7}$,
etc.\ (Later on, other wave functions were developed to describe more general
states in the hierarchy of rational fractional filling factors at which
quantized Hall plateaus were observed
\cite{SMGBOOK,TAPASHbook,DasSarmabook,stonebook,sciam}.)

The Laughlin wave function naturally builds in good correlations among the
electrons because each particle sees an $m$-fold zero at the positions of all
the other particles. The wave function vanishes extremely rapidly if any two
particles approach each other, and this helps minimize the expectation value of
the Coulomb energy.

Since the kinetic energy is fixed we need only concern ourselves with the
expectation value of the potential energy for this variational wave function.
Despite the fact that there are no adjustable variational parameters (other than
$m$ which controls the density) the Laughlin wave functions have proven to be
very nearly exact for almost any realistic form of repulsive interaction. To
understand how this can be so, it is instructive to consider a model for which
this wave function actually is the exact ground state. Notice that the form of
the wave function guarantees that every pair of particles has relative angular
momentum greater than or equal to $m$. One should not make the mistake of
thinking that every pair has relative angular momentum precisely equal to $m$.
This would require the spatial separation between particles to be very nearly
the same for every pair, which is of course impossible.

Suppose that we write the Hamiltonian in terms of the Haldane pseudopotentials
\begin{equation}
V = \sum_{m'=0}^{\infty}\; \sum_{i<j} v_{m'}\; P_{m'}(ij)
\end{equation}
where $P_{m}(ij)$ is the projection operator which selects out states in which
particles $i$ and $j$ have relative angular momentum $m$. If $P_{m'}(ij)$ and
$P_{m^{''}}(jk)$ commuted with each other things would be simple to solve, but
this is not the case. However if we consider the case of a `hard-core potential'
defined by $v_{m'} = 0$ for $m' \geq m$, then clearly the $m$th Laughlin state
is an exact, zero energy eigenstate
\begin{equation}
V\psi_{m}[z] = 0. \label{eq:12103}
\end{equation}
This follows from the fact that
\begin{equation}
P_{m'}(ij)\psi_{m} = 0
\end{equation}
for any $m' < m$ since every pair has relative angular momentum of at least $m$.

Because the relative angular momentum of a pair can change only in discrete
(even integer) units, it turns out that this hard core model has an excitation
gap. For example for $m = 3$, any excitation out of the Laughlin ground state
necessarily weakens the nearly ideal correlations by forcing at least one pair
of particles to have relative angular momentum $1$ instead of $3$ (or larger).
This costs an excitation energy of order $v_{1}$.

This excitation gap is essential to the existence of dissipationless
$(\sigma_{xx} = \rho_{xx} = 0)$ current flow. In addition this gap means that
the Laughlin state is stable against perturbations. Thus the difference between
the Haldane pseudopotentials $v_{m}$ for the Coulomb interaction and the
pseudopotentials for the hard core model can be treated as a small perturbation
(relative to the excitation gap). Numerical studies show that for realistic
pseudopotentials the overlap between the true ground state and the Laughlin
state is extremely good.

To get a better understanding of the correlations built into the Laughlin wave
function it is useful to consider the snapshot in fig.~(\ref{fig:snapshot})
which shows a typical configuration of particles in the Laughlin ground state
(obtained from a Monte Carlo sampling of $|\psi|^{2}$) compared to a random
(Poisson) distribution. Focussing first on the large scale features we see that
density fluctuations at long wavelengths are severely suppressed in the
Laughlin state. This is easily understood in terms of the plasma analogy and the
desire for local neutrality. A simple estimate for the density fluctuations
$\rho_{\vec{q}}$ at wave vector $\vec{q}$ can be obtained by noting that the
fake plasma potential energy can be written (ignoring a constant associated with
self-interactions being included)
\begin{equation}
U_{\mathrm{class}} = \frac{1}{2L^{2}} \sum_{\vec{q}\neq 0} \frac{2\pi
m^{2}}{q^{2}}\; \rho_{\vec{q}}\rho_{-\vec{q}}
\end{equation}
where $L^{2}$ is the area of the system and $\frac{2\pi}{q^{2}}$ is the Fourier
transform of the logarithmic potential (easily derived from
$\nabla^{2}\left(-\ln{(r})\right) = -2\pi\; \delta^{2}(\vec{r}\,)\,$). At long
wavelengths $(q^{2} \ll n)$ it is legitimate to treat $\rho_{\vec{q}}$ as a
collective coordinate of an elastic continuum. The distribution $e^{-\beta
U_{\mathrm{class}}}$ of these coordinates is a gaussian and so obeys (taking
into account the fact that $\rho_{-\vec{q}} = (\rho_{\vec{q}})^{*}$)
\begin{equation}
\langle\rho_{\vec{q}}\rho_{-\vec{q}}\rangle = L^{2} \frac{q^{2}}{4\pi m}.
\label{eq:12105}
\end{equation}
We clearly see that the long-range (fake) forces in the (fake) plasma strongly
suppress long wavelength density fluctuations. We will return more to this point
later when we study collective density wave excitations above the Laughlin
ground state.

The density fluctuations on short length scales are best studied in real space.
The radial correlation $g(r)$ function is a convenient object to consider. $g(r)$
tells us the density at $r$ given that there is a particle at
the origin 
\begin{equation}
g(r) = \frac{N(N-1)}{n^{2}Z} \int d^{2}z_{3} \ldots \int d^{2}z_{N}\; \left|\psi
(0,r,z_{3},\ldots ,z_{N})\right|^{2}
\end{equation}
where $Z \equiv \langle\psi|\psi\rangle$, $n$ is the density (assumed uniform)
and the remaining factors account for all the different pairs of particles that
could contribute. The factors of density are included in the denominator so that
$\lim_{r\rightarrow\infty} g(r) = 1$.

\begin{figure}
\centerline{\epsfxsize=5cm
 \epsffile{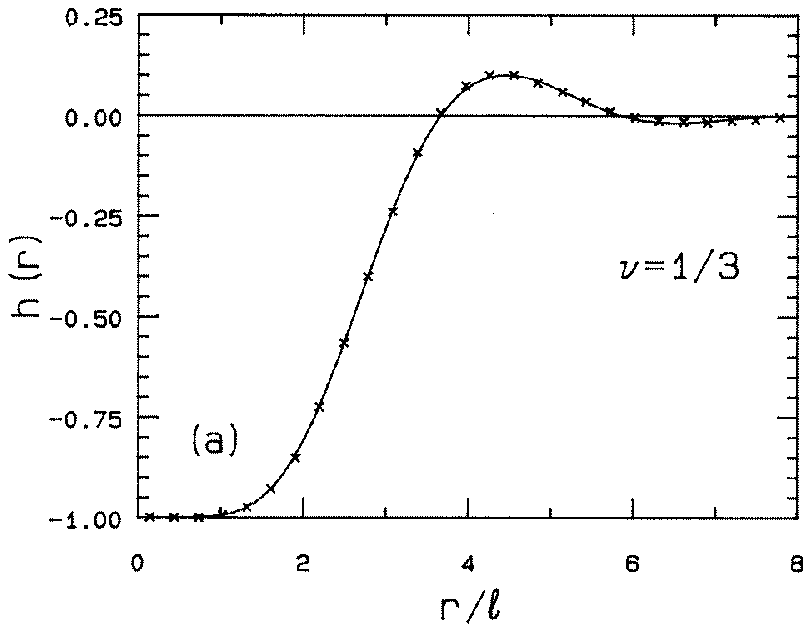}\hfill \epsfxsize=5cm\epsffile{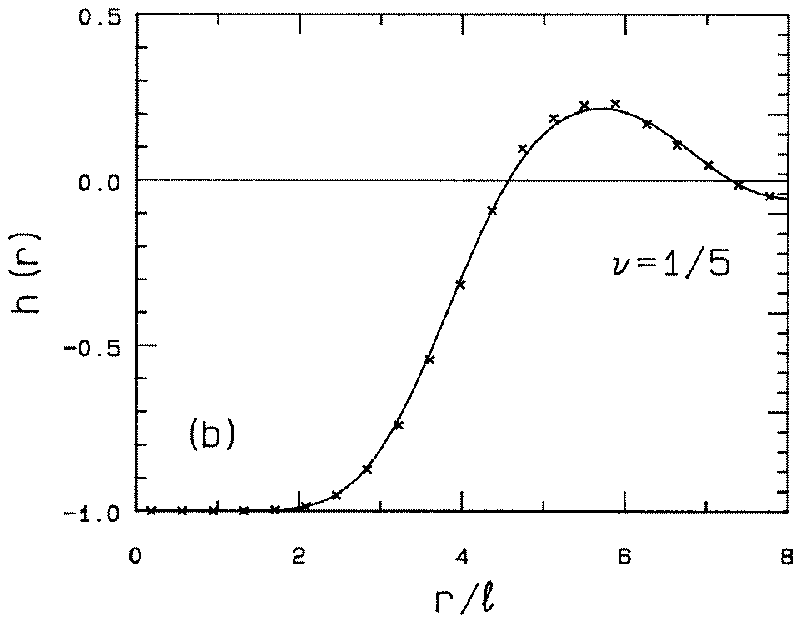}}
\caption[]{Plot of the two-point correlation function $h(r) \equiv 1-g(r)$ for
the Laughlin plasma with $\nu^{-1}= m = 3$ (left panel) and $m=5$ (right panel).
Notice that, unlike the result for $m=1$ given in eq.~(\ref{eq:12108}), $g(r)$
exhibits the oscillatory behavior characteristic of a strongly coupled plasma
with short-range solid-like local order.}
\label{fig:2pointqhe}
\end{figure}

Because the $m=1$ state is a single Slater determinant $g(z)$ can be computed
exactly
\begin{equation}
g(z) = 1 - e^{-\frac{1}{2}|z|^{2}}. 
\label{eq:12108}
\end{equation}
Fig.~(\ref{fig:2pointqhe})
shows numerical estimates of $h(r) \equiv 1-g(r)$ for the cases $m=3$ and $5$.
Notice that for the $\nu = 1/m$ state $g(z) \sim |z|^{2m}$ for small distances.
Because of the strong suppression of density fluctuations at long wavelengths,
$g(z)$ converges exponentially rapidly to unity at large distances. For $m > 1$,
$g$ develops oscillations indicative of solid-like correlations and, the plasma
actually freezes\footnote{That is, Monte Carlo simulation of $|\Psi|^2$ shows
that the particles are most likely to be found in a crystalline configuration
which breaks translation symmetry. Again we emphasize that this is a statement
about the Laughlin variational wave function, not necessarily a statement about
what the electrons actually do. It turns out that for $m \ge\,\sim 7$ the Laughlin
wave function is no longer the best variational wave function. One can write
down wave functions describing Wigner crystal states which have lower
variational energy than the Laughlin liquid.} at $m \approx 65$. The Coulomb
interaction energy can be expressed in terms of $g(z)$ as\footnote{This
expression assumes a strictly zero thickness electron gas. Otherwise one must
replace $\frac{e^2}{\epsilon|z|}$ by $\frac{e^{2}}{\epsilon}
\int_{-\infty}^{+\infty}ds \frac{\left| F(s)\right|^{2}}{\sqrt{|z|^{2} +
s^{2}}}$ where $F$ is the wavefunction factor describing the quantum well bound
state.}
\begin{equation}
\frac{\langle\psi|V|\psi\rangle}{\langle\psi|\psi\rangle} = \frac{nN}{2} \int
d^{2}z\; \frac{e^{2}}{\epsilon|z|}\; \left[ g(z) - 1\right] \label{eq:12109}
\end{equation}
where the $(-1)$ term accounts for the neutralizing background and $\epsilon$ is
the dielectric constant of the host semiconductor. We can interpret $g(z) - 1$
as the density of the `exchange-correlation hole' surrounding each particle.

The correlation energies per particle for $m=3$ and $5$ are \cite{levesque84}
\begin{equation}
\frac{1}{N}\;
\frac{\langle\psi_{3}|V|\psi_{3}\rangle}{\langle\psi_{3}|\psi_{3}\rangle} =
-0.4100\pm 0.0001
\end{equation}
and
\begin{equation}
\frac{1}{N}\;
\frac{\langle\psi_{5}|V|\psi_{5}\rangle}{\langle\psi_{5}|\psi_{5}\rangle} =
-0.3277\pm 0.0002
\end{equation}
in units of $e^{2}/\epsilon\ell$ which is $\approx 161~\mathrm{K}$ for $\epsilon
= 12.8$ (the value in GaAs), $B = 10\mathrm{T}$. For the filled Landau level
($m=1$) the exchange energy is $-\sqrt{\frac{\pi}{8}}$ as can be seen from
eqs.~(\ref{eq:12108}) and (\ref{eq:12109}).

\boxedtext{\begin{exercise}
Find the radial distribution function for a one-dimensional spinless free
electron gas of density $n$ by writing the ground state wave function as a
single Slater determinant and then integrating out all but two of the
coordinates. Use this first quantization method even if you already know how to
do this calculation using second quantization. Hint: Take advantage of the
following representation of the determinant of a $N \times N$ matrix $M$ in terms
of permutations $P$ of $N$ objects.
\[
\mathrm{Det}\; M = \sum_{P} (-1)^{P} \prod_{j=1}^{N} M_{jP_{j}}.
\]
\label{ex:9807}
\end{exercise}

\begin{exercise}
Using the same method derive eq.~(\ref{eq:12108}).
\label{ex:9808}
\end{exercise}}

\section{Neutral Collective Excitations}

So far we have studied one particular variational wave function and found that
it has good correlations built into it as graphically illustrated in
Fig.~\ref{fig:snapshot}. To further bolster the case that this wave function
captures the physics of the fractional Hall effect we must now demonstrate that
there is finite energy cost to produce excitations above this ground state. In
this section we will study the neutral collective excitations. We will examine
the charged excitations in the next section.

It turns out that the neutral excitations are phonon-like excitations similar to
those in solids and in superfluid helium. We can therefore use a simple
modification of Feynman's theory of the excitations in superfluid helium
\cite{feynman72,GMP}.

By way of introduction let us start with the simple harmonic oscillator. The
ground state is of the form
\begin{equation}
\psi_{0}(x) \sim e^{-\alpha x^{2}}.
\end{equation}
Suppose we did not know the excited state and tried to make a variational
ansatz for it. Normally we think of the variational method as applying only to
ground states. However it is not hard to see that the first excited state energy
is given by
\begin{equation}
\epsilon_{1} = \mathrm{min}\,
\left\{\frac{\langle\psi|H|\psi\rangle}{\langle\psi|\psi\rangle}\right\}
\end{equation}
provided that we do the minimization over the set of states $\psi$ which are
constrained to be orthogonal to the ground state $\psi_{0}$. One simple way to
produce a variational state which is automatically orthogonal to the ground
state is to change the parity by multiplying by the first power of the
coordinate
\begin{equation}
\psi_{1}(x) \sim x\; e^{-\alpha x^{2}}. \label{eq:1201}
\end{equation}
Variation with respect to $\alpha$ of course leads (in this special case) to the
\textit{exact} first excited state.

With this background let us now consider the case of phonons in superfluid
${}^{4}\hbox{He}$. Feynman argued that because of the Bose statistics of the
particles, there are no low-lying single-particle excitations. This is in stark
contrast to a fermi gas which has a high density of low-lying excitations around
the fermi surface. Feynman argued that the only low-lying excitations in
${}^{4}\hbox{He}$ are collective density oscillations that are well-described by
the following family of variational wave functions (that has no adjustable
parameters) labeled by the wave vector
\begin{equation}
\psi_{\vec{k}} = \frac{1}{\sqrt{N}}\; \rho_{\vec{k}}\; \Phi_{0} \label{eq:1202}
\end{equation}
where $\Phi_{0}$ is the exact ground state and
\begin{equation}
\rho_{\vec{k}} \equiv \sum_{j=1}^{N} e^{-i\vec{k}\cdot\vec{r}_{j}}
\label{eq:1203}
\end{equation}
is the Fourier transform of the density. The physical picture behind this is
that at long wavelengths the fluid acts like an elastic continuum and
$\rho_{\vec{k}}$ can be treated as a generalized oscillator normal-mode
coordinate. In this sense eq.~(\ref{eq:1202}) is then analogous to
eq.~(\ref{eq:1201}). To see that $\psi_{\vec{k}}$ is orthogonal to the ground
state we simply note that
\begin{eqnarray}
\langle\Phi_{0}|\psi_{\vec{k}}\rangle &=& \frac{1}{\sqrt{N}}\;
\langle\Phi_{0}|\rho_{\vec{k}}|\Phi_{0}\rangle\nonumber\\
&=& \frac{1}{\sqrt{N}} \int d^{3}R\; e^{-i\vec{k}\cdot\vec{R}}\;
\langle\Phi_{0}|\rho(\vec{r}\,)|\Phi_{0}\rangle . \label{eq:1204}
\end{eqnarray}
where
\begin{equation}
\rho(\vec{r}\,) \equiv \sum_{j=1}^{N} \delta^{3}(\vec{r}_{j} - \vec{R})
\end{equation}
is the density operator. If $\Phi_{0}$ describes a translationally invariant
liquid ground state then the Fourier transform of the mean density vanishes for
$k\neq 0$.

\begin{figure}
\centerline{\epsfxsize=6cm
 \epsffile{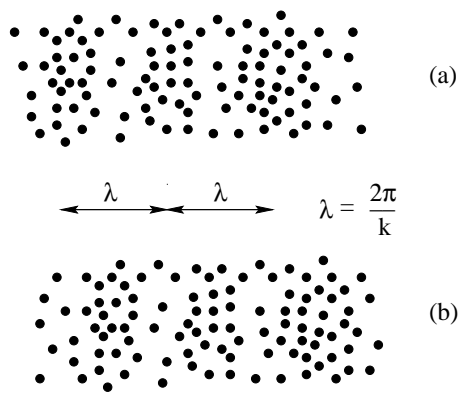}}
\caption[]{(a) Configuration of particles in which the Fourier transform of the
density at wave vector $k$ is non-zero. (b) The Fourier amplitude will have a
similar magnitude for this configuration but a different phase.}
\label{fig:densitywave}
\end{figure}

There are several reasons why $\psi_{\vec{k}}$ is a good variational wave
function, especially for small $k$. First, it contains the ground state as a
factor. Hence it contains all the special correlations built into the ground
state to make sure that the particles avoid close approaches to each other
without paying a high price in kinetic energy. Second, $\psi_{\vec{k}}$ builds
in the features we expect on physical grounds for a density wave. To see this,
consider evaluating $\psi_{\vec{k}}$ for a configuration of the particles like
that shown in fig.~(\ref{fig:densitywave}a)
which has a density modulation at wave vector $\vec{k}$. This is not a
configuration that maximizes $|\Phi_{0}|^{2}$, but as long as the density
modulation is not too large and the particles avoid close approaches,
$|\Phi_{0}|^{2}$ will not fall too far below its maximum value. More
importantly, $|\rho_{\vec{k}}|^{2}$ will be much larger than it would for a more
nearly uniform distribution of positions. As a result $|\psi_{\vec{k}}|^{2}$ will be
large and this will be a likely configuration of the particles in the excited
state. For a configuration like that in fig.~(\ref{fig:densitywave}b), the phase
of $\rho_{\vec{k}}$ will shift but $|\psi_{\vec{k}}|^{2}$ will have the same
magnitude. This is analogous to the parity change in the harmonic oscillator
example. Because all different phases of the density wave are equally likely,
$\rho_{\vec{k}}$ has a mean density which is uniform (translationally
invariant).

To proceed with the calculation of the variational estimate for the excitation
energy $\Delta(k)$ of the density wave state we write
\begin{equation}
\Delta(k) = \frac{f(k)}{s(k)} \label{eq:1205}
\end{equation}
where
\begin{equation}
f(k) \equiv \left\langle\psi_{\vec{k}}|(H - E_{0})|\psi_{\vec{k}}\right\rangle,
\label{eq:1206}
\end{equation}
with $E_{0}$ being the exact ground state energy and
\begin{equation}
s(k) \equiv \langle\psi_{\vec{k}}|\psi_{\vec{k}}\rangle = \frac{1}{N}\;
\langle\Phi_{0}|\rho_{\vec{k}}^{\dagger}\rho_{\vec{k}}^{\phantom{\dagger}}|\Phi_
{0}\rangle.
\label{eq:1207}
\end{equation}
We see that the norm of the variational state $s(k)$ turns out to be the static
structure factor of the ground state. It is a measure of the mean square density
fluctuations at wave vector $\vec{k}$. Continuing the harmonic oscillator
analogy, we can view this as a measure of the zero-point fluctuations of the
normal-mode oscillator coordinate $\rho_{\vec{k}}$. For superfluid
${}^{4}\hbox{He}$ $s(k)$ can be directly measured by neutron scattering and can
also be computed theoretically using quantum Monte Carlo methods
\cite{ceperley95}. We will return to this point shortly.

\boxedtext{\begin{exercise}
Show that for a uniform liquid state of density $n$, the static structure factor
is related to the Fourier transform of the radial distribution function by
\[
s(k) = N\; \delta_{\vec{k},\vec{0}} + 1 + n \int d^{3}r\;
e^{i\vec{k}\cdot\vec{r}}\; \left[g(r) - 1\right]
\]
\label{ex:static2g(r)}
\end{exercise}}

The numerator in eq.~(\ref{eq:1206}) is called the oscillator strength and can
be written
\begin{equation}
f(k) = \frac{1}{N}\; \left\langle\Phi_{0}|\rho_{\vec{k}}^{\dagger}
[H,\rho_{\vec{k}}^{\phantom{\dagger}}]|\Phi_{0}\right\rangle. \label{eq:1208}
\end{equation}
For uniform systems with parity symmetry we can write this as a double
commutator
\begin{equation}
f(k) = \frac{1}{2N}\; \left\langle\Phi_{0}\left|\left[\rho_{\vec{k}}^{\dagger}, [H,
\rho_{\vec{k}}^{\phantom{\dagger}}]\right]\right|\Phi_{0}\right\rangle
\label{eq:1209}
\end{equation}
from which we can derive the justifiably famous oscillator strength sum rule
\begin{equation}
f(k) = \frac{\hbar^{2}k^{2}}{2M}. \label{eq:1210}
\end{equation}
where $M$ is the (band) mass of the particles.\footnote{Later on in
Eq.~(\ref{eq:1217}) we will express the oscillator strength in terms of a
frequency integral. Strictly speaking if this is integrated up to very high
frequencies including interband transitions, then $M$ is replaced by the bare
electron mass.} Remarkably (and conveniently) this is a universal result
independent of the form of the interaction potential between the particles. This
follows from the fact that only the kinetic energy part of the Hamiltonian fails
to commute with the density.

\boxedtext{\begin{exercise}
Derive eq.~(\ref{eq:1209}) and then eq.~(\ref{eq:1210}) from
eq.~(\ref{eq:1208}) for a system of interacting particles.
\label{ex:oscstrength}
\end{exercise}}

We thus arrive at the Feynman-Bijl formula for the collective mode excitation
energy
\begin{equation}
\Delta(k) = \frac{\hbar^{2}k^{2}}{2M}\; \frac{1}{s(k)}. \label{eq:1211}
\end{equation}
We can interpret the first term as the energy cost if a single particle
(initially at rest) were to absorb all the momentum and the second term is a
renormalization factor describing momentum (and position) correlations among the
particles. One of the remarkable features of the Feynman-Bijl formula is that it
manages to express a \textit{dynamical} quantity $\Delta (k)$, which is a
property of the excited state spectrum, solely in terms of a \textit{static}
property of the ground state, namely $s(k)$. This is a very powerful and useful
approximation.

Returning to eq.~(\ref{eq:1202}) we see that $\psi_{\vec{k}}$ describes a linear
superposition of states in which one single particle has had its momentum
boosted by $\hbar\vec{k}$. We do not know which one however. The summation in
eq.~(\ref{eq:1203}) tells us that it is equally likely to be particle 1
\textit{or} particle 2 \textit{or} \dots, etc. This state should not be confused
with the state in which boost is applied to particle 1 \textit{and} particle 2
\textit{and} \dots, etc. This state is described by a product
\begin{equation}
\Phi_{\vec{k}} \equiv \left(\prod_{j=1}^{N} e^{i\vec{k}\cdot\vec{r}_{j}}\right)\;
\Phi_{0}
\end{equation}
which can be rewritten
\begin{equation}
\Phi_{\vec{k}} = \exp{\left\{ iN\vec{k} \cdot \left(\frac{1}{N} \sum_{j=1}^{N}
\vec{r}_{j}\right)\right\}}\; \Phi_{0}
\label{eq:12125}
\end{equation}
showing that this is an exact energy eigenstate (with energy $N\;
\frac{\hbar^{2}k^{2}}{2M}$) in which the center of mass momentum has been
boosted by $N\hbar\vec{k}$.

In superfluid ${}^{4}\hbox{He}$ the structure factor vanishes linearly at small
wave vectors
\begin{equation}
s(k) \sim \xi k \label{eq:1212}
\end{equation}
so that $\Delta(k)$ is linear as expected for a sound mode
\begin{equation}
\Delta(k) = \left(\frac{\hbar^{2}}{2M}\; \frac{1}{\xi}\right)\; k
\label{eq:1213}
\end{equation}
from which we see that the sound velocity is given by
\begin{equation}
c_{\mathrm{s}} = \frac{\hbar}{2M}\; \frac{1}{\xi}. \label{eq:1214}
\end{equation}
This phonon mode should not be confused with the ordinary hydrodynamic sound
mode in classical fluids. The latter occurs in a collision dominated regime
$\omega\tau \ll 1$ in which collision-induced pressure provides the restoring
force. The phonon mode described here by $\psi_{\vec{k}}$ is a low-lying
eigenstate of the quantum Hamiltonian.

\begin{figure}
\centerline{\epsfxsize=6cm
 \epsffile{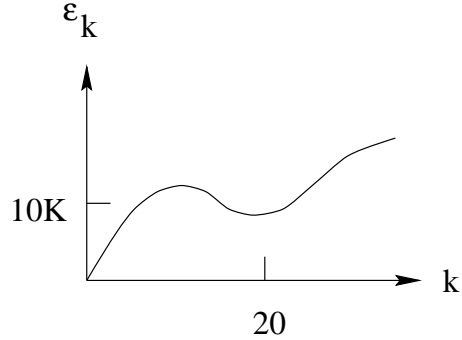}}
\caption[]{Schematic illustration of the phonon dispersion in superfluid liquid
${}^{4}$He. For small wave vectors the dispersion is linear, as is expected for
a gapless Goldstone mode. The roton minimum due to the peak in the static
structure factor occurs at a wave vector $k$ of approximately 20 in units of
inverse \AA. The roton energy is approximately $10$ in units of Kelvins.}
\label{fig:Heroton}
\end{figure}

At larger wave vectors there is a peak in the static structure factor caused by
the solid-like oscillations in the radial distribution function $g(r)$ similar
to those shown in Fig.~\ref{fig:2pointqhe} for the Laughlin liquid. This peak in
$s(k)$ leads to the so-called roton minimum in $\Delta(k)$ as illustrated in
fig.~(\ref{fig:Heroton}).

To better understand the Feynman picture of the collective excited states recall
that the dynamical structure factor is defined (at zero temperature) by
\begin{equation}
S(q,\omega) \equiv \frac{2\pi}{N}\;
\left\langle\Phi_{0}\left|\rho_{\vec{q}}^{\dagger}\;\, \delta\left(\omega -
\frac{H - E_{0}}{\hbar}\right)
\rho_{\vec{q}}^{\phantom{\dagger}}\right|\Phi_{0}\right\rangle. \label{eq:1215}
\end{equation}
The static structure factor is the zeroth frequency moment
\begin{equation}
s(q) = \int_{-\infty}^{\infty} \frac{d\omega}{2\pi}\; S(q,\omega) =
\int_{0}^{\infty} \frac{d\omega}{2\pi}\; S(q,\omega) \label{eq:1216} 
\end{equation}
(with the second equality valid only at zero temperature). Similarly the
oscillator strength in eq.~(\ref{eq:1206}) becomes (at zero temperature)
\begin{equation}
f(q) = \int_{-\infty}^{\infty} \frac{d\omega}{2\pi}\; \hbar\omega\; S(q,\omega)
= \int_{0}^{\infty} \frac{d\omega}{2\pi}\; \hbar\omega\; S(q,\omega).
\label{eq:1217}
\end{equation}

Thus we arrive at the result that the Feynman-Bijl formula can be rewritten
\begin{equation}
\Delta(q) = \frac{\int_{0}^{\infty} \frac{d\omega}{2\pi}\; \hbar\omega\;
S(q,\omega)}{\int_{0}^{\infty} \frac{d\omega}{2\pi}\; S(q,\omega)}.
\end{equation}
That is, $\Delta(q)$ is the mean excitation energy (weighted by the square of
the density operator matrix element). Clearly the mean exceeds the minimum and
so the estimate is variational as claimed. Feynman's approximation is equivalent
to the assumption that only a single mode contributes any oscillator strength so
that the zero-temperature dynamical structure factor contains only a single
delta function peak
\begin{equation}
S(q,\omega) = 2\pi\; s(q)\; \delta\left(\omega - \frac{1}{\hbar}\;
\Delta(q)\right). \label{eq:1218} 
\end{equation}
Notice that this approximate form satisfies both eq.~(\ref{eq:1216}) and
eq.~(\ref{eq:1217}) provided that the collective mode energy $\Delta(q)$ obeys
the Feynman-Bijl formula in eq.~(\ref{eq:1211}).

\boxedtext{\begin{exercise}
For a system with a homogeneous liquid ground state, the (linear response)
static susceptibility of the density to a perturbation $U = V_{\vec{q}}
\rho_{-\vec{q}}$ is defined by
\begin{equation}
\left\langle\rho_{\vec{q}}\right\rangle = \chi(q)V_{\vec{q}}.
\label{eq:linearresponse}
\end{equation}
Using first order perturbation theory show that the static susceptibility is
given in terms of the dynamical structure factor by
\begin{equation}
\chi(q) = -2\int_{0}^{\infty} \frac{d\omega}{2\pi} \frac{1}{\hbar\omega}
S(q,\omega).
\label{eq:static}
\end{equation}
Using the single mode approximation and the oscillator strength sum rule, derive
an expression for the collective mode dispersion in terms of $\chi(q)$. (Your
answer should \textbf{not} involve the static structure factor. Note also that
eq.(\ref{eq:linearresponse}) is not needed to produce the answer to this part.
Just work with eq.(\ref{eq:static}).)
\label{ex:9812}
\end{exercise}}

As we mentioned previously Feynman argued that in ${}^{4}\hbox{He}$ the Bose
symmetry of the wave functions guarantees that unlike in Fermi systems, there is
only a single low-lying mode, namely the phonon density mode. The paucity of
low-energy single particle excitations in boson systems is what helps make them
superfluid--there are no dissipative channels for the current to decay into.
Despite the fact that the quantum Hall system is made up of fermions, the
behavior is also reminiscent of superfluidity since the current flow is
dissipationless. Indeed, within the `composite boson' picture, one views the
FQHE ground state as a bose condensate \cite{compositeboson,sciam,sczhang}. Let
us therefore blindly make the single-mode approximation and see what happens.

{}From eq.~(\ref{eq:12105}) we see that the static structure factor for the
$m$th Laughlin state is (for small wave vectors only)
\begin{equation}
s(q) = \frac{L^{2}}{N}\; \frac{q^{2}}{4\pi m} = \frac{1}{2}\; q^{2}\ell^{2},
\label{eq:1219} 
\end{equation}
where we have used $L^{2}/N = 2\pi\ell^{2}m$. The Feynman-Bijl formula then
yields\footnote{We will continue to use the symbol $M$ here for the band mass of
the electrons to avoid confusion with the inverse filling factor $m$.}
\begin{equation}
\Delta(q) = \frac{\hbar^{2}q^{2}}{2M}\; \frac{2}{q^{2}\ell^{2}} =
\hbar\omega_{c}. \label{eq:1220} 
\end{equation}
This predicts that there is an excitation gap that is independent of wave vector
(for small $q$) and equal to the cyclotron energy. It is in fact correct that at
long wavelengths the oscillator strength is dominated by transitions in which a
single particle is excited from the $n=0$ to the $n=1$ Landau level. 
Furthermore, Kohn's theorem guarantees that the mode energy is precisely
$\hbar\omega_{c}$. Eq.~(\ref{eq:1220}) was derived specifically for the Laughlin
state, but it is actually quite general, applying to any translationally
invariant liquid ground state.

One might expect that the single mode approximation (SMA) will not work well in
an ordinary Fermi gas due to the high density of excitations around the Fermi
surface.\footnote{This expectation is only partly correct however as one
discovers when studying collective plasma oscillations in systems with
long-range Coulomb forces.} Here however the Fermi surface has been destroyed by
the magnetic field and the continuum of excitations with different kinetic
energies has been turned into a set of discrete inter-Landau-level excitations,
the lowest of which dominates the oscillator strength.

For filling factor $\nu =1$ the Pauli principle prevents any intra-level
excitations and the excitation gap is in fact $\hbar\omega_{c}$ as predicted by
the SMA. However for $\nu < 1$ there should exist intra-Landau-level excitations
whose energy scale is set by the interaction scale $e^{2}/\epsilon\ell$ rather
than the kinetic energy scale $\hbar\omega_{c}$. Indeed we can formally think of
taking the band mass to zero ($M \rightarrow 0$) which would send
$\hbar\omega_{c} \rightarrow \infty$ while keeping $e^{2}/\epsilon\ell$ fixed.
Unfortunately the SMA as it stands now is not very useful in this limit. What we
need is a variational wave function that represents a density wave but is
restricted to lie in the Hilbert space of the lowest Landau level. This can be
formally accomplished by replacing eq.~(\ref{eq:1202}) by
\begin{equation}
\psi_{\vec{k}} = \bar{\rho}_{\vec{k}}\; \psi_{m} \label{eq:1221} 
\end{equation}
where the overbar indicates that the density operator has been projected into
the lowest Landau level. The details of how this is accomplished are presented
in appendix~\ref{app:projection}.

The analog of eq.~(\ref{eq:1205}) is
\begin{equation}
\Delta(k) = \frac{\bar{f}(k)}{\bar{s}(k)} \label{eq:1222} 
\end{equation}
where $\bar{f}$ and $\bar{s}$ are the projected oscillator strength and
structure factor, respectively. As shown in appendix~\ref{app:projection}
\begin{eqnarray}
\bar{s}(k) &\equiv& \frac{1}{N}\;
\left\langle\psi_{m}|\bar{\rho}_{\vec{k}}^{\dagger}\;
\bar{\rho}_{\vec{k}}|\psi_{m}\right\rangle = s(k) - \left[ 1 - e^{-
\frac{1}{2}|k|^{2}\ell^{2}}\right]\nonumber\\
 &=& s(k) - s_{\nu=1}(k). \label{eq:1223}
\end{eqnarray}
This vanishes for the filled Landau level because the Pauli principle forbids
all intra-Landau-level excitations. For the $m$th Laughlin state
eq.~(\ref{eq:1219}) shows us that the leading term in $s(k)$ for small $k$ is
$\frac{1}{2}k^{2}\ell^{2}$. Putting this into eq.~(\ref{eq:1223}) we see that
the leading behavior for $\bar{s}(k)$ is therefore quartic
\begin{equation}
\bar{s}(k) \sim a(k\ell)^{4} + \ldots .
\end{equation}
We can not compute the coefficient $a$ without finding the $k^{4}$ correction to
eq.~(\ref{eq:1219}). It turns out that there exists a compressibility sum rule
for the fake plasma from which we can obtain the exact result \cite{GMP}
\begin{equation}
a = \frac{m-1}{8}.
\end{equation}

\begin{figure}
\centerline{\epsfxsize=10cm
 \epsffile{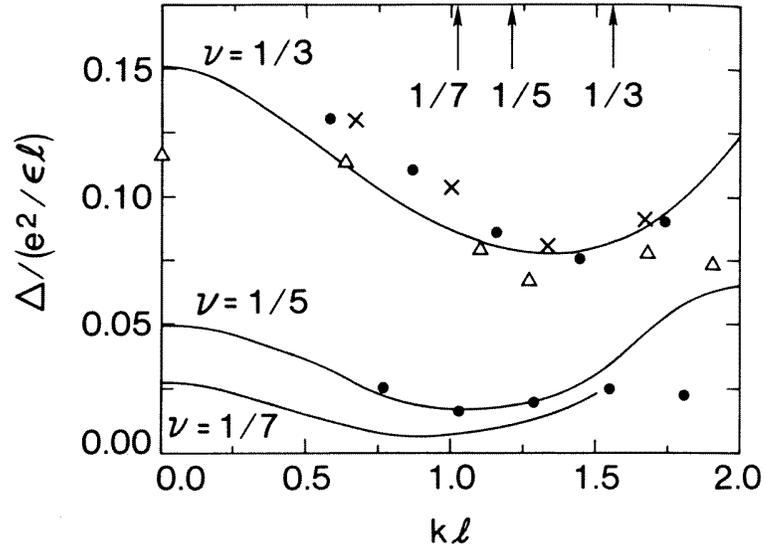}}
\caption[]{Comparison of the single mode approximation (SMA) prediction of the
collective mode energy for filling factors $\nu=1/3,1/5,1/7$ (solid lines) with
small-system numerical results for $N$ particles. Crosses indicate the $N=7,
\nu=1/3$ spherical system, triangles indicate the $N=6, \nu=1/3$ hexagonal unit
cell system results of Haldane and Rezayi \cite{haldane-rezayi}. Solid dots are
for $N=9,\nu=1/3$ and $N=7, \nu=1/5$ spherical system calculations of Fano et
al.~\cite{Fano} Arrows at the top indicate the magnitude of the reciprocal
lattice vector of the Wigner crystal at the corresponding filling factor. Notice
that unlike the phonon collective mode in superfluid helium shown in
fig.~(\ref{fig:Heroton}), the mode here is gapped.}
\label{fig:magnetoroton}
\end{figure}

The projected oscillator strength is given by eq.~(\ref{eq:1209}) with the
density operators replaced by their projections. In the case of
${}^{4}\hbox{He}$ only the kinetic energy part of the Hamiltonian failed to
commute with the density. It was for this reason that the oscillator strength
came out to be a universal number related to the mass of the particles. Within
the lowest Landau level however the kinetic energy is an irrelevant constant.
Instead, after projection the density operators no longer commute with each
other (see appendix~\ref{app:projection}). It follows from these commutation
relations that the projected oscillator strength is proportional to the strength
of the interaction term. The leading small $k$ behavior is \cite{GMP}
\begin{equation}
\bar{f}(k) = b\; \frac{e^{2}}{\epsilon\ell}(k\ell)^{4} + \ldots
\end{equation}
where $b$ is a dimensionless constant that depends on the details of the
interaction potential. The intra-Landau level excitation energy therefore has a
finite gap at small $k$
\begin{equation}
\Delta(k) = \frac{\bar{f}(k)}{\bar{s}(k)} \sim \frac{b}{a}\;
\frac{e^{2}}{\epsilon\ell} + \mathcal{O}(k^{2}) + \ldots
\end{equation}
This is quite different from the case of superfluid ${}^{4}\hbox{He}$ in which
the mode is gapless. However like the case of the superfluid, this
`magnetophonon' mode has a `magnetoroton' minimum at finite $k$ as illustrated
in fig.~(\ref{fig:magnetoroton}).
The figure also shows results from numerical exact diagonalization studies which
demonstrate that the single mode approximation is extremely accurate. Note that
the magnetoroton minimum occurs close to the position of the smallest reciprocal
lattice vector in the Wigner crystal of the same density. In the crystal the
phonon frequency would go exactly to zero at this point. (Recall that in a
crystal the phonon dispersion curves have the periodicity of the reciprocal
lattice.)

Because the oscillator strength is almost entirely in the cyclotron mode, the
dipole matrix element for coupling the collective excitations to light is very
small. They have however been observed in Raman scattering \cite{pinczukroton}
and found to have an energy gap in excellent quantitative agreement with the
single mode approximation.

Finally we remark that these collective excitations are characterized by a
well-defined wave vector $\vec{k}$ despite the presence of the strong magnetic
field. This is only possible because they are charge neutral which allows one to
define a gauge invariant conserved momentum \cite{Kallin}.

\section{Charged Excitations}

Except for the fact that they are gapped, the neutral magnetophonon excitations
are closely analogous to the phonon excitations in superfluid ${}^{4}\hbox{He}$.
We further pursue this analogy with a search for the analog of vortices in
superfluid films. A vortex is a topological defect which is the quantum version
of the familiar whirlpool. A reasonably good variational wave function for a
vortex in a two-dimensional film of ${}^{4}\hbox{He}$ is
\begin{equation}
\psi_{\vec{R}}^{\pm} = \left\{\prod_{j=1}^{N} f\left(|\vec{r}_{j} -
\vec{R}|\right)\; e^{\pm i\theta(\vec{r}_{j}-\vec{R})}\right\}\Phi_{0}.
\end{equation}
Here $\theta$ is the azimuthal angle that the particle's position makes relative
to $\vec{R}$, the location of the vortex center. The function $f$ vanishes as
$\vec{r}$ approaches $\vec{R}$ and goes to unity far away. The choice of sign in
the phase determines whether the vortex is right or left handed.

The interpretation of this wave function is the following. The vortex is a
topological defect because if any particle is dragged around a closed loop
surrounding $\vec{R}$, the phase of the wave function winds by $\pm 2\pi$. This
phase gradient means that current is circulating around the core. Consider a
large circle of radius $\xi$ centered on $\vec{R}$. The phase change of $2\pi$
around the circle occurs in a distance $2\pi\xi$ so the local gradient seen by
\textit{every} particle is $\hat{\theta}/\xi$. Recalling eq.~(\ref{eq:12125}) we
see that locally the center of mass momentum has been boosted by
$\pm\frac{\hbar}{\xi}\; \hat{\theta}$ so that the current density of the
whirlpool falls off inversely with distance from the core.\footnote{This slow
algebraic decay of the current density means that the total kinetic energy of a
single vortex diverges logarithmically with the size of the system. This in turn
leads to the Kosterlitz Thouless phase transition in which pairs of vortices
bind together below a critical temperature. As we will see below there is no
corresponding finite temperature transition in a quantum Hall system.} Near the
core $f$ falls to zero because of the `centrifugal barrier' associated with this
circulation. In a more accurate variational wave function the core would be
treated slightly differently but the asymptotic large distance behavior would be
unchanged.

What is the analog of all this for the lowest Landau level? For $\psi^{+}$ we
see that every particle has its angular momentum boosted by one unit. In the
lowest Landau level analyticity (in the symmetric gauge) requires us to replace
$e^{i\theta}$ by $z = x + iy$. Thus we are led to the Laughlin `quasi-hole'
wave function
\begin{equation}
\psi_{Z}^{+}[z] = \prod_{j=1}^{N} (z_{j} - Z)\; \psi_{m}[z] \label{eq:12145}
\end{equation}
where $Z$ is a complex number denoting the position of the vortex and $\psi_{m}$
is the Laughlin wave function at filling factor $\nu = 1/m$. The corresponding
antivortex (`quasi-electron' state) involves $z_{j}^{*}$ suitably projected (as
discussed in App.~\ref{app:projection}.):
\begin{equation}
\psi_{Z}^{-}[z] = \prod_{j=1}^{N} \left(2\frac{\partial}{\partial z_{j}} -
Z^{*}\right)\; \psi_{m}[z] \label{eq:12146}
\end{equation}
where as usual the derivatives act only on the polynomial part of $\psi_{m}$.
All these derivatives make $\psi^{-}$ somewhat difficult to work with. We will
therefore concentrate on the quasi-hole state $\psi^{+}$. The origin of the
names quasi-hole and quasi-electron will become clear shortly.

Unlike the case of a superfluid film, the presence of the vector potential
allows these vortices to cost only a finite energy to produce and hence the
electrical dissipation is always finite at any non-zero temperature. There is no
finite temperature transition into a superfluid state as in the Kosterlitz
Thouless transition. From a field theoretic point of view, this is closely
analogous to the Higg's mechanism \cite{compositeboson}.

Just as in our study of the Laughlin wave function, it is very useful to see how
the plasma analogy works for the quasi-hole state
\begin{equation}
|\psi_{Z}^{+}|^{2} = e^{-\beta U_{\mathrm{class}}}\; e^{-\beta V}
\end{equation}
where $U_{\mathrm{class}}$ is given by eq.~(\ref{eq:Uclass}), $\beta = 2/m$ as
before and
\begin{equation}
V \equiv m \sum_{j=1}^{N} \left(-\ln{|z_{j} - Z|}\right).
\end{equation}
Thus we have the classical statistical mechanics of a one-component plasma of
(fake) charge $m$ objects seeing a neutralizing jellium background plus a new
potential energy $V$ representing the interaction of these objects with an
`impurity' located at $Z$ and having unit charge.

Recall that the chief desire of the plasma is to maintain charge neutrality.
Hence the plasma particles will be repelled from $Z$. Because the plasma
particles have fake charge $m$, the screening cloud will have to have a net
reduction of $1/m$ particles to screen the impurity. But this means that the
quasi-hole has fractional fermion number! The (true) physical charge of the
object is a fraction of the elementary charge
\begin{equation}
q^{*} = \frac{e}{m}.
\end{equation}

\begin{figure}
\centerline{\epsfxsize=6cm
 \epsffile{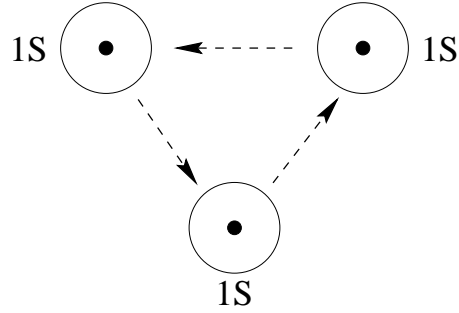}}
\caption[]{Illustration of an electron tunneling among the 1S orbitals of three
protons. The tunneling is exponentially slow for large separations which leads
to only exponentially small lifting of what would otherwise be a three-fold
degenerate ground state.}
\label{fig:1201}
\end{figure}

This is very strange! How can we possibly have an elementary excitation carrying
fractional charge in a system made up entirely of electrons? To understand this
let us consider an example of another quantum system that seems to have
fractional charge, but in reality doesn't. Imagine three protons arranged in an
equilateral triangle as shown in fig.~(\ref{fig:1201}).
Let there be one electron in the system. In the spirit of the tight-binding
model we consider only the 1S orbital on each of the three `lattice sites'. The
Bloch states are
\begin{equation}
\psi_{k} = \frac{1}{\sqrt{3}} \sum_{j=1}^{3} e^{ikj}\; |j\rangle
\end{equation}
where $|j\rangle$ is the 1S orbital for the $j$th atom. The equilateral triangle
is like a linear system of length 3 with periodic boundary conditions. Hence the
allowed values of the wavevector are $\left\{ k_{\alpha} = \frac{2\pi}{3}\alpha;\;\;
\alpha = -1,0,+1\right\}$. The energy eigenvalues are
\begin{equation}
\epsilon_{k_{\alpha}} = -E_{\mathrm{1S}} - 2J\; \cos{k_{\alpha}}
\end{equation}
where $E_{\mathrm{1S}}$ is the isolated atom energy and $-J$ is the hopping
matrix element related to the orbital overlap and is exponentially small for
large separations of the atoms.

The projection operator that measures whether or not the particle is on site $n$
is
\begin{equation}
P_{n} \equiv |n\rangle\; \langle n|.
\end{equation}
Its expectation value in any of the three eigenstates is
\begin{equation}
\left\langle\psi_{k_{\alpha}}|P_{n}|\psi_{k_{\alpha}}\right\rangle = \frac{1}{3}.
\end{equation}
This equation simply reflects the fact that as the particle tunnels from site to
site it is equally likely to be found on any site. Hence it will, on average, be
found on a particular site $n$ only 1/3 of the time. The average electron number
per site is thus 1/3. This however is a trivial example because the value of the
measured charge is always an integer. Two-thirds of the time we measure zero and
one third of the time we measure unity. This means that the charge
\textit{fluctuates}. One measure of the fluctuations is
\begin{equation}
\sqrt{\langle P_{n}^{2}\rangle - \langle P_{n}\rangle^{2}} = \sqrt{\frac{1}{3} -
\frac{1}{9}} = \frac{\sqrt{2}}{3},
\end{equation}
which shows that the fluctuations are larger than the mean value. This result is
most easily obtained by noting $P_{n}^{2} = P_{n}$.

A characteristic feature of this `imposter' fractional charge $\frac{e}{m}$ that
guarantees that it fluctuates is the existence in the spectrum of the
Hamiltonian of a set of $m$ nearly degenerate states. (In our toy example here,
$m=3$.) The characteristic time scale for the charge fluctuations is $\tau \sim
\hbar/\Delta\epsilon$ where $\Delta\epsilon$ is the energy splitting of the
quasi-degenerate manifold of states. In our tight-binding example $\tau \sim
\hbar/J$ is the characteristic time it takes an electron to tunnel from the 1S
orbital on one site to the next. As the separation between the sites increases
this tunneling time grows exponentially large and the charge fluctuations become
exponentially slow and thus easy to detect.

In a certain precise sense, the fractional charge of the Laughlin quasiparticles
behaves very differently from this. An electron added at low energies to a $\nu
= 1/3$ quantum Hall fluid breaks up into three charge 1/3 Laughlin
quasiparticles. These quasiparticles can move arbitrarily far apart from each
other\footnote{Recall that unlike the case of vortices in superfluids, these
objects are unconfined.} and yet no quasi-degenerate manifold of states appears.
The excitation gap to the first excited state remains finite. The only
degeneracy is that associated with the positions of the quasiparticles. If we
imagine that there are three impurity potentials that pin down the positions of
the three quasiparticles, then the state of the system is \textit{uniquely}
specified. Because there is no quasidegeneracy, we do not have to specify any
more information other than the positions of the quasiparticles. Hence in a deep
sense, they are true \textit{elementary particles} whose fractional charge is a
sharp quantum observable.

Of course, since the system is made up only of electrons, if we capture the
charges in some region in a box, we will always get an integer number of
electrons inside the box. However in order to close the box we have to locally
destroy the Laughlin state. This will cost (at a minimum) the excitation gap.
This may not seem important since the gap is small --- only a few Kelvin or so.
But imagine that the gap were an MeV or a GeV. Then we would have to build a
particle accelerator to `close the box' and probe the fluctuations in the
charge. These fluctuations would be analogous to the ones seen in quantum
electrodynamics at energies above $2m_{e}c^{2}$ where electron-positron pairs
are produced during the measurement of charge form factors by means of a
scattering experiment.

Put another way, the charge of the Laughlin quasiparticle fluctuates but only at
high frequencies $\sim \Delta/\hbar$. If this frequency (which is $\sim
50\hbox{GHz}$) is higher than the frequency response limit of our voltage
probes, we will see no charge fluctuations. We can formalize this by writing a
modified projection operator \cite{KivelsonGoldhaber} for the charge on some
site $n$ by
\begin{equation}
P_{n}^{(\Omega)} \equiv P^{\Omega}\; P_{n} P^{\Omega}
\end{equation}
where $P_{n} = |n\rangle\; \langle n|$ as before and 
\begin{equation}
P^{(\Omega)} \equiv \theta(\Omega - H + E_{0})
\end{equation}
is the operator that projects onto the subset of eigenstates with excitation
energies less than $\Omega$. $P_{n}^{(\Omega)}$ thus represents a measurement
with a high-frequency cutoff built in to represent the finite bandwidth of the
detector. Returning to our tight-binding example, consider the situation where
$J$ is large enough that the excitation gap $\Delta = \left(1 -
\cos{\frac{2\pi}{3}}\right) J$ exceeds the cutoff $\Omega$. Then
\begin{eqnarray}
P^{(\Omega)} &=& \sum_{\alpha=-1}^{+1} |\psi_{k_{\alpha}}\rangle\;
\theta(\Omega - \epsilon_{k_{\alpha}} + \epsilon_{k_{0}})\;
\langle\psi_{k_{\alpha}}|\nonumber\\
&=& |\psi_{k_{0}}\rangle\; \langle\psi_{k_{0}}|
\end{eqnarray}
is simply a projector on the ground state. In this case
\begin{equation}
P_{n}^{(\Omega)} = |\psi_{k_{0}}\rangle\; \frac{1}{3}\; \langle\psi_{k_{0}}|
\end{equation}
and
\begin{equation}
\left\langle\psi_{k_{0}}\left|[P_{n}^{(\Omega)}]^{2}\right|\psi_{k_{0}}\right\rangle
- \left\langle\psi_{k_{0}}|P_{n}^{(\Omega)}|\psi_{k_{0}}\right\rangle^{2} = 0.
\end{equation}
The charge fluctuations in the ground state are then zero (as measured by the
finite bandwidth detector).

The argument for the Laughlin quasiparticles is similar. We again emphasize that
one can not think of a single charge tunneling among three sites because the
excitation gap remains finite no matter how far apart the quasiparticle sites
are located. This is possible only because it is a correlated many-particle
system.

To gain a better understanding of fractional charge it is useful to compare this
situation to that in high energy physics. In that field of study one knows the
physics at low energies --- this is just the phenomena of our everyday world.
The goal is to study the high energy (short length scale) limit to see where
this low energy physics comes from. What force laws lead to our world? Probing
the proton with high energy electrons we can temporarily break it up into three
fractionally charged quarks, for example.

Condensed matter physics in a sense does the reverse. We know the phenomena at
`high' energies (i.e. room temperature) and we would like to see how the known
dynamics (Coulomb's law and non-relativistic quantum mechanics) leads to unknown
and surprising collective effects at low temperatures and long length scales.
The analog of the particle accelerator is the dilution refrigerator.

To further understand Laughlin quasiparticles consider the point of view of
`flatland' physicists living in the cold, two-dimensional world of a $\nu = 1/3$
quantum Hall sample. As far as the flatlanders are concerned the `vacuum' (the
Laughlin liquid) is completely inert and featureless. They discover however that
the universe is not completely empty. There are a few elementary particles
around, all having the same charge $q$. The flatland equivalent of Benjamin
Franklin chooses a unit of charge which not only makes $q$ negative but gives it
the fractional value $-1/3$. For some reason the Flatlanders go along with this.

Flatland cosmologists theorize that these objects are `cosmic strings',
topological defects left over from the `big cool down' that followed the
creation of the universe. Flatland experimentalists call for the creation of a
national accelerator facility which will reach the unprecedented energy scale of
10 Kelvin. With great effort and expense this energy scale is reached and the
accelerator is used to smash together three charged particles. To the
astonishment of the entire world a new short-lived particle is temporarily
created with the bizarre property of having integer charge!

\begin{figure}
\centerline{\epsfysize=10cm
 \epsffile{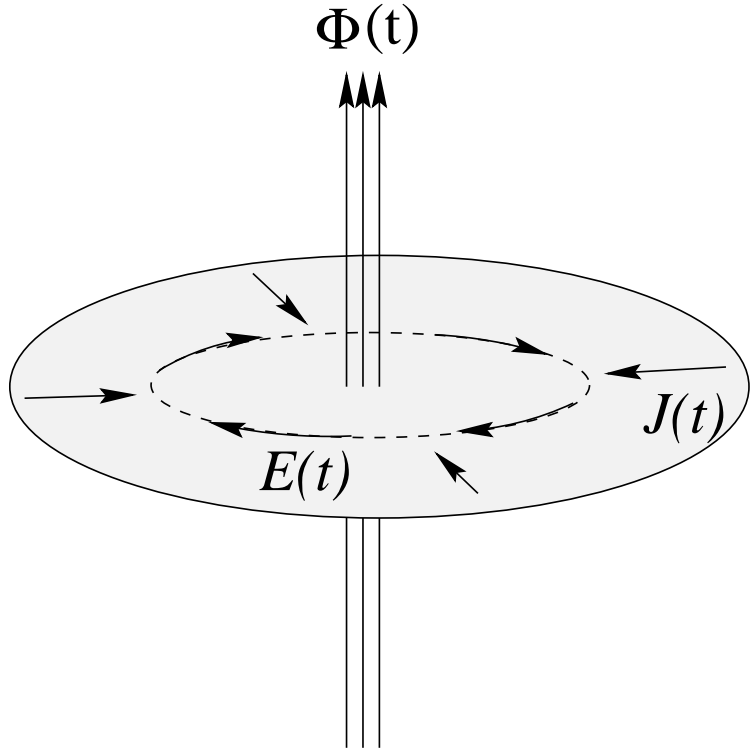}}
\caption[]{Construction of a Laughlin quasiparticle by adiabatically threading
flux $\Phi(t)$ through a point in the sample. Faraday induction gives an
azimuthal electric field $E(t)$ which in turn produces a radial current $J(t)$.
For each quantum of flux added, charge $\nu e$ flows into (or out of) the region
due to the quantized Hall conductivity $\nu e^{2}/h$. A flux tube containing an
integer number of flux quanta is invisible to the particles (since the Aharanov
phase shift is an integer multiple of $2\pi$) and so can be removed by a
singular gauge transformation.}
\label{fig:solenoid}
\end{figure}

There is another way to see that the Laughlin quasiparticles carry fractional
charge which is useful to understand because it shows the deep connection
between the sharp fractional charge and the sharp quantization of the Hall
conductivity. Imagine piercing the sample with an infinitely thin magnetic
solenoid as shown in fig.~(\ref{fig:solenoid}) 
and slowly increasing the magnetic flux $\Phi$ from 0 to $\Phi_{0} =
\frac{hc}{e}$ the quantum of flux. Because of the existence of a finite
excitation gap $\Delta$ the process is adiabatic and reversible if performed
slowly on a time scale long compared to $\hbar/\Delta$.

Faraday's law tells us that the changing flux induces an electric field obeying
\begin{equation}
\oint_{\Gamma} d\vec{r} \cdot \vec{E} = -\frac{1}{c}\;
\frac{\partial\Phi}{\partial t}
\end{equation}
where $\Gamma$ is any contour surrounding the flux tube. Because the electric
field contains only Fourier components at frequencies $\omega$ obeying
$\hbar\omega < \Delta$, there is no dissipation and $\sigma_{xx} = \sigma_{yy} =
\rho_{xx} = \rho_{yy} = 0$. The electric field induces a current density obeying
\begin{equation}
\vec{E} = \rho_{xy}\; \vec{J} \times \hat{z}
\end{equation}
so that
\begin{equation}
\rho_{xy} \oint_{\Gamma} \vec{J} \cdot (\hat{z} \times d\vec{r}) = -
\frac{1}{c}\; \frac{d\Phi}{dt}.
\end{equation}
The integral on the LHS represents the total current flowing into the region
enclosed by the contour. Thus the charge inside this region obeys
\begin{equation}
\rho_{xy}\; \frac{dQ}{dt} = -\frac{1}{c}\; \frac{d\Phi}{dt}.
\end{equation}
After one quantum of flux has been added the final charge is
\begin{equation}
Q = \frac{1}{c}\; \sigma_{xy} \Phi_{0} = \frac{h}{e}\; \sigma_{xy}.
\label{eq:1124165}
\end{equation}
Thus on the quantized Hall plateau at filling factor $\nu$ where $\sigma_{xy} =
\nu\; \frac{e^{2}}{h}$ we have the result
\begin{equation}
Q = \nu e.
\end{equation}
Reversing the sign of the added flux would reverse the sign of the charge.

The final step in the argument is to note that an infinitesimal tube containing
a quantum of flux is invisible to the particles. This is because the 
Aharonov-Bohm phase factor for traveling around the flux tube is unity.
\begin{equation}
\exp{\left\{ i \frac{e}{\hbar c} \oint_{\Gamma} \delta\vec{A} \cdot
d\vec{r}\right\}} = e^{\pm 2\pi i} = 1.
\end{equation}
Here $\delta\vec{A}$ is the additional vector potential due to the solenoid.
Assuming the flux tube is located at the origin and making the gauge choice
\begin{equation}
\delta\vec{A} = \Phi_{0}\; \frac{\hat{\theta}}{2\pi r},
\end{equation}
one can see by direct substitution into the Schr\"{o}dinger equation that the
only effect of the quantized flux tube is to change the phase of the wave
function by
\begin{equation}
\psi \rightarrow \psi \prod_{j} \frac{z_{j}}{|z_{j}|} = \psi \prod_{j}
e^{i\theta_{j}}.
\end{equation}
The removal of a quantized flux tube is thus a `singular gauge change' which has
no physical effect.

Let us reiterate. Adiabatic insertion of a flux quantum changes the state of the
system by pulling in (or pushing out) a (fractionally) quantized amount of
charge. Once the flux tube contains a quantum of flux it effectively becomes
invisible to the electrons and can be removed by means of a singular gauge
transformation.

Because the excitation gap is preserved during the adiabatic addition of the
flux, the state of the system is fully specified by the position of the
resulting quasiparticle. As discussed before there are no low-lying
quasi-degenerate states. This version of the argument highlights the essential
importance of the fact that $\sigma_{xx} = 0$ and $\sigma_{xy}$ is quantized.
The existence of the fractionally quantized Hall transport coefficients
guarantees the existence of fractionally charged elementary excitations

These fractionally charged objects have been observed directly by using an
ultrasensitive electrometer made from a quantum dot \cite{Vgoldman} and by the
reduced shot noise which they produce when they carry current \cite{shotnoise}.

Because the Laughlin quasiparticles are discrete objects they cost a non-zero
(but finite) energy to produce. Since they are charged they can be thermally
excited only in neutral pairs. The charge excitation gap is therefore
\begin{equation}
\Delta_{c} = \Delta_{+} + \Delta_{-}
\end{equation}
where $\Delta_{\pm}$ is the vortex/antivortex (quasielectron/quasihole)
excitation energy. In the presence of a transport current these thermally
excited charges can move under the influence of the Hall electric field and
dissipate energy. The resulting resistivity has the Arrhenius form
\begin{equation}
\rho_{xx} \sim \gamma \frac{h}{e^{2}}\; e^{-\beta\Delta_{c}/2}
\end{equation}
where $\gamma$ is a dimensionless constant of order unity. Note that the law of
mass action tells us that the activation energy is $\Delta_{c}/2$ not
$\Delta_{c}$ since the charges are excited in pairs. There is a close analogy
between the dissipation described here and the flux flow resistance caused by
vortices in a superconducting film.

Theoretical estimates of $\Delta_{c}$ are in good agreement with experimental
values determined from transport measurements \cite{gapmeasures}. Typical values
of $\Delta_{c}$ are only a few percent of $e^{2}/\epsilon\ell$ and hence no
larger than a few Kelvin. In a superfluid time-reversal symmetry guarantees that
vortices and antivortices have equal energies. The lack of time reversal
symmetry here means that $\Delta_{+}$ and $\Delta_{-}$ can be quite different.
Consider for example the hard-core model for which the Laughlin wave function
$\psi_{m}$ is an exact zero energy ground state as shown in
eq.~(\ref{eq:12103}). Equation~(\ref{eq:12145}) shows that the quasihole state
contains $\psi_{m}$ as a factor and hence is also an exact zero energy
eigenstate for the hard-core interaction. Thus the quasihole costs zero energy.
On the other hand eq.~(\ref{eq:12146}) tells us that the derivatives reduce the
degree of homogeneity of the Laughlin polynomial and therefore the energy of the
quasielectron \textit{must} be non-zero in the hard-core model. At filling factor
$\nu = 1/m$ this asymmetry has no particular significance since the
quasiparticles must be excited in pairs.

\begin{figure}
\centerline{\epsfxsize=6cm
 \epsffile{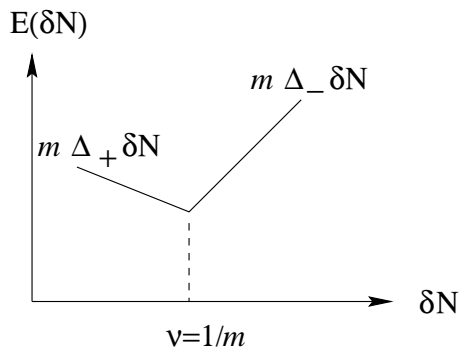}}
\caption[]{Energy cost for inserting $\delta N$ electrons into the Laughlin
state near filling factor $\nu=1/m$. The slope of the line is the chemical
potential. Its discontinuity at $\nu=1/m$ measures the charge excitation gap.}
\label{fig:energyslope}
\end{figure}

Consider now what happens when the magnetic field is increased slightly or the
particle number is decreased slightly so that the filling factor is slightly
smaller than $1/m$. The lowest energy way to accommodate this is to inject $m$
quasiholes into the Laughlin state for each electron that is removed (or for
each $m \Phi_{0}$ of flux that is added). The system energy (ignoring disorder
and interactions in the dilute gas of quasiparticles) is
\begin{equation}
E_{+} = E_{m} - \delta N\; m\Delta_{+}
\end{equation}
where $E_{m}$ is the Laughlin ground state energy and $-\delta N$ is the number
of added holes. Conversely for filling factors slightly greater than $1/m$ the
energy is (with $+\delta N$ being the number of added electrons) 
\begin{equation}
E_{-} = E_{m} + \delta N\; m\Delta_{-}.
\end{equation}
This is illustrated in fig.~(\ref{fig:energyslope}).
The slope of the lines in the figure determines the chemical potential
\begin{equation}
\mu_{\pm} = \frac{\partial E_{\pm}}{\partial\delta N} = \mp m\Delta_{\pm}.
\end{equation}
The chemical potential suffers a jump discontinuity of $m(\Delta_{+} + 
\Delta_{-}) = m\Delta_{c}$ just at filling factor $\mu = 1/m$. This jump in the
chemical potential is the signature of the charge excitation gap just as it is
in a semiconductor or insulator. Notice that this form of the energy is very
reminiscent of the energy of a type-II superconductor as a function of the
applied magnetic field (which induces vortices and therefore has an energy cost
$\Delta E \sim |B|$).

Recall that in order to have a quantized Hall plateau of finite width it is
necessary to have disorder present. For the integer case we found that disorder
localizes the excess electrons allowing the transport coefficients to not change
with the filling factor. Here it is the fractionally-charged quasiparticles that
are localized by the disorder.\footnote{Note again the essential importance of
the fact that the objects are `elementary particles'. That is, there are no
residual degeneracies once the positions are pinned down.} Just as in the
integer case the disorder may fill in the gap in the density of states but the
DC value of $\sigma_{xx}$ can remain zero because of the localization. Thus the
fractional plateaus can have finite width.

If the density of quasiparticles becomes too high they may delocalize and
condense into a correlated Laughlin state of their own. This gives rise to a
hierarchical family of Hall plateaus at rational fractional filling factors $\nu
= p/q$ (generically with $q$ odd due to the Pauli principle). There are several
different but entirely equivalent ways of constructing and viewing this
hierarchy which we will not delve into here \cite{SMGBOOK,TAPASHbook,DasSarmabook}.

\section{FQHE Edge States}

\begin{figure}
\centerline{\epsfysize=14cm
 \epsffile{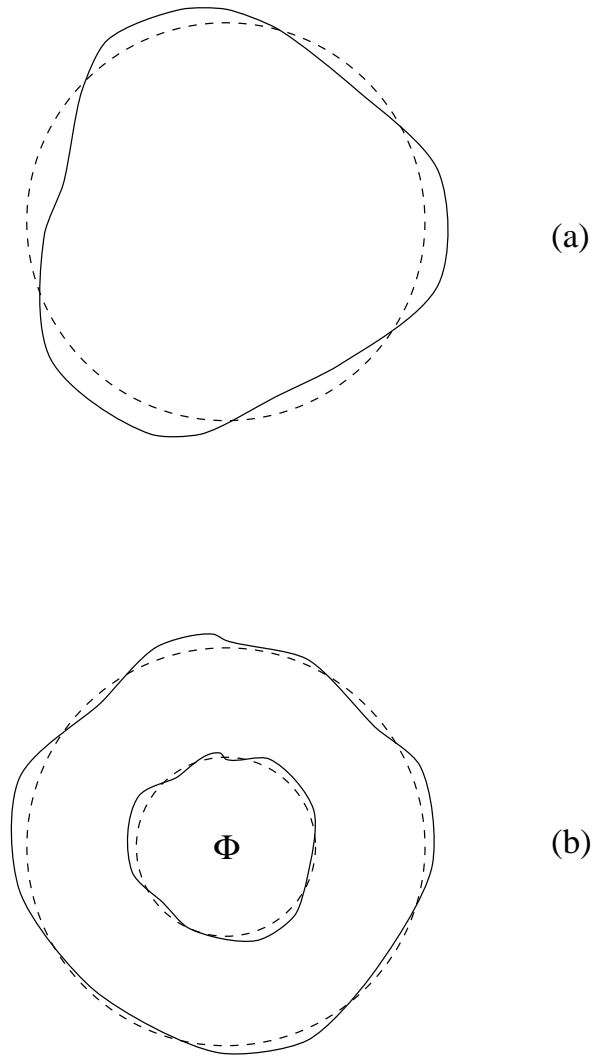}}
\caption[]{Area-preserving shape distortions of the incompressible quantum Hall
state. (a) IQHE Laughlin liquid `droplet' at $\nu=1$. (b) FQHE annulus at
$\nu=1/m$ formed by injecting a large number $n$ of flux quanta at the origin to
create $n$ quasiholes. There are thus two edge modes of opposite chirality.
Changing $n$ by one unit transfers fractional charge $\nu e$ from one edge to
the other by expanding or shrinking the size of the central hole. Thus the edge
modes have topological sectors labeled by the `winding number' $n$ and one can
view the gapless edge excitations as a gas of fractionally charged Laughlin
quasiparticles.}
\label{fig:edgewaves}
\end{figure}

We learned in our study of the integer QHE that gapless edge excitations exist
even when the bulk has a large excitation gap. Because the bulk is
incompressible the only gapless neutral excitations must be area-preserving
shape distortions such as those illustrated for a disk geometry in
fig.~(\ref{fig:edgewaves}a).
Because of the confining potential at the edges these shape distortions have a
characteristic velocity produced by the $\vec{E} \times \vec{B}$ drift. It is
possible to show that this view of the gapless neutral excitations is precisely
equivalent to the usual Fermi gas particle-hole pair excitations that we
considered previously in our discussion of edge states. Recall that we argued
that the contour line of the electrostatic potential separating the occupied
from the empty states could be viewed as a real-space analog of the Fermi
surface (since position and momentum are equivalent in the Landau gauge). The
charged excitations at the edge are simply ordinary electrons added or removed
from the vicinity of the edge.

In the case of a fractional QHE state at $\nu = 1/m$ the bulk gap is caused by
Coulomb correlations and is smaller but still finite. Again the only gapless
excitations are area-preserving shape distortions. Now however the charge of
each edge can be varied in units of $e/m$. Consider the annulus of Hall fluid
shown in fig.~(\ref{fig:edgewaves}b). The extension of the Laughlin wave
function $\psi_{m}$ to this situation is
\begin{equation}
\psi_{mn}[z] = \left(\prod_{j=1}^{N} z_{j}^{n}\right)\; \psi_{m}.
\end{equation}
This simply places a large number $n \gg 1$ of quasiholes at the origin.
Following the plasma analogy we see that this looks like a highly charged
impurity at the origin which repels the plasma, producing the annulus shown in
fig.~(\ref{fig:edgewaves}b). Each time we increase $n$ by one unit, the
annulus expands. We can view this expansion as increasing the electron number at
the outer edge by $1/m$ and reducing it by $1/m$ at the inner edge. (Thereby
keeping the total electron number integral as it must be.)

\begin{figure}
\centerline{\epsfysize=14cm
 \epsffile{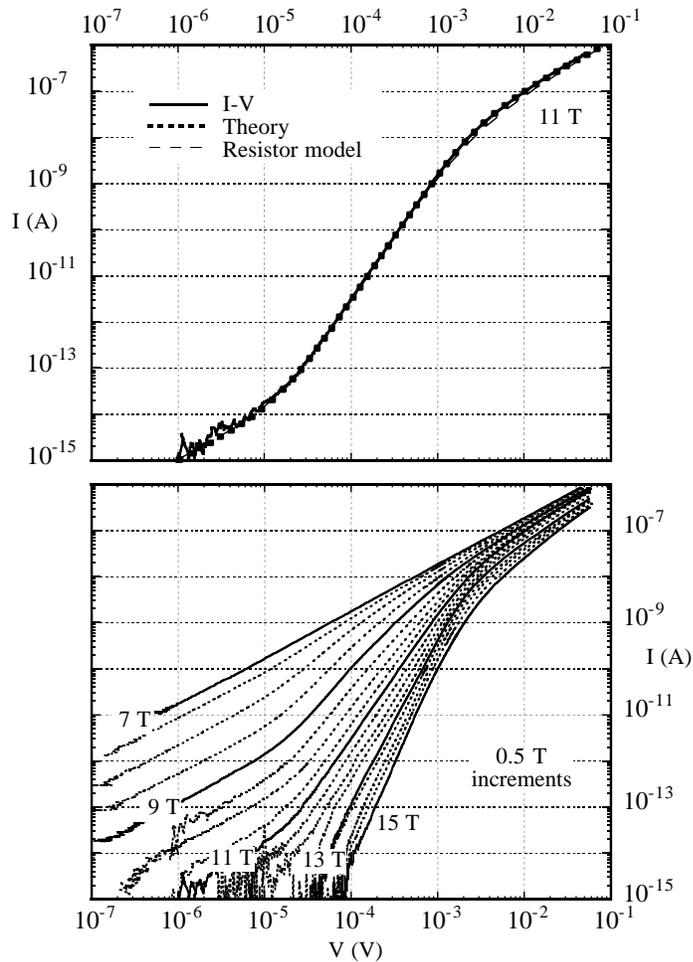}}
\caption[]{Non-linear current voltage response for tunneling an electron into a
FQHE edge state. Because the electron must break up into $m$ fractionally
charged quasiparticles, there is an orthogonality catastrophe leading to a
power-law density of states. The flattening at low currents is due to the finite
temperature. The upper panel shows the $\nu=1/3$ Hall plateau. The theory
\cite{KaneFisher,Chamon} works extremely well on the 1/3 quantized Hall plateau,
but the unexpectedly smooth variation of the exponent with magnetic field away
from the plateau shown in the lower panel is not yet fully understood. (After M.
Grayson \etal, Ref.~\cite{grayson}.}
\label{fig:changdata}
\end{figure}

It is appropriate to view the Laughlin quasiparticles, which are gapped in the
bulk, as being liberated at the edge. The gapless shape distortions in the Hall
liquid are thus excitations in a `gas' of fractionally charged quasiparticles.
This fact produces a profound alteration in the tunneling density of states to
inject an electron into the system. An electron which is suddenly added to an
edge (by tunneling through a barrier from an external electrode) will have very
high energy unless it breaks up into $m$ Laughlin quasiparticles. This leads to
an `orthogonality catastrophe' which simply means that the probability for this
process is smaller and smaller for final states of lower and lower energy. As a
result the current-voltage characteristic for the tunnel junction becomes
non-linear \cite{KaneFisher,Chamon,Wen}
\begin{equation}
I \sim V^{m}.
\end{equation}
For the filled Landau level $m=1$ the quasiparticles have charge $q = em = e$
and are ordinary electrons. Hence there is no orthogonality catastrophe and the
I-V characteristic is linear as expected for an ordinary metallic tunnel
junction. The non-linear tunneling for the $m=3$ state is shown in
fig.~(\ref{fig:changdata}).

\section{Quantum Hall Ferromagnets}
\label{sec:qhf}

\subsection{Introduction}
\label{subsec:QHEIntroduction}

\begin{figure}
\centerline{\epsfxsize=10cm
 \epsffile{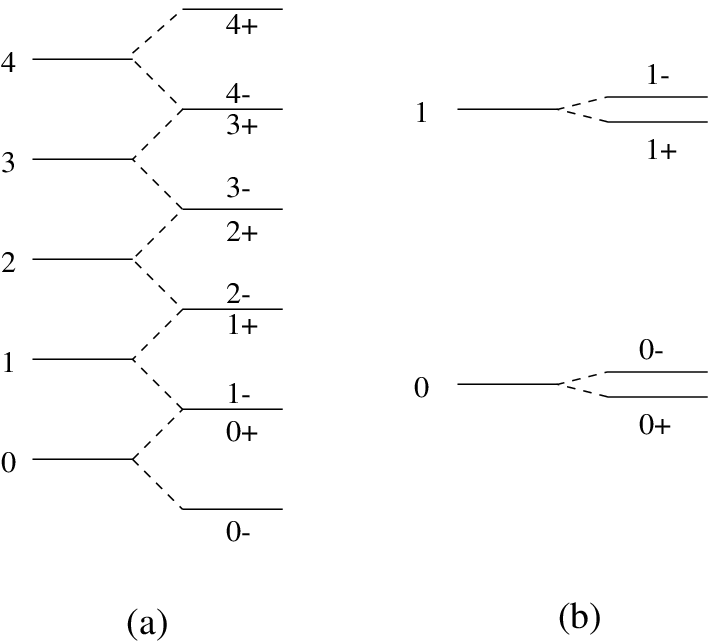}}
\caption[]{(a) Landau energy levels for an electron in free space. Numbers label
the Landau levels and $+ (-)$ refers to spin up (down). Since the $g$ factor is
2, the Zeeman splitting is exactly equal to the Landau level spacing,
$\hbar\omega_{c}$ and there are extra degeneracies as indicated. (b) Same for an
electron in GaAs. Because the effective mass is small and $g\approx -0.4$, the
degeneracy is strongly lifted and the spin assignments are reversed.}
\label{fig:zeeman}
\end{figure}

Naively one might imagine that electrons in the QHE have their spin dynamics
frozen out by the Zeeman splitting $g\mu_{\mathrm{B}}B$. In free space with $g = 2$
(neglecting QED corrections) the Zeeman splitting is exactly equal to the
cyclotron splitting $\hbar\omega_{c} \sim 100~\mathrm{K}$ as illustrated in
fig.~(\ref{fig:zeeman}~a). 
Thus at low temperatures we would expect for filling factors $\nu < 1$ all the
spins would be fully aligned. It turns out however that this naive expectation
is incorrect in GaAs for two reasons. First, the small effective mass $(m^{*} =
0.068)$ in the conduction band of GaAs increases the cyclotron energy by a
factor of $m/m^{*} \sim 14$. Second, spin-orbit scattering tumbles the spins
around in a way which reduces their effective coupling to the external magnetic
field by a factor of $-5$ making the $g$ factor $-0.4$. The Zeeman energy is
thus some 70 times smaller than the cyclotron energy and typically has a value
of about 2K, as indicated in fig.~(\ref{fig:zeeman}~b).

This decoupling of the scales of the orbital and spin energies means that it is
possible to be in a regime in which the orbital motion is fully quantized
($k_{\mathrm{B}}T \ll \hbar\omega_{c}$) but the low-energy spin fluctuations are not
completely frozen out ($k_{\mathrm{B}}T \sim g^{*}\mu_{\mathrm{B}}B$). The spin dynamics
in this regime are extremely unusual and interesting because the system is an
itinerant magnet with a quantized Hall coefficient. As we shall see, this leads
to quite novel physical effects.

The introduction of the spin degree of freedom means that we are dealing with
the QHE in multicomponent systems. This subject has a long history going back
to an early paper by Halperin \cite{BIHhelv} and has been reviewed extensively
\cite{GMbook,JPEbook,TAPASHbook}. In addition to the spin degree of freedom
there has been considerable recent interest in other multicomponent systems in
which spin is replaced by a pseudo-spin representing the layer index in double
well QHE systems or the electric subband index in wide single well systems.
Experiments on these systems are discussed by Shayegan in this volume
\cite{ShayeganLesHouches} and have also been reviewed in \cite{JPEbook}.

Our discussion will focus primarily on ferromagnetism near filling factor $\nu =
1$. In the subsequent section we will address analogous effects for pseudo-spin
degrees of freedom in multilayer systems.

\subsection{Coulomb Exchange}
\label{subsec:coulomb}

We tend to think of the integer QHE as being associated with the gap due to the
kinetic energy and ascribe importance to the Coulomb interaction only in the
fractional QHE. However study of ferromagnetism near integer filling factor $\nu
= 1$ has taught us that Coulomb interactions play an important role there as
well \cite{Sondhi}.

Magnetism occurs not because of direct magnetic forces, but rather because of a
combination of electrostatic forces and the Pauli principle. In a fully
ferromagnetically aligned state all the spins are parallel and hence the spin
part of the wave function is exchange symmetric
\begin{equation}
|\psi\rangle = \Phi(z_{1}, \ldots ,z_{N})\;
|\uparrow\uparrow\uparrow\uparrow\uparrow\ldots\uparrow\rangle .
\label{eq:052601}
\end{equation}
The spatial part $\Phi$ of the wave function must therefore be fully
antisymmetric and vanish when any two particles approach each other. This means
that each particle is surrounded by an `exchange hole' which thus lowers the
Coulomb energy per particle as shown in eq.~(\ref{eq:12109}). For filling factor
$\nu = 1$
\begin{equation}
\frac{\langle V\rangle}{N} = -\sqrt{\frac{\pi}{8}}\; \frac{e^{2}}{\epsilon\ell}
\sim 200\mathrm{K}
\end{equation}
This energy scale is two orders of magnitude larger than the Zeeman splitting
and hence strongly stabilizes the ferromagnetic state. Indeed at $\nu = 1$ the
ground state is spontaneously fully polarized at zero temperature even in the
absence of the Zeeman term. Ordinary ferromagnets like iron are generally only
partially polarized because of the extra kinetic energy cost of raising the
fermi level for the majority carriers. Here however the kinetic energy has been
quenched by the magnetic field and all states in the lowest Landau level are
degenerate. For $\nu = 1$ the large gap to the next Landau level means that we
know the spatial wave function $\Phi$ essentially exactly. It is simply the
single Slater determinant representing the fully filled Landau level. That is,
it is $m = 1$ Laughlin wave function. This simple circumstance makes this
perhaps the world's best understood ferromagnet.

\subsection{Spin Wave Excitations}
\label{subsec:spinwave}

It turns out that the low-lying `magnon' (spin wave) excited states can also be
obtained exactly. Before doing this for the QHE system let us remind ourselves
how the calculation goes in the lattice Heisenberg model for $N$ local moments
in an insulating ferromagnet
\begin{eqnarray}
H &=& -J \sum_{\langle ij\rangle} \vec{S}_{i} \cdot \vec{S}_{j} - \Delta
\sum_{j} S_{j}^{z}\nonumber\\
&=& -J \sum_{\langle ij\rangle} \left\{ S_{i}^{z} S_{j}^{z} + \frac{1}{2} \left(
S_{i}^{+} S_{j}^{-} + S_{i}^{-} S_{j}^{+}\right)\right\} - \Delta \sum_{j}
S_{j}^{z}
\end{eqnarray}
The ground state for $J > 0$ is the fully ferromagnetic state with total spin $S
= N/2$. Let us choose our coordinates in spin space so that $S_{z} = N/2$.
Because the spins are fully aligned the spin-flip terms in $H$ are ineffective
and (ignoring the Zeeman term)
\begin{equation}
H\; |\uparrow\uparrow\uparrow\ldots\uparrow\rangle = -\frac{J}{4} N_{b}\;
|\uparrow\uparrow\uparrow\ldots\uparrow\rangle
\end{equation}
where $N_{b}$ is the number of near-neighbor bonds and we have set $\hbar = 1$.
There are of course $2S + 1 = N + 1$ other states of the same total spin which
will be degenerate in the absence of the Zeeman coupling. These are generated by
successive applications of the total spin lowering operator
\begin{eqnarray}
S^{-} &\equiv& \sum_{j=1}^{N} S_{j}^{-}\\
S^{-}\; |\uparrow\uparrow\uparrow\ldots\uparrow\rangle &=&
|\downarrow\uparrow\uparrow\ldots\uparrow\rangle +
|\uparrow\downarrow\uparrow\ldots\uparrow\rangle\nonumber\\
&+&|\uparrow\uparrow\downarrow\ldots\uparrow\rangle + \dots
\end{eqnarray}

It is not hard to show that the one-magnon excited states are created by a
closely related operator
\begin{equation}
S_{\vec{q}}^{-} = \sum_{j=1}^{N} e^{-i\vec{q}\cdot\vec{R}_{j}}\; S_{j}^{-}
\end{equation}
where $\vec{q}$ lies inside the Brillouin zone and is the magnon wave
vector.\footnote{We use the phase factor $e^{-i\vec{q}\cdot\vec{R}_{j}}$ here
rather than $e^{+i\vec{q}\cdot\vec{R}_{j}}$ simply to be consistent with
$S_{\vec{q}}^{-}$ being the Fourier transform of $S_{j}^{-}$.} Denote these
states by
\begin{equation}
|\psi_{\vec{q}}\rangle = S_{\vec{q}}^{-}\; |\psi_{0}\rangle
\label{eq:052608}
\end{equation}
where $|\psi_{0}\rangle$ is the ground state. Because there is one flipped spin
in these states the transverse part of the Heisenberg interaction is able to
move the flipped spin from one site to a neighboring site
\begin{eqnarray}
H|\psi_{\vec{q}}\rangle &=& \left(E_{0} + \Delta + \frac{Jz}{2}\right)\;
|\psi_{\vec{q}}\rangle\nonumber\\
&&-\frac{J}{2} \sum_{\vec{\delta}} \sum_{j=1}^{N}
e^{-i\vec{q}\cdot\vec{R}_{j}}\; S_{j+\vec{\delta}}^{-}\; |\psi_{0}\rangle\\
H|\psi_{\vec{q}}\rangle &=& (E_{0} + \epsilon_{\vec{q}})\;
|\psi_{\vec{q}}\rangle
\end{eqnarray}
where $z$ is the coordination number, $\vec{\delta}$ is summed over near
neighbor lattice vectors and the magnon energy is
\begin{equation}
\epsilon_{\vec{q}} \equiv \frac{Jz}{2}\; \left\{1 - \frac{1}{z} \sum_{\vec{\delta}}
e^{-i\vec{q}\cdot\vec{\delta}}\right\} + \Delta
\end{equation}
For small $\vec{q}$ the dispersion is quadratic and for a 2D square lattice 
\begin{equation}
\epsilon_{\vec{q}} \sim \frac{Ja^{2}}{4} q^{2} + \Delta
\end{equation}
where $a$ is the lattice constant.

This is very different from the result for the antiferromagnet which has a
linearly dispersing collective mode. There the ground and excited states can
only be approximately determined because the ground state does not have all the
spins parallel and so is subject to quantum fluctuations induced by the
transverse part of the interaction. This physics will reappear when we study
non-collinear states in QHE magnets away from filling factor $\nu = 1$.

The magnon dispersion for the ferromagnet can be understood in terms of bosonic
`particle' (the flipped spin) hopping on the lattice with a tight-binding model
dispersion relation. The magnons are bosons because spin operators on different
sites commute. They are not free bosons however because of the hard core
constraint that (for spin 1/2) there can be no more than one flipped spin per
site. Hence multi-magnon excited states can not be computed exactly. Some nice
renormalization group arguments about magnon interactions can be found in
\cite{ReadandSachdev}.

The QHE ferromagnet is itinerant and we have to develop a somewhat different
picture. Nevertheless there will be strong similarities to the lattice
Heisenberg model. The exact ground state is given by eq.~(\ref{eq:052601}) with
\begin{equation}
\Phi(z_{1},\ldots,z_{N}) = \prod_{i<j} (z_{i} - z_{j})\; e^{-
\frac{1}{4}\sum_{k}|z_{k}|^{2}}.
\end{equation}
To find the spin wave excited states we need to find the analog of
eq.~(\ref{eq:052608}). The Fourier transform of the spin lowering operator for
the continuum system is
\begin{equation}
S_{\vec{q}}^{-} \equiv \sum_{j=1}^{N} e^{-i\vec{q}\cdot\vec{r}_{j}}\; S_{j}^{-}
\label{eq:052614}
\end{equation}
where $\vec{r}_{j}$ is the position operator for the $j$th particle. Recall from
eq.~(\ref{eq:1221}) that we had to modify Feynman's theory of the collective
mode in superfluid helium by projecting the density operator onto the Hilbert
space of the lowest Landau level. This suggests that we do the same in
eq.~(\ref{eq:052614}) to obtain the projected spin flip operator. In contrast to
the good but approximate result we obtained for the collective density mode,
this procedure actually yields the \textit{exact} one-magnon excited state (much
like we found for the lattice model).

Using the results of appendix~\ref{app:projection}, the projected spin lowering
operator is
\begin{equation}
\bar{S}_{q}^{-} = e^{-\frac{1}{4}|q|^{2}} \sum_{j=1}^{N} \tau_{q}(j)\; S_{j}^{-}
\label{eq:052615}
\end{equation}
where $q$ is the complex number representing the dimensionless wave vector
$\vec{q}\ell$ and $\tau_{q}(j)$ is the magnetic translation operator for the
$j$th particle. The commutator of this operator with the Coulomb interaction
Hamiltonian is 
\begin{eqnarray}
{}[H,\bar{S}_{q}^{-}] &=& \frac{1}{2} \sum_{k\neq 0} v(k)\; 
\left[\bar{\rho}_{-k}\bar{\rho}_{k},\bar{S}_{q}^{-}\right]\nonumber\\
&=& \frac{1}{2} \sum_{k\neq 0} v(k)\; \left\{\bar{\rho}_{-k}
\left[\bar{\rho}_{k},\bar{S}_{q}^{-}\right] + 
\left[\bar{\rho}_{-k},\bar{S}_{q}^{-}\right]\; \bar{\rho}_{k}\right\}.
\end{eqnarray}
We will shortly be applying this to the fully polarized ground state
$|\psi\rangle$. As discussed in appendix~\ref{app:projection}, no density wave
excitations are allowed in this state and so it is annihilated by
$\bar{\rho}_{k}$. Hence we can without approximation drop the second term above
and replace the first one by
\begin{equation}
{}[H,\bar{S}_{q}^{-}]\; |\psi\rangle = \frac{1}{2} \sum_{k\neq 0} v(k)\;
\left[\bar{\rho}_{-k},\left[\bar{\rho}_{k},\bar{S}_{q}^{-}\right]\right]\;
|\psi\rangle
\end{equation}
Evaluation of the double commutator following the rules in
appendix~\ref{app:projection} yields 
\begin{equation}
{}[H,\bar{S}_{q}^{-}]\; |\psi\rangle = \epsilon_{q}\; \bar{S}_{q}^{-}\;
|\psi\rangle
\end{equation}
where
\begin{equation}
\epsilon_{q} \equiv 2\sum_{k\neq 0} e^{-\frac{1}{2}|k|^{2}}\; v(k)\;
\sin^{2}{\left(\frac{1}{2} q \wedge k\right)}.
\label{eq:1123195}
\end{equation}
Since $|\psi\rangle$ is an eigenstate of $H$, this proves that 
$\bar{S}_{q}^{-}\; |\psi\rangle$ is an exact excited state of $H$ with
excitation energy $\epsilon_{q}$. In the presence of the Zeeman coupling
$\epsilon_{q} \rightarrow \epsilon_{q} + \Delta$.

This result tells us that, unlike the case of the density excitation, the
single-mode approximation is exact for the case of the spin density excitation.
The only assumption we made is that the ground state is fully polarized and has
$\nu = 1$.

\begin{figure}
\centerline{\epsfxsize=6cm
 \epsffile{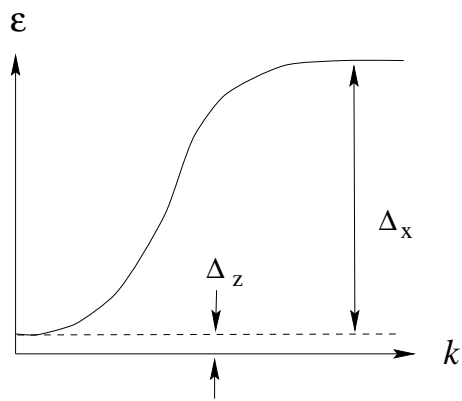}}
\caption[]{Schematic illustration of the QHE ferromagnet spinwave dispersion.
There is a gap at small $k$ equal to the Zeeman splitting,
$\Delta_{\mathrm{Z}}$. At large wave vectors, the energy saturates at the
Coulomb exchange energy scale $\Delta_{x} + \Delta_{\mathrm{Z}} \sim 100$K.}
\label{fig:LLLspinwave}
\end{figure}

For small $q$ the dispersion starts out quadratically
\begin{equation}
\epsilon_{q} \sim Aq^{2}
\label{eq:1124198}
\end{equation}
with
\begin{equation}
A \equiv \frac{1}{4} \sum_{k\neq 0} e^{-\frac{1}{2}|k|^{2}}\; v(k)\; |k|^{2}
\end{equation}
as can be seen by expanding the sine function to lowest order. For very large
$q$ $\sin^{2}$ can be replaced by its average value of $\frac{1}{2}$ to yield
\begin{equation}
\epsilon_{q} \sim \sum_{k\neq 0} v(k)\; e^{-\frac{1}{2}|k|^{2}}.
\end{equation}
Thus the energy saturates at a constant value for $q \rightarrow \infty$ as
shown in fig.~(\ref{fig:LLLspinwave}).
(Note that in the lattice model the wave vectors are restricted to the first
Brillouin zone, but here they are not.)

\begin{figure}
\centerline{\epsfxsize=6cm
 \epsffile{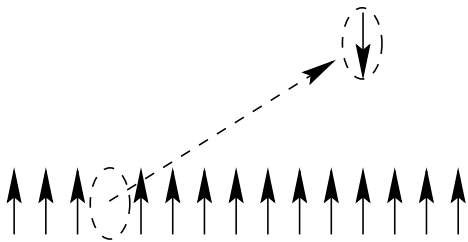}}
\caption[]{Illustration of the fact that the spin flip operator causes
translations when projected into the lowest Landau level. For very large wave
vectors the particles is translated completely away from the exchange hole and
loses all its favorable Coulomb exchange energy.}
\label{fig:phpairflip}
\end{figure}

While the derivation of this exact result for the spin wave dispersion is
algebraically rather simple and looks quite similar (except for the LLL
projection) to the result for the lattice Heisenberg model, it does not give a
very clear physical picture of the nature of the spin wave collective mode. This
we can obtain from eq.~(\ref{eq:052615}) by noting that $\tau_{q}(j)$ translates
the particle a distance $\vec{q} \times \hat{z}\ell^{2}$. Hence the spin wave
operator $\bar{S}_{q}^{-}$ flips the spin of one of the particles and translates
it spatially leaving a hole behind and creating a particle-hole pair carrying
net momentum proportional to their separation as illustrated in
fig.~(\ref{fig:phpairflip}).
For large separations the excitonic Coulomb attraction between the particle and
hole is negligible and the energy cost saturates at a value related to the
Coulomb exchange energy of the ground state given in eq.~(\ref{eq:12109}). The
exact dispersion relation can also be obtained by noting that scattering
processes of the type illustrated by the dashed lines in
fig.~(\ref{fig:phpairflip}) mix together Landau gauge states
\begin{equation}
c_{k - q_{y},\downarrow}^{\dagger}\; c_{k,\uparrow}^{\phantom{\dagger}}\;
|\uparrow\uparrow\uparrow\uparrow\uparrow\uparrow\rangle
\end{equation}
with different wave vectors $k$. Requiring that the state be an eigenvector of
translation uniquely restricts the mixing to linear combinations of the form
\begin{equation}
\sum_{k} e^{-ikq_{x}\ell^{2}}\; c_{k-q_{y},\downarrow}^{\dagger}\;
c_{k,\uparrow}^{\phantom{\dagger}}\;
|\uparrow\uparrow\uparrow\uparrow\uparrow\uparrow\rangle.
\end{equation}
Evaluation of the Coulomb matrix elements shows that this is indeed an exact
eigenstate.

\subsection{Effective Action}
\label{subsec:effectiveaction}

It is useful to try to reproduce these microscopic results for the spin wave
excitations within an effective field theory for the spin degrees of freedom.
Let $\vec{m}(\vec{r}\,)$ be a vector field obeying $\vec{m} \cdot \vec{m} = 1$
which describes the local orientation of the order parameter (the
magnetization). Because the Coulomb forces are spin independent, the potential
energy cost can not depend on the orientation of $\vec{m}$ but only on its
gradients. Hence we must have to leading order in a gradient expansion
\begin{equation}
U = \frac{1}{2} \rho_{s} \int d^{2}r\; \partial_{\mu}m^{\nu}\;
\partial_{\mu}m^{\nu} - \frac{1}{2} n \Delta \int d^{2}r\; m^{z}
\label{eq:1124203}
\end{equation}
where $\rho_{s}$ is a phenomenological `spin stiffness' which in two dimensions
has units of energy and $n \equiv \frac{\nu}{2\pi\ell^{2}}$ is the particle
density. We will learn how to evaluate it later.

We can think of this expression for the energy as the leading terms in a
functional Taylor series expansion. Symmetry requires that (except for the
Zeeman term) the expression for the energy be invariant under uniform global
rotations of $\vec{m}$. In addition, in the absence of disorder, it must be
translationally invariant. Clearly the expression in (\ref{eq:1124203})
satisfies these symmetries. The only zero-derivative term of the appropriate
symmetry is $m^{\mu} m^{\mu}$ which is constrained to be unity everywhere. There
exist terms with more derivatives but these are irrelevant to the physics at
very long wavelengths. (Such terms have been discussed by Read and Sachdev
\cite{ReadandSachdev}.)

To understand how time derivatives enter the effective action we have to recall
that spins obey a first-order (in time) precession equation under the influence
of the local exchange field.\footnote{That is, the Coulomb exchange energy which
tries to keep the spins locally parallel. In a Hartree-Fock picture we could
represent this by a term of the form $-\vec{h}(\vec{r}\,) \cdot \vec{s}(\vec{r}\,)$
where $\vec{h}(\vec{r}\,)$ is the self-consistent field.} Consider as a toy model
a single spin in an external field $\vec{\Delta}$.
\begin{equation}
H = -\hbar\Delta^{\alpha} S^{\alpha}
\end{equation}
The Lagrangian describing this toy model
needs to contain a first order time derivative and so must have
the form (see discussion in appendix~\ref{app:BerryPhase})
\begin{equation}
\mathcal{L} = \hbar S\; \left\{- \dot{m}^{\mu} \mathcal{A}^{\mu}[\vec{m}] +
\Delta^{\mu} m^{\mu} + \lambda (m^{\mu}m^{\mu} - 1)\right\}
\label{eq:1124204}
\end{equation}
where $S = \frac{1}{2}$ is the spin length and $\lambda$ is a Lagrange
multiplier to enforce the fixed length constraint. The unknown vector
$\vec{\mathcal{A}}$ can be determined by requiring that it reproduce the correct
precession equation of motion. The precession equation is
\begin{eqnarray}
\frac{d}{dt} S^{\mu} &=& \frac{i}{\hbar} [H,S^{\mu}] = -i\Delta^{\alpha}
[S^{\alpha},S^{\mu}]\nonumber\\
&=& \epsilon^{\alpha\mu\beta} \Delta^{\alpha} S^{\beta}\\
\dot{\vec{S}} &=& -{\vec \Delta}  \times \vec{S}
\label{eq:precession}
\end{eqnarray}
which corresponds to \textit{counterclockwise} precession around the magnetic
field.

We must obtain the same equation of motion from 
the Euler-Lagrange equation for the Lagrangian in eq.~(\ref{eq:1124204})
\begin{equation}
\frac{d}{dt}\; \frac{\delta\mathcal{L}}{\delta\dot{m}^{\mu}} -
\frac{\delta\mathcal{L}}{\delta m^{\mu}} = 0
\end{equation}
which may be written as
\begin{equation}
\Delta^{\mu} + 2\lambda m^{\mu} = F^{\mu\nu} \dot{m}^{\nu}
\label{eq:060301}
\end{equation}
where
\begin{equation}
F^{\mu\nu} \equiv \partial_{\mu} \mathcal{A}_{\nu} - \partial_{\nu}
\mathcal{A}_{\mu}
\end{equation}
and $\partial_{\mu}$ means $\frac{\partial}{\partial m^{\mu}}$ (\textit{not} the
derivative with respect to some spatial coordinate). Since
$F^{\mu\nu}$ is antisymmetric let us guess a solution of the form
\begin{equation}
F^{\mu\nu} = \epsilon^{\alpha\mu\nu} m^{\alpha}.
\label{eq:060303}
\end{equation}
Using this in 
eq.~(\ref{eq:060301}) yields
\begin{equation}
\Delta^{\mu} + 2\lambda m^{\mu} = \epsilon^{\alpha\mu\nu} m^{\alpha}
\dot m^{\nu}.
\end{equation}
Applying $\epsilon^{\gamma\beta\mu}m^\beta$ to both sides and 
using the identity
\begin{equation}
\epsilon^{\nu\alpha\beta} \epsilon^{\nu\lambda\eta} = \delta_{\alpha\lambda}
\delta_{\beta\eta} - \delta_{\alpha\eta} \delta_{\beta\lambda}
\end{equation}
we obtain
\begin{equation}
-(\vec\Delta\times\vec m)^\gamma = \dot m^\gamma - m^\gamma(\dot m^\beta m^\beta).
\end{equation}
The last term on the right vanishes due to the length constraint.  Thus we find that our ansatz
in eq.~(\ref{eq:060303}) does indeed make the Euler-Lagrange equation correctly reproduce
eq.~(\ref{eq:precession}).

\begin{figure}
\centerline{\epsfxsize=6cm
 \epsffile{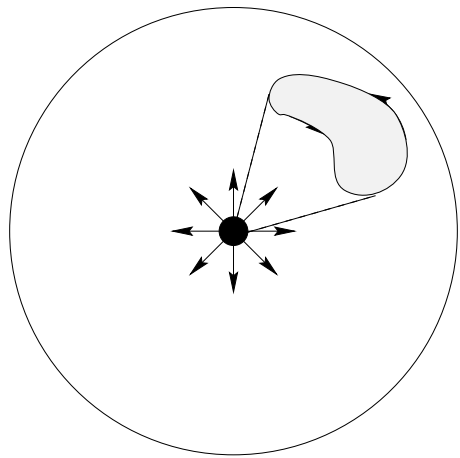}}
\caption[]{Magnetic monopole in spin space. Arrows indicate the curl of the
Berry connection $\vec{\nabla} \times \vec{\mathcal{A}}$ emanating from the
origin. Shaded region indicates closed path $\vec{m}(t)$ taken by the spin order
parameter during which it acquires a Berry phase proportional to the monopole
flux passing through the shaded region.}
\label{fig:monopole}
\end{figure}

Eq.~(\ref{eq:060303}) is equivalent to
\begin{equation}
\vec{\nabla}_{m} \times \vec{\mathcal{A}} [\vec{m}] = \vec{m}
\label{eq:060308}
\end{equation}
indicating that $\vec{\mathcal{A}}$ is the vector potential of a unit magnetic
monopole sitting at the center of the unit sphere on which $\vec{m}$ lives as
illustrated in fig.~(\ref{fig:monopole}).  Note (the always confusing point) that we are
interpreting $\vec m$ as the coordinate of a fictitious particle living on the unit sphere
(in spin space) surrounding the monopole.

Recalling eq.~(\ref{eq:classicalLagrangian}), we see that
the Lagrangian for a single spin in eq.~(\ref{eq:1124204}) is equivalent to
the Lagrangian of a massless object of charge $-S$, 
located at position $\vec m$, moving on the unit sphere containing a
magnetic monopole. The Zeeman term represents a constant electric field
$-\vec{\Delta}$ producing a force $\vec{\Delta}S$ on the particle. The
Lorentz force caused by the monopole causes the particle to orbit the sphere at
constant `latitude'. Because no kinetic term of the form $\dot{m}^{\alpha}
\dot{m}^{\alpha}$ enters the Lagrangian, the charged particle is massless and so
lies only in the lowest Landau level of the monopole field. Note the similarity
here to the previous discussion of the high field limit and the semiclassical
percolation picture of the integer Hall effect. For further details the reader
is directed to appendix~\ref{app:BerryPhase} and to Haldane's discussion of
monopole spherical harmonics \cite{HaldaneSMGbook}.

If the `charge' moves slowly around a closed counterclockwise path $\vec{m}(t)$
during the time interval $[0,T]$ as illustrated in fig.~(\ref{fig:monopole}),
the quantum amplitude
\begin{equation}
e^{\frac{i}{\hbar}\int_{0}^{T}dt\mathcal{L}}
\end{equation}
contains a Berry's phase \cite{Berry} contribution proportional to the
`magnetic flux' enclosed by the path
\begin{equation}
e^{-iS\int_{0}^{T}dt\dot{m}^{\nu}\mathcal{A}^{\nu}} =
e^{-iS\oint\vec{\mathcal{A}}\cdot d\vec{m}}.
\end{equation}
As discussed in appendix~\ref{app:BerryPhase}, this is a purely geometric phase in
the sense that it depends only on the geometry of the path and not the rate at
which the path is traversed (since the expression is time reparameterization
invariant). Using Stokes theorem and eq.~(\ref{eq:060308}) we can write the
contour integral as a surface integral
\begin{equation}
e^{-iS\oint\vec{\mathcal{A}}\cdot d\vec{m}} = e^{-iS\int
d\vec{\Omega}\cdot\vec{\nabla}\times\vec{\mathcal{A}}} = e^{-iS\Omega}
\end{equation}
where $d\vec{\Omega} = \vec{m}d\Omega$ is the directed area (solid angle)
element and $\Omega$ is the total solid angle subtended by the contour as viewed
from the position of the monopole. Note from fig.~(\ref{fig:monopole}) that
there is an ambiguity on the sphere as to which is the inside and which is the
outside of the contour. Since the total solid angle is $4\pi$ we could equally
well have obtained\footnote{The change in the sign from $+i$ to $-i$ is due to
the fact that the contour switches from being counterclockwise to clockwise if
viewed as enclosing the $4\pi - \Omega$ area instead of the $\Omega$ area.}
\begin{equation}
e^{+iS(4\pi-\Omega)}.
\end{equation}
Thus the phase is ambiguous unless $S$ is an integer or half-integer. This
constitutes a `proof' that the quantum spin length must be quantized.

Having obtained the correct Lagrangian for our toy model we can now readily
generalize it to the spin wave problem using the potential energy in
eq.~(\ref{eq:1124203})
\begin{eqnarray}
\mathcal{L} &=& -\hbar Sn \int d^{2}r\; \Biggl\{\dot{m}^{\mu}(\vec{r}\,)\;
\mathcal{A}^{\mu}[\vec{m}] - \Delta m^{z}(\vec{r}\,)\Biggr\}\nonumber\\
&&-\frac{1}{2} \rho_{s} \int d^{2}r\; \partial_{\mu} m^{\nu} \partial_{\mu}
m^{\nu} + \int d^{2}r\; \lambda(\vec{r}\,)\; (m^{\mu} m^{\mu} - 1).
\label{eq:1124219}
\end{eqnarray}
The classical equation of motion can be analyzed just as for the toy model,
however we will take a slightly different approach here. Let us look in the low
energy sector where the spins all lie close to the $\hat{z}$ direction. Then we
can write
\begin{eqnarray}
\vec{m} &=& \left(m^{x}, m^{y}, \sqrt{1-m^{x}m^{x}-m^{y}m^{y}}\right)\nonumber\\
&\approx& \left(m^{x}, m^{y}, 1 - \frac{1}{2} m^{x}m^{x} - \frac{1}{2}
m^{y}m^{y}\right).
\end{eqnarray}
Now choose the `symmetric gauge'
\begin{equation}
\vec{\mathcal{A}} \approx \frac{1}{2} (-m^{y}, m^{x}, 0)
\end{equation}
which obeys eq.~(\ref{eq:060308}) for $\vec{m}$ close to $\hat{z}$.

Keeping only quadratic terms in the Lagrangian we obtain
\begin{eqnarray}
\mathcal{L} &=& -\hbar Sn \int d^{2}r\; \bigg\{\frac{1}{2} (\dot{m}^{y} m^{x} -
\dot{m}^{x} m^{y}) \nonumber\\
&&- \Delta \left(1 - \frac{1}{2} m^{x}m^{x} - \frac{1}{2}
m^{y}m^{y}\right)\bigg\}\nonumber\\
&&-\frac{1}{2} \rho_{s} \int d^{2}r\; (\partial_{\mu}m^{x} \partial_{\mu}m^{x} +
\partial_{\mu}m^{y} \partial_{\mu}m^{y}).
\end{eqnarray}
This can be conveniently rewritten by defining a complex field
\begin{displaymath}
\psi \equiv m^{x} + im^{y}
\end{displaymath}
\begin{eqnarray}
\mathcal{L} &=&-Sn \hbar \int d^{2}r\; \bigg\{\frac{1}{4} 
\left[\psi^{*}\left(-i\frac{\partial}{\partial t}\right)\psi - 
\psi\left(-i\frac{\partial}{\partial t}\right)\psi^{*}\right] \nonumber\\
&&- \Delta \left(1 - \frac{1}{2} \psi^{*}\psi\right)\bigg\}
-\frac{1}{2} \rho_{s} \int d^{2}r\; \partial_{\mu}\psi^{*} \partial_{\mu}\psi
\end{eqnarray}
The classical equation of motion is the Schr\"{o}dinger like equation
\begin{equation}
+i\hbar\frac{\partial\psi}{\partial t} = -\frac{\rho_{s}}{nS}
\partial_{\mu}^{2}\psi + \hbar\Delta\psi.
\end{equation}
This has plane wave solutions with quantum energy
\begin{equation}
\epsilon_{k} = \hbar\Delta + \frac{\rho_{s}}{nS} k^{2}.
\label{eq:1105228}
\end{equation}
We can fit the phenomenological stiffness to the exact dispersion relation in
eq.~(\ref{eq:1124198}) to obtain
\begin{equation}
\rho_{s} = \frac{nS}{4} \sum_{k\neq 0} e^{-\frac{1}{2}|k|^{2}}\; v(k) |k|^{2}.
\label{eq:060322}
\end{equation}

\boxedtext{\begin{exercise}
Derive eq.~(\ref{eq:060322}) from first principles by evaluating the loss of
exchange energy when the Landau gauge $\nu = 1$ ground state is distorted to
make the spin tumble in the $x$ direction
\begin{equation}
|\psi\rangle = \prod_{k} \left(\cos{\frac{\theta_{k}}{2}}
c_{k\uparrow}^{\dagger} + \sin{\frac{\theta_{k}}{2}}
c_{k\downarrow}^{\dagger}\right) |0\rangle
\end{equation}
where $\theta_{k} = -\gamma k\ell^{2}$ and $\gamma =
\frac{\partial\theta}{\partial x}$ is the (constant) spin rotation angle
gradient (since $x = -k\ell^{2}$ in this gauge).
\label{ex:9809}
\end{exercise}}

\subsection{Topological Excitations}
\label{subsec:topological}

\begin{figure}
\centerline{\epsfxsize=10cm
 \epsffile{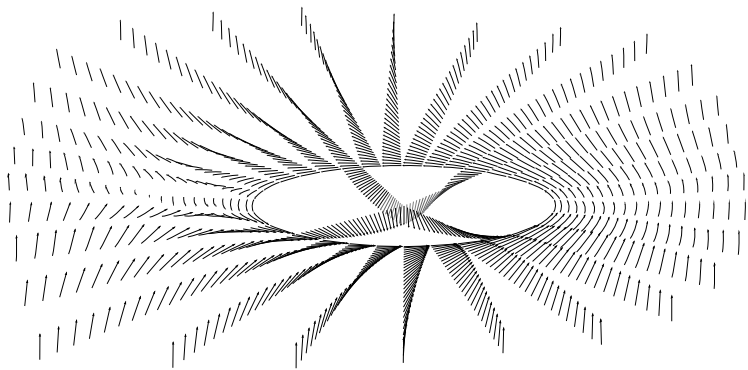}}
\caption[]{Illustration of a skyrmion spin texture. The spin is down at the
origin and gradually turns up at infinite radius. At intermediate distances, the
XY components of the spin exhibit a vortex-like winding. Unlike a $U(1)$ vortex,
there is no singularity at the origin.}
\label{fig:skyrmion}
\end{figure}

So far we have studied neutral collective excitations that take the form of spin
waves. They are neutral because as we have seen from eq.~(\ref{eq:052615}) they
consist of a particle-hole pair. For very large momenta the spin-flipped particle
is translated a large distance $\vec{q} \times \hat{z}\ell^{2}$ away from its
original position as discussed in appendix~\ref{app:projection}. This looks
locally like a charged excitation but it is very expensive because it loses all
of its exchange energy. It is sensible to inquire if it is possible to make a
cheaper charged excitation. This can indeed be done by taking into account the
desire of the spins to be locally parallel and producing a smooth topological
defect in the spin orientation
\cite{LeeandKane,Sondhi,Ilong,Tsvelik,Rodriguez,Abolfath,Apel,GreenTsvelik}
known as a skyrmion by analogy with related objects in the Skyrme model of
nuclear physics \cite{skyrme}. Such an object has the beautiful form exhibited
in fig.~(\ref{fig:skyrmion}).
Rather than having a single spin suddenly flip over, this object gradually turns
over the spins as the center is approached. At intermediate distances the spins
have a vortex-like configuration. However unlike a $U(1)$ vortex, there is no
singularity in the core region because the spins are able to rotate downwards
out of the xy plane.

In nuclear physics the Skyrme model envisions that the vacuum is a `ferromagnet'
described by a four component field $\Phi^{\mu}$ subject to the constraint
$\Phi^{\mu} \Phi^{\mu} = 1$. There are three massless (i.e. linearly dispersing)
spin wave excitations corresponding to the three directions of oscillation about
the ordered direction. These three massless modes represent the three (nearly)
massless pions $\pi^{+}, \pi^{0}, \pi^{-}$. The nucleons (proton and neutron)
are represented by skyrmion spin textures. Remarkably, it can be shown (for an
appropriate form of the action) that these objects are \textit{fermions} despite
the fact that they are in a sense made up of a coherent superposition of (an
infinite number of) \textit{bosonic} spin waves.

We shall see a very similar phenomenology in QHE ferromagnets. At filling factor
$\nu$, skyrmions have charge $\pm \nu e$ and fractional statistics much like
Laughlin quasiparticles. For $\nu = 1$ these objects are fermions. Unlike
Laughlin quasiparticles, skyrmions are extended objects, and they involve many
flipped (and partially flipped) spins. This property has profound implications
as we shall see.

Let us begin our analysis by understanding how it is that spin textures can
carry charge. It is clear from the Pauli principle that it is \textit{necessary}
to flip at least some spins to locally increase the charge density in a $\nu =
1$ ferromagnet. What is the \textit{sufficient} condition on the spin
distortions in order to have a density fluctuation? Remarkably it turns out to
be possible, as we shall see, to uniquely express the charge density solely in
terms of gradients of the local spin orientation.

Consider a ferromagnet with local spin orientation $\vec{m}(\vec{r}\,)$ which is
static. As each electron travels we assume that the strong exchange field keeps
the spin following the local orientation $\vec{m}$. If the electron has velocity
$\dot{x}^{\mu}$, the rate of change of the local spin orientation it sees is
$\dot{m}^{\nu} = \dot{x}^{\mu} \frac{\partial}{\partial x^{\mu}} m^{\nu}$. This
in turn induces an additional Berry's phase as the spin orientation varies. Thus
the single-particle Lagrangian contains an additional first order time
derivative in addition to the one induced by the magnetic field coupling to the
orbital motion
\begin{equation}
\mathcal{L}_{0} = -\frac{e}{c} \dot{x}^{\mu} A^{\mu} + \hbar S
\dot{m}^{\nu} \mathcal{A}^{\nu}[\vec{m}].
\end{equation}
Here $A^{\mu}$ refers to the electromagnetic vector potential and
$\mathcal{A}^{\nu}$ refers to the monopole vector potential obeying
eq.~(\ref{eq:060308}) and we have set the mass to zero (i.e. dropped the
$\frac{1}{2}M\; \dot{x}^{\mu} \dot{x}^{\mu}$ term). This can be rewritten
\begin{equation}
\mathcal{L}_{0} = -\frac{e}{c} \dot{x}^{\mu} (A^{\mu} + a^{\mu})
\end{equation}
where (with $\Phi_{0}$ being the flux quantum)
\begin{equation}
a^{\mu} \equiv -\Phi_{0}S \left(\frac{\partial}{\partial x^{\mu}}
m^{\nu}\right)\; \mathcal{A}^{\nu}[\vec{m}]
\end{equation}
represents the `Berry connection', an additional vector potential which
reproduces the Berry phase. The additional fake magnetic flux due to the curl of
the Berry connection is
\begin{eqnarray}
b &=& \epsilon^{\alpha\beta} \frac{\partial}{\partial x^{\alpha}}
a^{\beta}\nonumber\\
&=& -\Phi_{0} S\epsilon^{\alpha\beta} \frac{\partial}{\partial x^{\alpha}}\;
\left(\frac{\partial}{\partial x^{\beta}} m^{\nu}\right)
\mathcal{A}^{\nu}[\vec{m}]\nonumber\\
&=& -\Phi_{0} S\epsilon^{\alpha\beta} \left\{\left(\frac{\partial}{\partial
x^{\alpha}}\; \frac{\partial}{\partial x^{\beta}} m^{\nu}\right)
\mathcal{A}^{\nu}[\vec{m}]\right.\nonumber\\
&&\left.+\left(\frac{\partial}{\partial x^{\beta}} m^{\nu}\right)\;
\frac{\partial m^{\gamma}}{\partial x^{\alpha}}\; \frac{\partial
\mathcal{A}^{\nu}}{\partial m^{\gamma}}\right\}.
\end{eqnarray}
The first term vanishes by symmetry leaving
\begin{equation}
b = -\Phi_{0} S\epsilon^{\alpha\beta} \frac{\partial m^{\nu}}{\partial
x^{\beta}}\; \frac{\partial m^{\gamma}}{\partial x^{\alpha}}\; \frac{1}{2}
F^{\nu\gamma}
\end{equation}
where $F^{\nu\gamma}$ is given by eq.~(\ref{eq:060303}) and we have taken
advantage of the fact that the remaining factors are antisymmetric under the
exchange $\nu \leftrightarrow \gamma$. Using eq.~(\ref{eq:060303}) and setting
$S = \frac{1}{2}$ we obtain
\begin{equation}
b = -\Phi_{0} \tilde{\rho}
\label{eq:060329}
\end{equation}
where
\begin{eqnarray}
\tilde{\rho} &\equiv& \frac{1}{8\pi} \epsilon^{\alpha\beta} \epsilon^{abc} m^{a}
\partial_{\alpha} m^{b} \partial_{\beta} m^{c}\nonumber\\
&=& \frac{1}{8\pi} \epsilon^{\alpha\beta} \vec{m} \cdot \partial_{\alpha}\vec{m}
\times \partial_{\beta}\vec{m}
\end{eqnarray}
is (for reasons that will become clear shortly) called the \textit{topological
density} or the Pontryagin density.

Imagine now that we adiabatically deform the uniformly magnetized spin state
into some spin texture state. We see from eq.~(\ref{eq:060329}) that the orbital
degrees of freedom see this as adiabatically adding additional flux
$b(\vec{r}\,)$. Recall from eq.~(\ref{eq:1124165}) and the discussion of the
charge of the Laughlin quasiparticle, that extra charge density is associated
with extra flux in the amount
\begin{eqnarray}
\delta\rho &=& \frac{1}{c} \sigma_{xy} b\\
\delta\rho &=& \nu e\tilde{\rho}.
\end{eqnarray}
Thus we have the remarkable result that the changes in the electron charge
density are proportional to the topological density.

Our assumption of adiabaticity is valid as long as the spin fluctuation
frequency is much lower than the charge excitation gap. This is an excellent
approximation for $\nu = 1$ and still good on the stronger fractional Hall
plateaus.

It is interesting that the fermionic charge density in this model can be
expressed solely in terms of the vector boson field $\vec{m}(\vec{r}\,)$, but
there is something even more significant here. The skyrmion spin texture has
total topological charge
\begin{equation}
Q_{\mathrm{top}} \equiv \frac{1}{8\pi} \int d^{2}r\; \epsilon^{\alpha\beta}
\vec{m} \cdot \partial_{\alpha}\vec{m} \times \partial_{\beta}\vec{m}
\label{eq:060333}
\end{equation}
which is always an integer. In fact for \textit{any} smooth spin texture in
which the spins at infinity are all parallel, $Q_{\mathrm{top}}$ is always an
integer. Since it is impossible to continuously deform one integer into another,
$Q_{\mathrm{top}}$ is a topological invariant. That is, if $Q_{\mathrm{top}} =
\pm 1$ because a skyrmion (anti-skyrmion) is present, $Q_{\mathrm{top}}$ is
stable against smooth continuous distortions of the field $\vec{m}$. For example
a spin wave could pass through the skyrmion and $Q_{\mathrm{top}}$ would remain
invariant. Thus this charged object is topologically stable and has fermion
number (i.e., the number of fermions (electrons) that flow into the region when the object
is formed)
\begin{equation}
N = \nu Q_{\mathrm{top}}.
\end{equation}
For $\nu = 1$, $N$ is an integer ($\pm 1$ say) and has the fermion number of an
electron. It is thus continuously connected to the single flipped spin example
discussed earlier.

We are thus led to the remarkable conclusion that the spin degree of freedom
couples to the electrostatic potential. Because skyrmions carry charge, we can
affect the spin configuration using electric rather than magnetic fields!

\begin{figure}
\centerline{\epsfxsize=10cm
 \epsffile{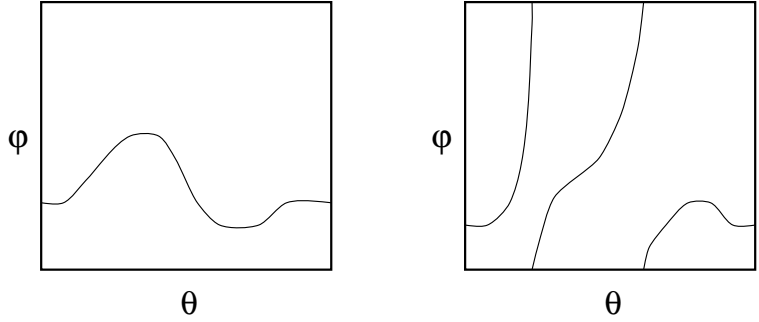}}
\caption[]{Illustration of mappings $\varphi(\theta)$ with: zero winding
number (left) and winding number $+2$ (right).}
\label{fig:toruswinding}
\end{figure}

To understand how $Q_{\mathrm{top}}$ always turns out to be an integer, it is
useful to consider a simpler case of a one-dimensional ring. We follow here the
discussion of \cite{Rajaraman}. Consider the unit circle (known to topologists
as the one-dimensional sphere $S_{1}$). Let the angle $\theta\;
\epsilon[0,2\pi]$ parameterize the position along the curve. Consider a
continuous, suitably well-behaved, complex function $\psi(\theta) =
e^{i\varphi(\theta)}$ defined at each point on the circle and obeying $|\psi| =
1$. Thus associated with each point $\theta$ is another unit circle giving the
possible range of values of $\psi(\theta)$. The function $\psi(\theta)$ thus
defines a trajectory on the torus $S_{1} \times S_{1}$ illustrated in
fig.~(\ref{fig:toruswinding}).
The possible functions $\psi(\theta)$ can be classified into different homotopy
classes according to their winding number $n \in \mathbf{Z}$
\begin{eqnarray}
n &\equiv& \frac{1}{2\pi} \int_{0}^{2\pi} d\theta\; \psi^{*}\left(-i
\frac{d}{d\theta}\right)\psi\nonumber\\
&=& \frac{1}{2\pi} \int_{0}^{2\pi} d\theta\;
\frac{d\varphi}{d\theta} = \frac{1}{2\pi} \left[\varphi(2\pi) -
\varphi(0)\right].
\label{eq:s1winding}
\end{eqnarray}
Because the points $\theta = 0$ and $\theta = 2\pi$ are identified as the same
point
\begin{equation}
\psi(0) = \psi(2\pi) \Rightarrow \varphi(2\pi) - \varphi(0) = 2\pi \times
\mbox{ integer}
\end{equation}
and so $n$ is an integer.
Notice  the crucial role played by the fact that the `topological density'
$\frac{1}{2\pi}\; \frac{d\varphi}{d\theta}$ is the Jacobian for converting from
the coordinate $\theta$ in the domain to the coordinate $\varphi$ in the range.
It is this fact that makes the integral in eq.~(\ref{eq:s1winding}) independent of 
the detailed local form of the mapping $\varphi(\theta)$
and depend only on the overall winding number.
As we shall shortly see,
this same feature will also turn out to be true for the Pontryagin density.

Think of the function $\varphi(\theta)$ as defining the path of an elastic band
wrapped around the torus. Clearly the band can be stretched, pulled and
distorted in any smooth way without any effect on $n$. The only way to change
the winding number from one integer to another is to discontinuously break the
elastic band, unwind (or wind) some extra turns, and then rejoin the cut pieces.

\begin{figure}
\centerline{\epsfxsize=10cm
 \epsffile{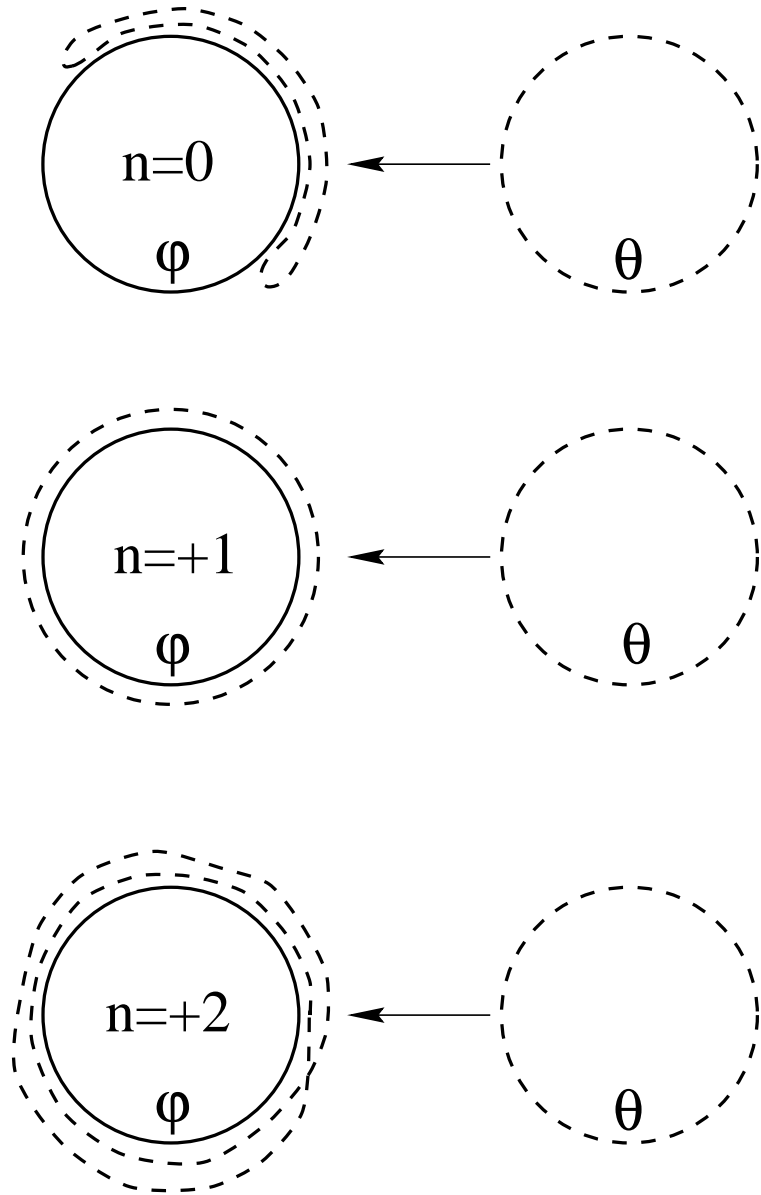}}
\caption[]{A different representation of the mappings from $\theta$ to
$\varphi$. The dashed line represents the domain $\theta$ and the solid line
represents the range $\varphi$. The domain is `lifted up' by the mapping and
placed on the range. The winding number $n$ is the number of times the dashed
circle wraps the solid circle (with a possible minus sign depending on the
orientation).}
\label{fig:homotopySISI}
\end{figure}

Another way to visualize the homotopy properties of mappings from $S_{1}$ to
$S_{1}$ is illustrated in fig.~(\ref{fig:homotopySISI}).
The solid circle represents the domain $\theta$ and the dashed circle
represents the range $\varphi$. It is useful to imagine the $\theta$ circle as
being an elastic band (with points on it labeled by coordinates running from
$0$ to $2\pi$) which can be `lifted up' to the $\varphi$ circle in such a
way that each point of $\theta$ lies just outside the image point
$\varphi(\theta)$. The figure illustrates how the winding number $n$ can be
interpreted as the number of times the domain $\theta$ circle wraps around the
range $\varphi$ circle. (Note:
even though the elastic band is `stretched' and may wrap around
the $\varphi$ circle more than once, its coordinate labels still only run from $0$ to $2\pi$.)
This interpretation is the one which we will generalize
for the case of skyrmions in 2D ferromagnets.

We can think of the equivalence class of mappings having a given winding number
as an element of a group called the homotopy group $\pi_{1}(S_{1})$. The group
operation is addition and the winding number of the sum of two functions,
$\varphi(\theta) \equiv \varphi_{1}(\theta) + \varphi_{2}(\theta)$, is the sum
of the two winding numbers $n = n_{1} + n_{2}$. Thus $\pi_{1}(S_{1})$ is
isomorphic to $\mathbf{Z}$, the group of integers under addition.

Returning now to the ferromagnet we see that the unit vector order parameter
$\vec{m}$ defines a mapping from the plane $R_{2}$ to the two-sphere $S_{2}$
(i.e. an ordinary sphere in three dimensions having a two-dimensional surface).
Because we assume that $\vec{m} = \hat{z}$ for all spatial points far from the
location of the skyrmion, we can safely use a projective map to `compactify'
$R_{2}$ into a sphere $S_{2}$. In this process all points at infinity in $R_{2}$
are mapped into a single point on $S_{2}$, but since $\vec{m}(\vec r)$ is the
same for all these different points, no harm is done. We are thus interested in
the generalization of the concept of the winding number to the mapping $S_{2}
\rightarrow S_{2}$. The corresponding homotopy group $\pi_{2}(S_{2})$ is also
equivalent to $\mathbf{Z}$ as we shall see. 

\begin{figure}
\centerline{\epsfxsize=6cm
 \epsffile{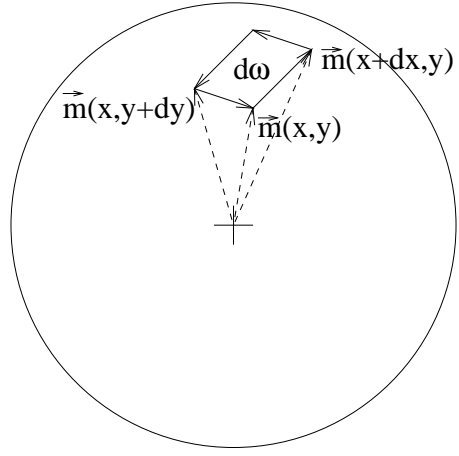}}
\caption[]{Infinitesimal circuit in spin space associated with an infinitesimal circuit in
real space via the mapping $\vec m(\vec r)$.}
\label{fig:mapping}
\end{figure}

Consider the following four points in the plane and their images (illustrated in
fig.~(\ref{fig:mapping})) under the mapping
\begin{eqnarray}
(x,y) &\longrightarrow& \vec{m}(x,y)\nonumber\\
(x+dx,y) &\longrightarrow& \vec{m}(x+dx,y)\nonumber\\
(x,y+dy) &\longrightarrow& \vec{m}(x,y+dy)\nonumber\\
(x+dx,y+dy) &\longrightarrow& \vec{m}(x+dx,y+dy).
\end{eqnarray}
The four points in the plane define a rectangle of area $dxdy$. The four points
on the order parameter (spin) 
sphere define an approximate parallelogram whose area (solid angle) is
\begin{eqnarray}
d\omega &\approx& \left[\vec{m}(x+dx,y) - \vec{m}(x,y)\right] \times
\left[\vec{m}(x,y+dy) - \vec{m}(x,y)\right] \cdot \vec{m}(x,y)\nonumber\\
&\approx& \frac{1}{2} \epsilon^{\mu\nu}\; \vec{m} \cdot \partial_{\mu}\vec{m}
\times \partial_{\nu} \vec{m}\; dxdy\nonumber\\
&=& 4\pi \tilde{\rho}\; dxdy.
\end{eqnarray}
Thus the Jacobian converting area in the plane into solid angle on the sphere is
$4\pi$ times the Pontryagin density $\tilde{\rho}$. This means that the total
topological charge given in eq.~(\ref{eq:060333}) must be an integer since it
counts the number of times the compactified plane is wrapped around the order
parameter sphere by the mapping. The `wrapping' is done by lifting each point
$\vec{r}$ in the compactified plane up to the corresponding point
$\vec{m}(\vec{r}\,)$ on the sphere just as was described for $\pi_{1}(S_{1})$ in
fig.~(\ref{fig:homotopySISI}).

For the skyrmion illustrated in fig.~(\ref{fig:skyrmion}) the order parameter
function $\vec{m}(\vec{r}\,)$ was chosen to be the standard form that minimizes
the gradient energy \cite{Rajaraman}
\alpheqn{
\begin{eqnarray}
m^{x} &=& \frac{2\lambda r\; \cos{(\theta - \varphi)}}{\lambda^{2} +
r^{2}}\label{eq:060339a}\\
m^{y} &=& \frac{2\lambda r\; \sin{(\theta - \varphi)}}{\lambda^{2} +
r^{2}}\label{eq:060339b}\\
m^{z} &=& \frac{r^{2} - \lambda^{2}}{\lambda^{2} + r^{2}}
\label{eq:060339c}
\end{eqnarray}}
\reseteqn

\noindent where $(r,\theta)$ are the polar coordinates in the plane, $\lambda$
is a constant that controls the size scale, and $\varphi$ is a constant that
controls the XY spin orientation. (Rotations about the Zeeman axis leave the
energy invariant.) From the figure it is not hard to see that the skyrmion
mapping wraps the compactified plane around the order parameter sphere exactly
once. The sense is such that $Q_{\mathrm{top}} = -1$.

\boxedtext{\begin{exercise}
Show that the topological density can be written in polar spatial coordinates as
\[
\tilde{\rho} = \frac{1}{4\pi r} \vec{m} \cdot \frac{\partial\vec{m}}{\partial r}
\times \frac{\partial\vec{m}}{\partial\theta}.
\]
Use this result to show
\[
\tilde{\rho} = -\frac{1}{4\pi} \left(\frac{2\lambda}{\lambda^{2} + r^{2}}\right)^{2}
\]
and hence
\[
Q_{\mathrm{top}} = -1
\] for the skyrmion mapping in eqs.~(\ref{eq:060339a}--\ref{eq:060339c}).
\label{ex:9810}
\end{exercise}}

It is worthwhile to note that it is possible to write down simple microscopic
variational wave functions for the skyrmion which are closely related to the
continuum field theory results obtained above. Consider the following state in
the plane \cite{Ilong}
\begin{equation}
\psi_{\lambda} = \prod_j \left(\begin{array}{c}
z_{j} \\ 
\lambda\end{array}\right)_{j} \Psi_{1},
\label{eq:skyrmicro}
\end{equation}
where $\Psi_{1}$ is the $\nu = 1$ filled Landau level state $(\cdot)_{j}$ refers
to the spinor for the $j$th particle, and $\lambda$ is a fixed length scale.
This is a skyrmion because it has its spin purely down at the origin (where
$z_{j} = 0$) and has spin purely up at infinity (where $|z_{j}| \gg \lambda$).
The parameter $\lambda$ is simply the size scale of the skyrmion
\cite{Sondhi,Rajaraman}. At radius $\lambda$ the spinor has equal weight for up
and down spin states (since $|z_{j}| = \lambda$) and hence the spin lies in the
XY plane just as it does for the solution in eq.~(\ref{eq:060339c}). Notice that
in the limit $\lambda \longrightarrow 0$ (where the continuum effective action
is invalid but this microscopic wave function is still sensible) we recover a
fully spin polarized filled Landau level with a charge-1 Laughlin quasihole at
the origin. Hence the number of flipped spins interpolates continuously from
zero to infinity as $\lambda$ increases.

In order to analyze the skyrmion wave function in eq.~(\ref{eq:skyrmicro}), we
use the Laughlin plasma analogy. Recall from our discussion in
sec.~\ref{subsec:nuequalsone} that in this analogy the norm of $\psi_{\lambda}$,
$Tr_{\{\sigma\}} \int D[z]\; |\Psi[z]|^{2}$ is viewed as the partition function of a
Coulomb gas. In order to compute the density distribution we simply need to take
a trace over the spin
\begin{equation}
Z = \int D[z]\; e^{-2\left\{\sum_{i>j} - \log{|z_{i}-z_{j}|} - \frac{1}{2}
\sum_{k} \log{(|z_{k}|^{2}+\lambda^{2})} + \frac{1}{4} \sum_{k}
|z_{k}|^{2}\right\}}. 
\label{eqm20}
\end{equation}
This partition function describes the usual logarithmically interacting Coulomb
gas with uniform background charge plus a spatially varying impurity background
charge $\Delta\rho_b(r)$,
\begin{eqnarray}
\Delta\rho_{b}(r) &\equiv& -\frac{1}{2\pi} \nabla^{2} V(r) =
+\frac{\lambda^{2}}{\pi(r^{2}+\lambda^{2})^{2}}, \label{eqm30}\\
V(r) &=& -\frac{1}{2} \log{(r^{2}+\lambda^{2})}.
\label{eqm40}
\end{eqnarray}

For large enough scale size $\lambda \gg \ell$, local neutrality of the plasma
\cite{jasonho} forces the electrons to be expelled from the vicinity of the
origin and implies that the excess electron number density is precisely
$-\Delta\rho_{b}(r)$, so that eq.~(\ref{eqm30}) is in agreement with the standard
result \cite{Rajaraman} for the topological density given in ex.~\ref{ex:9810}. 

Just as it was easy to find an explicit wave function for the Laughlin
quasi-hole but proved difficult to write down an analytic wave function for the
Laughlin quasi-electron, it is similarly difficult to make an explicit wave
function for the anti-skyrmion. Finally, we note that by replacing $z \choose
\lambda$ by $z^{n} \choose \lambda^{n}$, we can generate a skyrmion with a
Pontryagin index $n$.

\boxedtext{
\begin{exercise}
The argument given above for the charge density of the microscopic skyrmion
state wave function used local neutrality of the plasma and hence is valid only
on large length scales and thus requires $\lambda \gg \ell$. Find the complete
microscopic analytic solution for the charge density valid for arbitrary
$\lambda$, by using the fact that the proposed manybody wave function is nothing
but a Slater determinant of the single particle states $\phi_m(z)$,
\begin{equation}
\phi_m(z) = \frac {z^m}{\sqrt{2\pi 2^{m+1} m!
\left(m+1+\frac{\lambda^{2}}{2}\right)}} {z \choose \lambda} e^{-\frac
{|z^{2}|}{4}}.
\label{eqm50}
\end{equation}
Show that the excess electron number density is then
\begin{equation}
\Delta n^{(1)}(z) \equiv \sum_{m=0}^{N-1} |\phi_m(z)|^{2} -\frac{1}{2\pi},
\label{eqm60}
\end{equation}
which yields
\begin{equation}
\Delta n^{(1)}(z) = \frac{1}{2\pi} \left(\frac{1}{2} \int_{0}^{1} d\alpha\;
\alpha^{\frac{\lambda^{2}}{2}} e^{-\frac{|z|^{2}}{2} (1-\alpha)}
(|z|^{2}+\lambda^{2}) - 1 \right).
\label{eqm70}
\end{equation}
Similarly, find the spin density distribution $S^z(r)$ and show that it also
agrees with the field-theoretic expression in eq.~(\ref{eq:060339c}) in the
large $\lambda$ limit.
\label{ex:9811}
\end{exercise}}

The skyrmion solution in eqs.~(\ref{eq:060339a}--\ref{eq:060339c}) minimizes the
gradient energy
\begin{equation}
E_{0} = \frac{1}{2} \rho_{s} \int d^{2}r\; \partial_{\mu}m^{\nu}
\partial_{\mu}m^{\nu}.
\end{equation}
Notice that the energy cost is scale invariant since this expression contains
two integrals and two derivatives. Hence the total gradient energy is
independent of the scale factor $\lambda$ and for a single skyrmion is given by
\cite{Sondhi,Rajaraman}
\begin{equation}
E_{0} = 4\pi\rho_{s} = \frac{1}{4} \epsilon_{\infty}
\label{eq:1105255}
\end{equation}
where $\epsilon_{\infty}$ is the asymptotic large $q$ limit of the spin wave
energy in eq.~(\ref{eq:1123195}). Since this spin wave excitation produces a
widely separated particle-hole pair, we see that the energy of a widely
separated skyrmion-antiskyrmion pair $\left(\frac{1}{4} + \frac{1}{4}\right)
\epsilon_{\infty}$ is only half as large. Thus skyrmions are considerably
cheaper to create than simple flipped spins.\footnote{This energy advantage is
reduced if the finite thickness of the inversion layer is taken into account.
The skyrmion may in some cases turn out to be disadvantageous in higher Landau
levels.}

Notice that eq.~(\ref{eq:1105255}) tells us that the charge excitation gap,
while only half as large as naively expected, is finite as long as the spin
stiffness $\rho_{s}$ is finite. Thus we can expect a dissipationless Hall
plateau. Therefore, as emphasized by Sondhi \etal\ \cite{Sondhi}, the Coulomb
interaction plays a central role in the $\nu = 1$ integer Hall effect. Without
the Coulomb interaction the charge gap would simply be the tiny Zeeman gap. With
the Coulomb interaction the gap is large even in the limit of zero Zeeman energy
because of the spontaneous ferromagnetic order induced by the spin stiffness.

At precisely $\nu = 1$ skyrmion/antiskyrmion pairs will be thermally activated
and hence exponentially rare at low temperatures. On the other hand, because
they are the cheapest way to inject charge into the system, there will be a
finite density of skyrmions even in the ground state if $\nu \neq 1$. Skyrmions
also occur in ordinary 2D magnetic films but since they do not carry charge (and
are energetically expensive since $\rho_{s}$ is quite large) they readily freeze
out and are not particularly important.

The charge of a skyrmion is sharply quantized but its number of flipped spins
depends on its area $\sim \lambda^{2}$. Hence if the energy were truly scale
invariant, the number of flipped spins could take on any value. Indeed one of
the early theoretical motivations for skyrmions was the discovery in numerical
work by Rezayi \cite{Sondhi,Rezayi} that adding a single charge to a filled
Landau level converted the maximally ferromagnetic state into a spin singlet. In
the presence of a finite Zeeman energy the scale invariance is lost and there is
a term in the energy that scales with $\Delta\lambda^{2}$ and tries to minimize
the size of the skyrmion. Competing with this however is a Coulomb term which we
now discuss.

The Lagrangian in eq.~(\ref{eq:1124219}) contains the correct leading order
terms in a gradient expansion. There are several possible terms which are fourth
order in gradients, but a particular one dominates over the others at long
distances. This is the Hartree energy associated with the charge density of the
skyrmion
\begin{equation}
V_{\mathrm{H}} = \frac{1}{2\epsilon} \int d^{2}r\; \int d^{2}r'\;
\frac{\delta\rho(\vec{r}\,)\; \delta\rho(\vec{r}^{\,\prime})}{|\vec{r} -
\vec{r}^{\,\prime}|}
\label{eq:fourthHartree}
\end{equation}
where
\begin{equation}
\delta\rho = \frac{\nu e}{8\pi}\; \epsilon^{\alpha\beta}\; \vec{m} \cdot
\partial_{\alpha}\vec{m} \times \partial_{\beta}\vec{m}
\end{equation}
and $\epsilon$ is the dielectric constant. The long range of the Coulomb
interaction makes this effectively a three gradient term that distinguishes it
from the other possible terms at this order. Recall that the Coulomb interaction
already entered in lower order in the computation of $\rho_{s}$. That however
was the exchange energy while the present term is the Hartree energy. The
Hartree energy scales like $\frac{e^{2}}{\epsilon\lambda}$ and so prefers to
expand the skyrmion size. The competition between the Coulomb and Zeeman
energies yields an optimal number of approximately four flipped spins according
to microscopic Hartree Fock calculations \cite{FertigHF}.

\begin{figure}
\centerline{\epsfxsize=6cm
 \epsffile{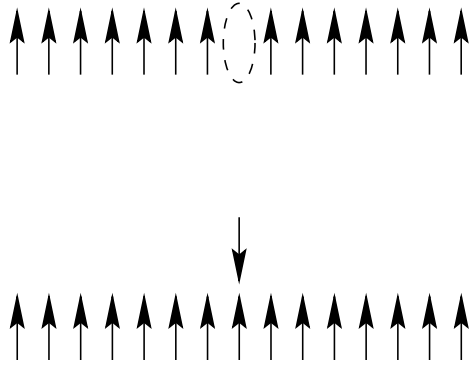}}
\caption[]{Illustration of the spin configurations for non-interacting electrons
at filling factor $\nu=1$ in the presence of a hole (top) and an extra electron
(bottom).}
\label{fig:non-int}
\end{figure}

Thus a significant prediction for this model is that each charge added (or
removed) from a filled Landau level will flip several $(\sim 4)$ spins. This is
very different from what is expected for non-interacting electrons. As
illustrated in fig.~(\ref{fig:non-int})
removing an electron leaves the non-interacting system still polarized. The
Pauli principle forces an added electron to be spin reversed and the
magnetization drops from unity at $\nu = 1$ to zero at $\nu = 2$ where both spin
states of the lowest Landau level are fully occupied.

\begin{figure}
\centerline{\epsfxsize=10cm
 \epsffile{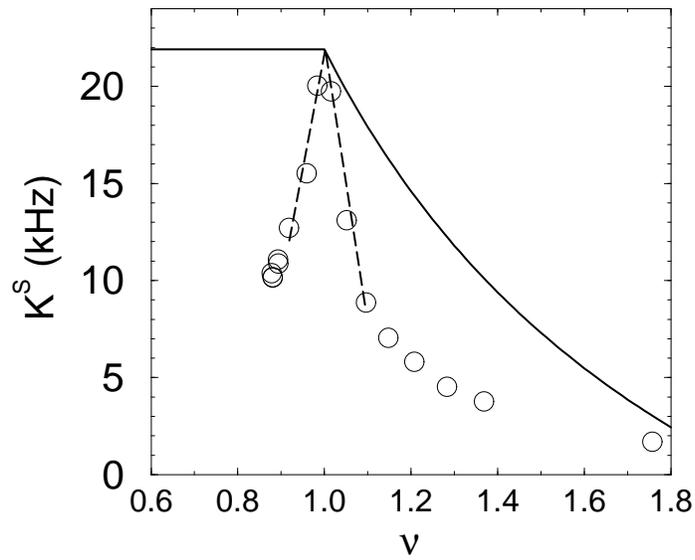}}
\caption[]{NMR Knight shift measurement of the electron spin polarization near
filling factor $\nu=1$. Circles are the data of Barrett \etal~\cite{Barrett}.
The dashed line is a guide to the eye. The solid line is the prediction for
non-interacting electrons. The peak represents 100\% polarization at $\nu=1$.
The steep slope on each side indicates that many ($\sim 4$) spins flip over for
each charge added (or subtracted). The observed symmetry around $\nu=1$ is due
to the particle-hole symmetry between skyrmions and antiskyrmions not present in
the free-electron model.}
\label{fig:knightdata}
\end{figure}

Direct experimental evidence for the existence of skyrmions was first obtained
by Barrett \etal\ \cite{Barrett} using a novel optically pumped NMR technique.
The Hamiltonian for a nucleus is \cite{Slichter}
\begin{equation}
H_{N} = -\Delta_{N}I^{z} + \Omega\vec{I} \cdot \vec{s}
\end{equation}
where $\vec{I}$ is the nuclear angular momentum, $\Delta_{N}$ is the nuclear
Zeeman frequency (about 3 orders of magnitude smaller than the electron Zeeman
frequency), $\Omega$ is the hyperfine coupling and $\vec{s}$ is the electron
spin density at the nuclear site. If, as a first approximation we replace
$\vec{s}$ by its average value
\begin{equation}
H_{N} \approx \left(-\Delta_{N} + \Omega\langle s^{z}\rangle\right)\; I^{z}
\end{equation}
we see that the precession frequency of the nucleus will be shifted by an amount
proportional to the magnetization of the electron gas. The magnetization deduced
using this so-called Knight shift is shown in fig.~(\ref{fig:knightdata}).
The electron gas is 100\% polarized at $\nu = 1$, but the polarization drops
off sharply (and symmetrically) as charge is added or subtracted. This is in
sharp disagreement with the prediction of the free electron model as shown in
the figure. The initial steep slope of the data allows one to deduce that 3.5--4
spins reverse for each charge added or removed. This is in excellent
quantitative agreement with Hartree-Fock calculations for the skyrmion model
\cite{FertigHF}.

Other evidence for skyrmions comes from the large change in Zeeman energy with
field due to the large number of flipped spins. This has been observed in
transport \cite{Eisensteinskyrme} and in optical spectroscopy \cite{BennettGoldberg}. Recall
that spin-orbit effects in GaAs make the electron $g$ factor $-0.4$. Under
hydrostatic pressure $g$ can be tuned towards zero which should greatly enhance
the skyrmion size. Evidence for this effect has been seen \cite{pressuretune}.

\section{Skyrmion Dynamics}
\label{sec:skyrmion}

\begin{figure}
\centerline{\rotatebox{-90}{\epsfysize=10cm \epsffile{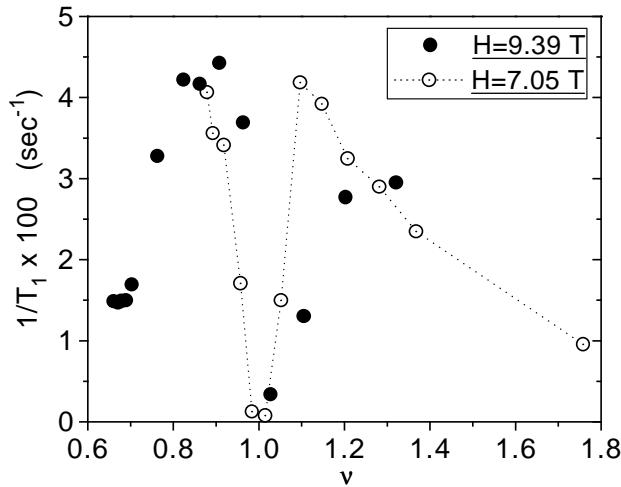}}}
\caption[]{NMR nuclear spin relaxation rate $1/T_1$ as a function of filling
factor. After Tycko \etal~\cite{Tycko}. The relaxation rate is very small at
$\nu=1$, but rises dramatically away from $\nu=1$ due to the presence of
skyrmions.}
\label{fig:relaxrate}
\end{figure}

NMR \cite{Barrett} and nuclear specific heat \cite{Bayot} data indicate that
skyrmions dramatically enhance the rate at which the nuclear spins relax. This
nuclear spin relaxation is due to the transverse terms in the hyperfine
interaction which we neglected in discussing the Knight shift
\begin{equation}
\frac{1}{2} \Omega\; (I^{+}s^{-} + I^{-}s^{+}) = \frac{1}{2} \Omega\; \left\{
I^{+} \sum_{\vec{q}} S_{\vec{q}}^{-} + \mbox{h.c.}\right\}.
\label{eq:060807}
\end{equation}
The free electron model would predict that it would be impossible for an
electron and a nucleus to undergo mutual spin flips because the Zeeman energy
would not be conserved. (Recall that $\Delta_{N} \sim 10^{-3}\Delta$.) The spin
wave model shows that the problem is even worse than this. Recall from
eq.~(\ref{eq:1123195}) that the spin Coulomb interaction makes spin wave energy
much larger than the electron Zeeman gap except at very long wavelengths. The
lowest frequency spin wave excitations lie above 20--50~GHz while the nuclei
precess at 10--100~MHz. Hence the two sets of spins are unable to couple
effectively. At $\nu = 1$ this simple picture is correct. The nuclear relaxation
time $T_{1}$ is extremely long (tens of minutes to many hours depending on the
temperature) as shown in fig.~(\ref{fig:relaxrate}).
However the figure also shows that for $\nu \neq 1$ the relaxation rate
$1/T_{1}$ rises dramatically and $T_{1}$ falls to $\sim 20~\mbox{seconds}$. In
order to understand this dramatic variation we need to develop a theory of spin
dynamics in the presence of skyrmions.

The $1/T_{1}$ data is telling us that for $\nu \neq 1$ at least some of the
electron spin fluctuations are orders of magnitude lower in frequency than the
Zeeman splitting and these low frequency modes can couple strongly to the
nuclei. One way this might occur is through the presence of disorder. We see
from eq.~(\ref{eq:060807}) that NMR is a local probe which couples to spin flip
excitations at all wave vectors. Recall from eq.~(\ref{eq:052615}) that lowest
Landau level projection implies that $\overline{S_{\vec{q}}^{-}}$ contains a
translation operator $\tau_{q}$. In the presence of strong disorder the Zeeman
and exchange cost of the spin flips could be compensated by translation to a
region of lower potential energy. Such a mechanism was studied in
\cite{Antoniou} but does not show sharp features in $1/T_{1}$ around $\nu = 1$.

We are left only with the possibility that the dynamics of skyrmions somehow
involves low frequency spin fluctuations. For simplicity we will analyze this
possibility ignoring the effects of disorder, although this may not be a valid
approximation.

Let us begin by considering a ferromagnetic $\nu = 1$ state containing a single
skyrmion of the form parameterized in eqs.~(\ref{eq:060339a}--\ref{eq:060339c}).
There are two degeneracies at the classical level in the effective field theory:
The energy does not depend on the position of the skyrmion and it does not
depend on the angular orientation $\varphi$. These continuous degeneracies are
known as zero modes \cite{Rajaraman} and require special treatment of the
quantum fluctuations about the classical solution.

In the presence of one or more skyrmions, the quantum Hall ferromagnet is
\textit{non-colinear}. In an ordinary ferromagnet where all the spins are
parallel, global rotations about the magnetization axis only change the quantum phase
of the state --- they do not produce a new state.\footnote{Rotation about the
Zeeman alignment axis is accomplished by $R = e^{-\frac{i}{\hbar}\varphi S^{z}}$.
But a colinear ferromagnet ground state is an eigenstate of $S^{z}$, so rotation
leaves the state invariant up to a phase.} Because the skyrmion has
distinguishable orientation, each one induces a new $U(1)$ degree of freedom in
the system. In addition because the skyrmion has a distinguishable location,
each one induces a new translation degree of freedom. As noted above, both of
these are zero energy modes at the classical level suggesting that they might
well be the source of low energy excitations which couple so effectively to the
nuclei. We shall see that this is indeed the case, although the story is
somewhat complicated by the necessity of correctly quantizing these modes.

Let us begin by finding the effective Lagrangian for the translation mode
\cite{stonebook}. We take the spin configuration to be
\begin{equation}
\vec{m}(\vec{r},t) = \vec{m}_{0}\left(\vec{r} - \vec{R}(t)\right)
\end{equation}
where $\vec{m}_{0}$ is the static classical skyrmion solution and $\vec{R}(t)$
is the position degree of freedom. We ignore all other spin wave degrees of
freedom since they are gapped. (The gapless $U(1)$ rotation mode will be treated
separately below.) Eq.~(\ref{eq:1124219}) yields a Berry phase term
\begin{equation}
\mathcal{L}_{0} = -\hbar S \int d^{2}r\; \dot{m}^{\mu}
\mathcal{A}^{\mu}[\vec{m}]\; n(\vec{r}\,)
\label{eq:060808}
\end{equation}
where
\begin{equation}
\dot{m}^{\mu} = -\dot{R}^{\nu} \frac{\partial}{\partial r^{\nu}}
m_{0}^{\mu}(\vec{r} - \vec{R})
\end{equation}
and unlike in eq.~(\ref{eq:1124219}) we have taken into account our new-found
knowledge that the density is non-uniform
\begin{equation}
n(\vec{r}\,) = n_{0} + \frac{1}{8\pi} \epsilon^{\mu\nu}\; \vec{m} \cdot
\partial_{\mu}\vec{m} \times \partial_{\nu}\vec{m}.
\label{eq:060810}
\end{equation}
The second term in eq.~(\ref{eq:060810}) can be shown to produce an extra Berry
phase when two skyrmions are exchanged leading to the correct minus sign for
Fermi statistics (on the $\nu = 1$ plateau) but we will not treat it further.
Eq.~(\ref{eq:060808}) then becomes
\begin{equation}
\mathcal{L}_{0} = +\hbar\dot{R}^{\nu} a^{\nu}(\vec{r}\,)
\label{eq:060811}
\end{equation}
where the `vector potential'
\begin{equation}
a^{\nu}(\vec{r}\,) \equiv Sn_{0} \int d^{2}r\; (\partial_{\nu}m^{\mu}) {\cal A}^{\mu}
\end{equation}
has curl
\begin{eqnarray}
\epsilon^{\lambda\nu} \frac{\partial}{\partial R^{\lambda}}a^{\nu} &=& -
\epsilon^{\lambda\nu} \frac{\partial}{\partial r^{\lambda}}a^{\nu}\nonumber\\
&=& -Sn_{0}\; \epsilon^{\lambda\nu} \int d^{2}r\; \partial_{\lambda}
\left\{(\partial_{\nu}m^{\mu}) {\cal A}^{\mu}\right\}\nonumber\\
&=& -Sn_{0}\; \epsilon^{\lambda\nu} \int d^{2}r\; (\partial_{\nu}m^{\mu})\;
(\partial_{\lambda}m^{\gamma}) \frac{\partial {\cal A}^{\mu}}{\partial
m^{\gamma}}\nonumber\\
&=& -\frac{Sn_{0}}{2} \int d^{2}r\; \epsilon^{\lambda\nu}\; \partial_{\nu}m^{\mu}
\partial_{\lambda}m^{\gamma} F^{\gamma\mu}\nonumber\\
&=& -2\pi n_{0} Q_{\mathrm{top}}
\end{eqnarray}
Thus eq.~(\ref{eq:060811}) corresponds to the kinetic Lagrangian for a 
massless particle
of charge $-eQ_{\mathrm{top}}$ moving in a uniform magnetic field of strength $B
= \frac{\Phi_{0}}{2\pi\ell^{2}}$. But this of course is precisely what the
skyrmion is \cite{stonebook}. 

We have kept here only the lowest order adiabatic time derivative term in the
action.\footnote{There may exist higher-order time-derivative terms which give
the skyrmion a mass and there will also be damping due to radiation of spin
waves at higher velocities. \cite{fertigradiation}} This is justified by the
existence of the spin excitation gap and the fact that we are interested only in
much lower frequencies (for the NMR). 

If we ignore the disorder potential then the kinetic Lagrangian simply leads to
a Hamiltonian that yields quantum states in the lowest Landau level, all of
which are degenerate in energy and therefore capable of relaxing the nuclei
(whose precession frequency is extremely low on the scale of the electronic
Zeeman energy).

Let us turn now to the rotational degree of freedom represented by the
coordinate $\varphi$ in eqs.~(\ref{eq:060339a}--\ref{eq:060339c}). The full
Lagrangian is complicated and contains the degrees of freedom of the continuous
field $\vec m(\vec r)$. We need to introduce the collective coordinate $\varphi$
describing the orientation of the skyrmion as one of the degrees of freedom and
then carry out the Feynman path integration over the quantum fluctuations in all
the infinite number of remaining degrees of freedom.\footnote{Examples of how to
do this are discussed in various field theory texts, including Rajaraman
\cite{Rajaraman}.} This is a non-trivial task, but fortunately we do not
actually have to carry it out. Instead we will simply write down the answer. The
answer is some functional of the path for the single variable $\varphi(t)$. We
will express this functional (using a functional Taylor series expansion) in the
most general form possible that is consistent with the symmetries in the
problem. Then we will attempt to identify the meaning of the various terms in
the expansion and evaluate their coefficients (or assign them values
phenomenologically). After integrating out the high frequency spin wave
fluctuations, the lowest-order symmetry-allowed terms in the action are
\begin{equation}
\mathcal{L}_{\varphi} = \hbar K \dot{\varphi} + \frac{\hbar^{2}}{2U}
\dot{\varphi}^{2} + \ldots
\label{eq:060814}
\end{equation}
Again, there is a first-order term allowed by the lack of time-reversal symmetry
and we have included the leading non-adiabatic correction. The full action
involving $\vec{m}(\vec{r},t)$ contains only a first-order time derivative but a
second order term is allowed by symmetry to be generated upon integrating out
the high frequency fluctuations. We will not perform this explicitly but rather
treat $U$ as a phenomenological fitting parameter.

The coefficient $K$ can be computed exactly since it is simply the Berry phase
term. Under a slow rotation of all the spins through $2\pi$ the Berry phase is
(using eq.~(\ref{eq:berry22}) in appendix~\ref{app:BerryPhase})
\begin{equation}
\int d^{2}r\; n(\vec{r}\,)\; (-S2\pi)\; \left[1 - m_{0}^{z}(\vec{r}\,)\right] =
{\frac{1}{\hbar} \int_{0}^{T} \mathcal{L}_{\varphi}}  = 2\pi K.
\end{equation}
(The non-adiabatic term gives a $1/T$ contribution that vanishes in the
adiabatic limit $T \rightarrow \infty$.) Thus we arrive at the important
conclusion that $K$ is the expectation value of the number of overturned spins
for the classical solution $\vec{m}_{0}(\vec{r}\,)$. We emphasize that this is
the Hartree-Fock (i.e., `classical') skyrmion solution and therefore $K$ need
not be an integer.

The canonical angular momentum conjugate to $\varphi$ in eq.~(\ref{eq:060814})
is
\begin{equation}
L_{z} = \frac{\delta\mathcal{L}_{\varphi}}{\delta\dot{\varphi}} = \hbar K +
\frac{\hbar^{2}}{U} \dot{\varphi}
\end{equation}
and hence the Hamiltonian is
\begin{eqnarray}
H_{\varphi} &=& L_{z} \dot{\varphi} - \mathcal{L}_{\varphi}\nonumber\\
&=& \left(\hbar K + \frac{\hbar^{2}}{U} \dot{\varphi}\right) \dot{\varphi} -
\hbar K - \frac{\hbar^{2}}{2U} \dot{\varphi}^{2}\nonumber\\
&=& +\frac{\hbar^{2}}{2U} \dot{\varphi}^{2}=\frac{U}{2\hbar^{2}} (L_{z} - \hbar K)^{2}
\label{eq:1268old}
\end{eqnarray}
Having identified the Hamiltonian and expressed it in terms of the coordinate
and the canonical momentum conjugate to that coordinate, we quantize $H_\varphi$
 by simply making the substitution
\begin{equation}
L_z \longrightarrow -i\hbar\frac{\partial}{\partial\varphi}
\end{equation}
to obtain
\begin{equation}
H_\varphi=  +\frac{U}{2}\; 
\left(-i\frac{\partial}{\partial\varphi} - K\right)^{2}.
\label{eq:1268}
\end{equation}
This can be interpreted as the Hamiltonian of a (charged) XY quantum rotor
with moment of inertia $\hbar^{2}/U$ circling a solenoid containing $K$ flux
quanta. (The Berry phase term in eq.~(\ref{eq:060814}) is then interpreted as
the Aharonov-Bohm phase.) The eigenfunctions are
\begin{equation}
\psi_{m}(\varphi) = \frac{1}{\sqrt{2\pi}} e^{im\varphi}
\label{eq:060818}
\end{equation}
and the eigenvalues are
\begin{equation}
\epsilon_{m} = \frac{U}{2} (m - K)^{2}.
\label{eq:060819}
\end{equation}
The angular momentum operator $L_{z}$ is actually the operator giving the number
of flipped spins in the skyrmion. Because of the rotational symmetry about the
Zeeman axis, this is a good quantum number and therefore takes on integer
values (as required in any quantum system of finite size with rotational symmetry
about the $z$ axis). The ground state value of $m$ is the nearest integer to
$K$. The ground state angular velocity is
\begin{equation}
\dot{\varphi} = \left\langle\frac{\partial H_{\varphi}}{\partial
L_{z}}\right\rangle = \frac{U}{\hbar} (m - K).
\end{equation}
Hence if $K$ is not an integer the skyrmion is spinning around at a finite
velocity. In any case the actual orientation angle $\varphi$ for the skyrmion is
completely uncertain since from eq.~(\ref{eq:060818})
\begin{equation}
\left|\psi_{m}(\varphi)\right|^{2} = \frac{1}{2\pi}
\end{equation}
$\varphi$ has a flat probability distribution (due to quantum zero point
motion). We interpret this as telling us that the global U(1) rotation symmetry
broken in the classical solution is restored in the quantum solution because of
quantum fluctuations in the coordinate $\varphi$. This issue will arise again in
our study of the Skyrme lattice where we will find that for an infinite array of
skyrmions, the symmetry can sometimes remain broken.

Microscopic analytical \cite{Breywithoutsigma} and numerical \cite{FertigHF}
calculations do indeed find a family of low energy excitations with an
approximately parabolic relation between the energy and the number of flipped
spins just as is predicted by eq.~(\ref{eq:060819}). As mentioned earlier, $K
\sim 4$ for typical parameters. Except for the special case where $K$ is a half
integer the spectrum is non-degenerate and has an excitation gap on the scale of
$U$ which is in turn some fraction of the Coulomb energy scale $\sim
100~\mbox{K}$. In the absence of disorder even a gap of only 1~K would make
these excitations irrelevant to the NMR. We shall see however that this
conclusion is dramatically altered in the case where many skyrmions are present.

\subsection{Skyrme Lattices}

\begin{figure}
\centerline{\rotatebox{-90}{\epsfysize=10cm\epsffile{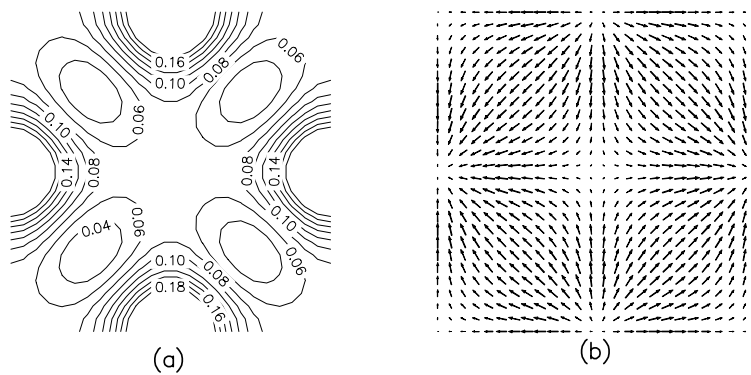}}}
\caption[]{Electronic structure of the skyrmion lattice as determined by
numerical Hartree-Fock calculations for filling factor $\nu=1.1$ and Zeeman
energy $0.015\frac{e^{2}}{\epsilon\ell}$. (a) Excess charge density (in units of
$1/(2\pi\ell^{2})$) and (b) Two-dimensional vector representation of the XY
components of the spin density. The spin stiffness makes the square lattice more
stable than the triangular lattice at this filling factor and Zeeman coupling.
Because of the $U(1)$ rotational symmetry about the Zeeman axis, this is simply
one representative member of a continuous family of degenerate Hartree-Fock
solutions. After Brey \etal~\cite{Breywithoutsigma}.}
\label{fig:skyrmelattice}
\end{figure}

For filling factors slightly away from $\nu = 1$ there will be a finite density
of skyrmions or antiskyrmions (all with the same sign of topological charge) in
the ground state \cite{Breyxtal,usPRL,GreenTsvelik}. Hartree-Fock calculations
\cite{Breyxtal} indicate that the ground state is a Skyrme crystal. Because the
skyrmions are charged, the Coulomb potential in eq.~(\ref{eq:fourthHartree}) is
optimized for the triangular lattice. This is indeed the preferred structure for
very small values of $|\nu - 1|$ where the skyrmion density is low. However at
moderate densities the square lattice is preferred. The Hartree-Fock ground
state has the angular variable $\varphi_{j}$ shifted by $\pi$ between
neighboring skyrmions as illustrated in fig.~(\ref{fig:skyrmelattice}).
This `antiferromagnetic' arrangement of the XY spin orientation minimizes the
spin gradient energy and would be frustrated on the triangular lattice. Hence it
is the spin stiffness that stabilizes the square lattice structure.

The Hartree-Fock ground state breaks both global translation and global $U(1)$
spin rotation symmetry. It is a kind of `supersolid' with both diagonal
\begin{equation}
G^{z} \equiv \left\langle s^{z}(\vec{r}\,)\;
s^{z}(\vec{r}^{\,\prime})\right\rangle
\end{equation}
and off-diagonal
\begin{equation}
G^{\perp} \equiv \left\langle s^{+}(\vec{r}\,)\;
s^{-}(\vec{r}^{\,\prime})\right\rangle
\end{equation}
long-range order. For the case of a single skyrmion we found that the $U(1)$
symmetry was broken at the Hartree-Fock (classical) level but fully restored by
quantum fluctuations of the zero mode coordinate $\varphi$. In the thermodynamic
limit of an infinite number of skyrmions coupled together, it is possible for
the global $U(1)$ rotational symmetry breaking to survive quantum
fluctuations.\footnote{Loosely speaking this corresponds to the infinite system
having an infinite moment of inertia (for global rotations) which allows a
quantum wave packet which is initially localized at a particular orientation
$\varphi$ not to spread out even for long times.} If this occurs then an
excitation gap is \textit{not} produced. Instead we have a new kind of gapless
spin wave Goldstone mode \cite{Senthil,skyrmelatticePRL}. This mode is gapless
despite the presence of the Zeeman field and hence has a profound effect on the
NMR relaxation rate. The gapless Goldstone mode associated with the broken
translation symmetry is the ordinary magneto-phonon of the Wigner crystal. This
too contributes to the nuclear relaxation rate.

In actual practice, disorder will be important. In addition, the NMR experiments
have so far been performed at temperatures which are likely well above the
lattice melting temperature. Nevertheless the zero temperature lattice
calculations to be discussed below probably capture the essential physics of
this non co-linear magnet. Namely, there exist spin fluctuations at frequencies
orders of magnitude below the Zeeman gap. At zero temperature these are coherent
Goldstone modes. Above the lattice melting temperature they will be overdamped
diffusive modes derived from the Goldstone modes. The essential physics will
still be that the spin fluctuations have strong spectral density at frequencies
far below the Zeeman gap.

It turns out that at long wavelengths the magnetophonon and $U(1)$ spin modes
are decoupled. We will therefore ignore the positional degrees of freedom when
analyzing the new $U(1)$ mode. We have already found the $U(1)$ Hamiltonian for
a single skyrmion in eq.~(\ref{eq:1268}). The simplest generalization to the
Skyrme lattice which is consistent with the symmetries of the problem is
\begin{equation}
H = \frac{U}{2} \sum_{j} (\hat{K}_{j} - K)^{2} - J \sum_{\langle ij\rangle}
\cos{(\varphi_{i} - \varphi_{j})}
\label{eq:061103}
\end{equation}
where $\hat{K}_{j} \equiv -i \frac{\partial}{\partial\varphi_{j}}$ is the
angular momentum operator. The global $U(1)$ symmetry requires that the
interactive term be invariant if all of the $\varphi_{j}$'s are increased by a
constant. In addition $H$ must be invariant under $\varphi_{j} \rightarrow
\varphi_{j} + 2\pi$ for any single skyrmion. We have assumed the simplest
possible near-neighbor coupling, neglecting the possibility of longer range
higher-order couplings of the form $\cos n(\varphi_{i} - \varphi_{j})$ which are
also symmetry allowed. The phenomenological coupling $J$ must be negative to be
consistent with the `antiferromagnetic' XY order found in the Hartree-Fock
ground state illustrated in fig.~(\ref{fig:skyrmelattice}).   However we will find it
convenient to instead make $J$ positive and compensate for this by a `gauge' change
$\varphi_j\rightarrow \varphi_j+\pi$ on one sublattice.  This is convenient 
because it makes the coupling `ferromagnetic' rather than `antiferromagnetic.'

Eq.~(\ref{eq:061103}) is the Hamiltonian for the quantum XY rotor model, closely
related to the boson Hubbard model \cite{Cha,Sorensen,Fisherboson}. Readers
familiar with superconductivity will recognize that this model is commonly used
to describe the superconductor-insulator transition in Josephson arrays
\cite{Cha,Sorensen}. The angular momentum eigenvalue of the $\hat{K}_{j}$
operator represents the number of bosons (Cooper pairs) on site $j$ and the $U$
term describes the charging energy cost when this number deviates from the
electrostatically optimal value of $K$. The boson number is non-negative while
$\hat{K}_{j}$ has negative eigenvalues. However we assume that $K \gg 1$ so that
the negative angular momentum states are very high in energy.

The $J$ term in the quantum rotor model is a mutual torque that transfers units
of angular momentum between neighboring sites. In the boson language the wave
function for the state with $m$ bosons on site $j$ contains a factor
\begin{equation}
\psi_{m}(\varphi_{j}) = e^{im\varphi_{j}}.
\end{equation}
The raising and lowering operators are thus\footnote{These operators have matrix
elements $\langle\psi_{m+1}|e^{+i\varphi}|\psi_{m}\rangle = 1$ whereas a boson
raising operator would have matrix element $\sqrt{m+1}$. For $K \gg 1$, $m \sim
K$ and this is nearly a constant. Arguments like this strongly suggest that the
boson Hubbard model and the quantum rotor model are essentially equivalent. In
particular their order/disorder transitions are believed to be in the same
universality class.} $e^{\pm i\varphi_{j}}$. This shows us that the cosine term
in eq.~(\ref{eq:061103}) represents the Josephson coupling that hops bosons
between neighboring sites.

For $U \gg J$ the system is in an insulating phase well-described by the wave
function
\begin{equation}
\psi(\varphi_{1},\varphi_{2},\ldots,\varphi_{N}) = \prod_{j} e^{im\varphi_{j}}
\end{equation}
where $m$ is the nearest integer to $K$. In this state every rotor has the same
fixed angular momentum and thus every site has the same fixed particle number in
the boson language. There is a large excitation gap
\begin{equation}
\Delta \approx U\left(1 - 2|m - K|\right)
\end{equation}
and the system is insulating.\footnote{An exception occurs if $|m - K| =
\frac{1}{2}$ where the gap vanishes. See \cite{Fisherboson}.}

Clearly $|\psi|^{2} \approx 1$ in this phase and it is therefore quantum
disordered. That is, the phases $\{\varphi_{j}\}$ are wildly fluctuating because
every configuration is equally likely. The phase fluctuations are nearly
uncorrelated
\begin{equation}
\langle e^{i\varphi_{j}}\; e^{-i\varphi_{k}}\rangle \sim e^{-|\vec{r}_{j}-
\vec{r}_{k}|/\xi}.
\end{equation}

For $J \gg U$ the phases on neighboring sites are strongly coupled together and
the system is a superconductor. A crude variational wave function that captures
the essential physics is 
\begin{equation}
\psi(\varphi_{1},\varphi_{2},\ldots,\varphi_{N}) \sim e^{\lambda\sum_{\langle
ij\rangle}\;\cos{(\varphi_{i}-\varphi_{j})}}
\end{equation}
where $\lambda$ is a variational parameter \cite{Rana}. This is the simplest
ansatz consistent with invariance under $\varphi_{j} \rightarrow \varphi_{j} +
2\pi$. For $J \gg U$, $\lambda \gg 1$ and $|\psi|^{2}$ is large only for spin
configurations with all of the XY spins locally parallel. Expanding the cosine
term in eq.~(\ref{eq:061103}) to second order gives a harmonic Hamiltonian which
can be exactly solved. The resulting gapless `spin waves' are the Goldstone
modes of the superconducting phase.

For simplicity we work with the Lagrangian rather than the Hamiltonian
\begin{equation}
\mathcal{L} = \sum_{j} \left[\hbar K\dot{\varphi}_{j} +
\frac{\hbar^{2}}{2U}\dot{\varphi}_{j}^{2}\right] + J\sum_{\langle ij\rangle}
\cos{(\varphi_{i} - \varphi_{j})}
\end{equation}
The Berry phase term is a
total derivative and can not affect the equations of motion.\footnote{In fact in
the quantum path integral this term has no effect except for time histories in
which a `vortex' encircles site $j$ causing the phase to wind
$\varphi_{j}(\hbar\beta) = \varphi_{j}(0) \pm 2\pi$. We explicitly ignore this
possibility when we make the harmonic approximation.} Dropping this term and
expanding the cosine in the harmonic approximation yields
\begin{equation}
\mathcal{L} = \frac{\hbar^{2}}{2U} \sum_{j} \dot{\varphi}_{j}^{2} - \frac{J}{2}
\sum_{\langle ij\rangle} (\varphi_{i} - \varphi_{j})^{2}.
\end{equation}
This `phonon' model has linearly dispersing gapless collective modes at small
wavevectors
\begin{equation}
\hbar\omega_{q} = \sqrt{UJ}\; qa
\end{equation}
where $a$ is the lattice constant. The parameters $U$ and $J$ can be fixed by
fitting to microscopic Hartree-Fock calculations of the spin wave velocity and
the magnetic susceptibility (`boson compressibility')
\cite{FertigHF,skyrmelatticePRL}. This in turn allows one to estimate the
regime of filling factor and Zeeman energy in which the $U(1)$ symmetry is not
destroyed by quantum fluctuations \cite{skyrmelatticePRL}.

Let us now translate all of this into the language of our non-colinear QHE
ferromagnet \cite{Senthil,skyrmelatticePRL}. Recall that the angular momentum
(the `charge') conjugate to the phase angle $\varphi$ is the spin angular
momentum of the overturned spins that form the skyrmion. In the quantum
disordered `insulating' phase, each skyrmion has a well defined integer-valued
`charge' (number of overturned spins) much like we found when we quantized the
$U(1)$ zero mode for the plane angle $\varphi$ of a single isolated skyrmion in
eq.~(\ref{eq:060818}). There is an excitation gap separating the energies of the
discrete quantized values of the spin.

The `superfluid' state with broken $U(1)$ symmetry is a totally new kind of spin
state unique to non-colinear magnets \cite{Senthil,skyrmelatticePRL}. Here the
phase angle is well-defined and the number of overturned spins is uncertain. The
off-diagonal long-range order of a superfluid becomes
\begin{equation}
\langle b_{j}^{\dagger} b_{k}^{\phantom{\dagger}}\rangle \rightarrow \langle
e^{i\varphi_{j}} e^{-i\varphi_{k}}\rangle
\end{equation}
or in the spin language\footnote{There is a slight complication here. Because
the XY spin configuration of the skyrmion has a vortex-like structure $\langle
s^{+}\rangle \equiv \langle s^{x} + is^{y}\rangle$ winds in phase around the
skyrmion so the `bose condensation' is not at zero wave vector.}
\begin{equation}
\left\langle s^{+}(\vec{r}\,) s^{-}(\vec{r}^{\,\prime})\right\rangle.
\end{equation}
Thus in a sense we can interpret a spin flip interaction between an electron and
a nucleus as creating a boson in the superfluid. But this boson has a finite
probability of `disappearing' into the superfluid `condensate' and hence the
system does not have to pay the Zeeman price to create the flipped spin. That
is, the superfluid state has an uncertain number of flipped spins (even though
$S_{\mathrm{tot}}^{z}$ commutes with $H$) and so the Zeeman energy cost is
uncertain.

In classical language the skyrmions locally have finite (slowly varying) x and y
spin components which act as effective magnetic fields around which the nuclear spins
precess and which thus cause $I^{z}$ to change with
time. The key here is that $s^{x}$ and $s^{y}$ can, because of the broken $U(1)$
symmetry, fluctuate very slowly (i.e. at MHz frequencies that the nuclei can
follow rather than just the very high Zeeman precession frequency).

Detailed numerical calculations \cite{skyrmelatticePRL} show that the Skyrme
lattice is very efficient at relaxing the nuclei and $1/T_{1}$ and is enhanced
by a factor of $\sim 10^{3}$ over the corresponding rate at zero magnetic field.
We expect this qualitative distinction to survive even above the Skryme lattice
melting temperature for the reasons discussed earlier.

Because the nuclear relaxation rate increases by orders of magnitude, the
equilibration time at low temperatures drops from hours to seconds. This means
that the nuclei come into thermal equilibrium with the electrons and hence the
lattice. The nuclei therefore have a well-defined temperature and contribute to
the specific heat. Because the temperature is much greater than the nuclear
Zeeman energy scale $\Delta \sim 1~\mbox{mK}$, each nucleus contributes only a
tiny amount $\sim k_{\mathrm{B}} \frac{\Delta^{2}}{T^{2}}$ to the specific heat.
On the other hand, the electronic specific heat per particle $\sim
k_{\mathrm{B}} \frac{T}{T_{\mathrm{fermi}}}$ is low and the electron density is
low. In fact there are about $10^{6}$ nuclei per quantum well electron and the
nuclei actually enhance the specific heat more than 5 orders of magnitude
\cite{Bayot}!

Surprisingly, at around 30~mK there is a \textit{further} enhancement of the
specific heat by an additional order of magnitude. This may be a signal of the
Skyrme lattice melting transition \cite{Bayot,skyrmelatticePRL,TimmMelting},
although the situation is somewhat murky at the present time. The peak can not
possibly be due to the tiny amount of entropy change in the Skyrme lattice
itself. Rather it is due to the nuclei in the thick AlAs barrier between the
quantum wells.\footnote{For somewhat complicated reasons it may be that the
barrier nuclei are efficiently dipole coupled to the nuclei in the quantum
wells (and therefore in thermal equilibrium) only due to the critical slowing
down of the electronic motion in the vicinity of the Skyrme lattice melting
transition.}

\section{Double-Layer Quantum Hall Ferromagnets}
\label{sec:doublelayer}

\subsection{Introduction}
\label{sec:intro}

We learned in our study of quantum Hall ferromagnets that the Coulomb
interaction plays an important role at Landau level filling factor $\nu=1$
because it causes the electron spins to spontaneously align ferromagnetically
and this in turn profoundly alters the charge excitation spectrum by producing a
gap.\footnote{Because the charged excitations are skyrmions, this gap is not as
large as naive estimates would suggest, but it is still finite as long as the
spin stiffness is finite.} A closely related effect occurs in double-layer
systems in which layer index is analogous to spin
\cite{murphyPRL,JPEbook,GMbook}. Building on our knowledge of the dynamics of
ferromagnets developed in the last section, we will use this analogy to explore
the rich physics of double-layer systems.

Novel fractional quantum Hall effects due to correlations \cite{gsnum} in
multicomponent systems were anticipated in early work by Halperin
\cite{BIHhelv} and the now extensive literature has been reviewed in
\cite{GMbook}. There have also been recent interesting studies of systems in
which the spin and layer degrees of freedom are coupled in novel ways
\cite{Pinczukdouble,DasSarmaSachdev}.

\begin{figure}
\centerline{\epsfxsize=6cm
 \epsffile{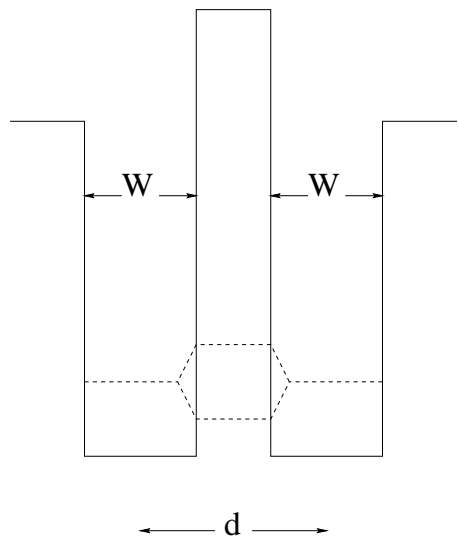}}
\caption[]{Schematic conduction band edge profile for a
double-layer two-dimensional electron gas system.  Typical widths
and separations are $W\sim d\sim 100\hbox{\AA}$ 
and are comparable to the spacing
between electrons within each inversion layer.}
\label{fig:wellschematic}
\end{figure}

As described in this volume by Shayegan \cite{ShayeganLesHouches}, modern MBE
techniques make it possible to produce double-layer (and multi-layer)
two-dimensional electron gas systems of extremely low disorder and high
mobility. As illustrated schematically in Fig.~(\ref{fig:wellschematic}),
these systems consist of a pair of 2D electron gases separated by a distance $d$
so small ($d\sim 100\mbox{\AA}$) as to be comparable to the typical spacing
between electrons in the same layer. A second type of system has also recently
been developed to a high degree of perfection \cite{mansour}. These systems
consist of single wide quantum wells in which strong mixing of the two lowest
electric subbands allows the electrons to localize themselves on opposites sides
of the well to reduce their correlation energy. We will take the point of view
that these systems can also be approximately viewed as double-well systems with
some effective layer separation and tunnel barrier height. 

As we have already learned, correlations are especially important in the strong
magnetic field regime because all electrons can be accommodated within the
lowest Landau level and execute cyclotron orbits with a common kinetic energy.
The fractional quantum Hall effect occurs when the system has a gap for making
charged excitations, \textit{i.e.} when the system is incompressible. Theory has
predicted \cite{BIHhelv,gsnum,amdreview} that at some Landau level filling
factors, gaps occur in double-layer systems only if interlayer interactions are
sufficiently strong. These theoretical predictions have been
confirmed \cite{expamd}. More recently work from several different points of
view \cite{wenandzee,ezawa,ahmz1,gapless,harfok,JasonHo} has suggested that
inter-layer correlations can also lead to unusual broken symmetry states with a
novel kind of spontaneous phase coherence between layers which are isolated from
each other except for inter-layer Coulomb interactions. It is this spontaneous
interlayer phase coherence which is responsible \cite{usPRL,Ilong,II,GMbook} for
a variety of novel features seen in the experimental data to be discussed below
\cite{murphyPRL,JPEbook}.

\subsection{Pseudospin Analogy}

We will make the simplifying assumption that the Zeeman energy is large enough
that fluctuations of the (true) spin order can be ignored, leaving out the
possibility of mixed spin and pseudospin correlations
\cite{Pinczukdouble,DasSarmaSachdev}. We will limit our attention to the lowest
electric subband of each quantum well (or equivalently, the two lowest bands of
a single wide well). Hence we have a two-state system that can be labeled by a
pseudospin 1/2 degree of freedom. Pseudospin up means that the electron is in
the (lowest electric subband of the) upper layer and pseudospin down means that
the electron is in the (lowest electric subband of the) lower layer. 

Just as in our study of ferromagnetism we will consider states with total
filling factor $\nu \equiv \nu_{\uparrow} + \nu_{\downarrow} =1$. A state
exhibiting interlayer phase coherence and having the pseudospins
ferromagnetically aligned in the direction defined by polar angle $\theta$ and
azimuthal angle $\varphi$ can be written in the Landau gauge just as for
ordinary spin
\begin{equation}
|\psi\rangle = \prod_{k}\left\{\cos(\theta/2) c^{\dagger}_{k\uparrow} 
+ \sin(\theta/2)e^{i\varphi} c^{\dagger}_{k\downarrow}\right\}|0\rangle.
\end{equation}
Every $k$ state contains one electron and hence this state has $\nu=1$
as desired.  Note however that the layer index for each electron is
uncertain.  The amplitude to find a particular electron in the upper
layer is $\cos(\theta/2)$ and the amplitude to find it in the lower layer
is $\sin(\theta/2)e^{i\varphi}$.  Even if the two layers are
completely independent with no tunneling between them, quantum mechanics
allows for the somewhat peculiar possibility that we are uncertain
which layer the electron is in.

For the case of ordinary spin we found that the Coulomb interaction produced an
exchange energy which strongly favored having the spins locally parallel. Using
the fact that the Coulomb interaction is completely spin independent (it is only
the Pauli principle that indirectly induces the ferromagnetism) we wrote down
the spin rotation invariant effective theory in eq.~(\ref{eq:1124219}). Here we
do not have full SU(2) invariance because the interaction between electrons in
the same layer is clearly stronger than the interaction between electrons in
opposite layers. Thus for example, if all the electrons are in the upper (or
lower) layer, the system will look like a charged capacitor and have higher
energy than if the layer occupancies are equal. Hence to leading order in
gradients we expect the effective action to be modified slightly
\begin{eqnarray}
\mathcal{L} = &-&\int d^{2}r\; \left\{\hbar S n\, \dot{m}^{\mu}(\vec{r}\,)
\mathcal{A}^{\mu}[\vec{m}] -\lambda(\vec{r}\,) (m^{\mu} m^{\mu} - 1)
\right\}\nonumber\\
&-&\int d^{2}r\, \left\{\frac{1}{2} \rho_{s} \partial_{\mu} m^{\nu}
\partial_{\mu} m^{\nu} + \beta m^{z} m^{z} - \Delta m^{z} - nt m^{x}\right\}.
\label{eq:easyplane}
\end{eqnarray}
The spin stiffness $\rho_{s}$ represents the SU(2) invariant part of the
exchange energy and is therefore somewhat smaller than the value computed in
eq.~(\ref{eq:060322}). The coefficient $\beta$ is a measure of the capacitive
charging energy.\footnote{We have taken the charging energy to be a local
quantity characterized by a fixed, wave vector independent capacitance. This is
appropriate only if $m^{z}(\vec{r}\,)$ represents the local charge imbalance
between the layers coarse-grained over a scale larger than the layer separation.
Any wave vector dependence of the capacitance will be represented by higher
derivative terms which we will ignore.} The analog of the Zeeman energy $\Delta$
represents an external electric field applied along the MBE growth direction
which unbalances the charge densities in the two layers. The coefficient $t$
represents the amplitude for the electrons to tunnel between the two layers. It
prefers the pseudospin to be aligned in the $\hat x$ direction because this
corresponds to the spinor
\begin{equation}
\frac{1}{\sqrt{2}} \left(\begin{array}{c}
1\\
1
\end{array}\right)
\end{equation}
which represents the \textit{symmetric} (i.e. bonding) linear combination of the
two well states. The state with the pseudospin pointing in the $-\hat{x}$
direction represents the \textit{antisymmetric} (i.e. antibonding) linear
combination which is higher in energy.

For the moment we will assume that both $t$ and $\Delta$ vanish, leaving only
the $\beta$ term which breaks the pseudospin rotational symmetry. The case
$\beta < 0$ would represent `Ising anisotropy'. Clearly the physically realistic
case for the capacitive energy gives $\beta > 0$ which represents so-called
`easy plane anisotropy.' The energy is minimized when $m^{z} = 0$ so that the
order parameter lies in the XY plane giving equal charge densities in the two
layers. Thus we are left with an effective XY model which should exhibit
long-range off-diagonal order\footnote{At finite temperatures $\Psi(\vec{r}\,)$
will vanish but will have long-range algebraically decaying correlations. Above
the Kosterlitz-Thouless phase transition temperature, the correlations will fall
off exponentially.}
\begin{equation}
\Psi(\vec{r}\,) = \langle m^{x}(\vec{r}\,) + i m^{y}(\vec{r}\,)\rangle.
\end{equation}
The order is `off-diagonal' because it corresponds microscopically to an
operator
\begin{equation}
\Psi(\vec{r}\,) = \langle s^{+}(\vec{r}\,)\rangle = \langle
\psi^{\dagger}_{\uparrow}(\vec{r}\,) \psi_{\downarrow}(\vec{r}\,)\rangle
\label{eq:odlro}
\end{equation}
which is not diagonal in the $s^{z}$ basis, much as in a superfluid where the
field operator changes the particle number and yet it condenses and acquires a
finite expectation value. 

One other comment worth making at this point is that eq.~(\ref{eq:odlro}) shows
that, unlike the order parameter in a superconductor or superfluid, this one
corresponds to a charge neutral operator. Hence it will be able to condense
despite the strong magnetic field (which fills charged condensates with vortices
and generally destroys the order).

In the next subsection we review the experimental evidence that long-range XY
correlations exist and that as a result, the system exhibits excitations which
are highly collective in nature. After that we will return to further analysis
and interpretation of the effective Lagrangian in eq.~(\ref{eq:easyplane}) to
understand those excitations.

\subsection{Experimental Background}
\label{sec:expback}

As illustrated by the dashed lines in fig.~(\ref{fig:wellschematic}), the lowest
energy eigenstates split into symmetric and antisymmetric combinations separated
by an energy gap $\Delta_{\mathrm{SAS}} = 2t$ which can, depending on the sample,
vary from essentially zero to many hundreds of Kelvins. The splitting can
therefore be much less than or greater than the interlayer interaction energy
scale, $E_{\mathrm{c}} \equiv e^{2}/\epsilon d$. Thus it is possible to make
systems which are in either the weak or strong correlation limits.

\begin{figure}
\centerline{\epsfxsize=10cm
 \epsffile{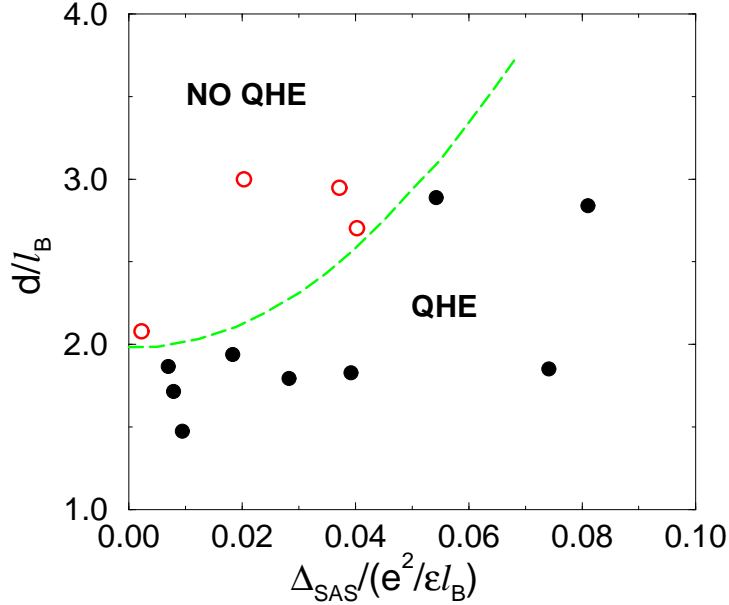}}
\caption[]{Phase diagram for the double layer QHE system (after
Murphy et al. \cite{murphyPRL}).  Only samples whose parameters
lie below the dashed line exhibit
a quantized Hall plateau and excitation gap.}
\label{fig:qhe-noqhe}
\end{figure}

When the layers are widely separated, there will be no correlations between them
and we expect no dissipationless quantum Hall state since each layer has
\cite{nuhalf} $\nu = 1/2$. For smaller separations, it is observed
experimentally that there is an excitation gap and a quantized Hall plateau
\cite{greg,mansour,murphyPRL}. This has either a trivial or a highly non-trivial
explanation, depending on the ratio $\Delta_{\mathrm{SAS}}/ E_{\mathrm{c}}$. For
large $\Delta_{\mathrm{SAS}}$ the electrons tunnel back and forth so rapidly
that it is as if there is only a single quantum well. The tunnel splitting
$\Delta_{\mathrm{SAS}}$ is then analogous to the electric subband splitting in a
(wide) single well. All symmetric states are occupied and all antisymmetric
states are empty and we simply have the ordinary $\nu = 1$ integer Hall effect.
Correlations are irrelevant in this limit and the excitation gap is close to the
single-particle gap $\Delta_{\mathrm{SAS}}$ (or $\hbar\omega_{\mathrm{c}}$,
whichever is smaller). What is highly non-trivial about this system is the fact
that the $\nu = 1$ quantum Hall plateau survives even when
$\Delta_{\mathrm{SAS}} \ll E_{\mathrm{c}}$. In this limit the excitation gap has
clearly changed to become highly collective in nature since the observed
\cite{mansour,murphyPRL} gap can be on the scale of 20K even when
$\Delta_{\mathrm{SAS}} \sim 1~\mathrm{K}$. Because of the spontaneously broken
XY symmetry \cite{wenandzee,ezawa,harfok,usPRL,Ilong}, the excitation gap
actually survives the limit $\Delta_{\mathrm{SAS}} \longrightarrow 0$! This
cross-over from single-particle to collective gap is quite analogous to that for
spin polarized single layers. There the excitation gap survives the limit of
zero Zeeman splitting so long as the Coulomb interaction makes the spin
stiffness non-zero. This effect in double-layer systems is visible in
fig.~(\ref{fig:qhe-noqhe})
which shows the QHE phase diagram obtained by Murphy \etal
\cite{murphyPRL,JPEbook} as a function of layer-separation and tunneling energy.
A $\nu=1$ quantum Hall plateau and gap is observed in the regime below the
dashed line. Notice that far to the right, the single particle tunneling energy
dominates over the coulomb energy and we have essentially a one-body integer QHE
state. However the QHE survives all the way into $\Delta_{\mathrm{SAS}} =0 $
provided that the layer separation is below a critical value
$d/\ell_{\mathrm{B}} \approx 2$. In this limit there is no tunneling and the gap
is purely many-body in origin and, as we will show, is associated with the
remarkable `pseudospin ferromagnetic' quantum state exhibiting spontaneous
interlayer phase coherence. 

\begin{figure}
\centerline{\epsfxsize=10cm
 \epsffile{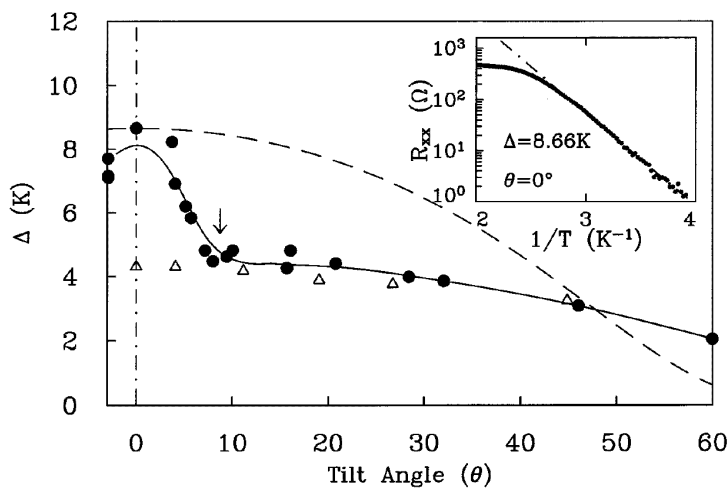}}
\caption[]{The charge activation energy gap, $\Delta$, as a function of tilt
angle in a weakly tunneling double-layer sample ($\Delta_{\mathrm{SAS}} = 0.8$K). The
solid circles are for filling $\nu=1$, open triangles for $\nu=2/3$. The arrow
indicates the critical angle $\theta_{\mathrm{c}}$. The solid line is a guide to the
eye. The dashed line refers to a simple estimate of the renormalization of the
tunneling amplitude by the parallel magnetic field. Relative to the actual
decrease, this one-body effect is very weak and we have neglected it. Inset:
Arrhenius plot of dissipation. The low temperature activation energy is $\Delta
= 8.66$K and yet the gap collapses at a much lower temperature scale of about
$0.4$K ($1/T\approx 2.5$). (After Murphy \textit{et al.} \cite{murphyPRL}).}
\label{fig:arrhenius}
\end{figure}

A second indication of the highly collective nature of the excitations can be
seen in the Arrhenius plots of thermally activated dissipation \cite{murphyPRL}
shown in the inset of fig.~(\ref{fig:arrhenius})
The low temperature activation energy $\Delta$ is, as already noted, much larger
than $\Delta_{\mathrm{SAS}}$. If $\Delta$ were nevertheless somehow a
single-particle gap, one would expect the Arrhenius law to be valid up to
temperatures of order $\Delta$. Instead one observes a fairly sharp leveling off
in the dissipation as the temperature increases past values as low as $\sim 0.05
\Delta$. This is consistent with the notion of a thermally induced collapse of
the order that had been producing the collective gap.

The third significant feature of the experimental data pointing to a
highly-ordered collective state is the strong response of the system to
relatively weak magnetic fields $B_{\parallel}$ applied in the plane of the 2D
electron gases. In fig.~(\ref{fig:arrhenius}) we see that the charge activation
gap drops dramatically as the magnetic field is tilted (keeping $B_{\perp}$
constant).

\begin{figure}
\centerline{\epsfxsize=8cm
 \epsffile{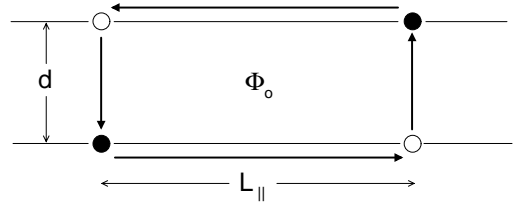}}
\caption[]{A process in a double-layer two-dimensional electron gas system which
encloses flux from the parallel component of the magnetic field.  One interpretation
of this process is that an electron tunnels from
the upper layer to the lower layer (near the left end of the figure).  The resulting
particle-hole pair then travels coherently to the right and is annihilated by a subsequent
tunneling event in the reverse direction.
The quantum amplitude for such paths is sensitive to the parallel component of the field.}
\label{fig:figloop}
\end{figure}

Within a model that neglects higher electric subbands, we can treat the electron
gases as strictly two-dimensional. This is important since $B_{\parallel}$ can
affect the system only if there are processes that carry electrons around closed
loops containing flux. A prototypical such process is illustrated in
fig.~(\ref{fig:figloop}).
An electron tunnels from one layer to the other at point A, and travels to point
B. Then it (or another indistinguishable electron) tunnels back and returns to
the starting point. The parallel field contributes to the quantum amplitude for
this process (in the 2D gas limit) a gauge-invariant Aharonov-Bohm phase factor
$\exp\left(2\pi i \Phi/\Phi_{0}\right)$ where $\Phi$ is the enclosed flux and
$\Phi_{0}$ is the quantum of flux.

Such loop paths evidently contribute significantly to correlations in the system
since the activation energy gap is observed to decrease very rapidly with
$B_{\parallel}$, falling by factors of order two or more until a critical field,
$B^{*}_{\parallel} \sim 0.8\mathrm{T}$, is reached at which the gap essentially
ceases changing \cite{murphyPRL}. To understand how remarkably small
$B^{*}_{\parallel}$ is, consider the following. We can define a length
$L_{\parallel}$ from the size of the loop needed to enclose one quantum of flux:
$L_{\parallel} B^{*}_{\parallel} d = \Phi_{0}$. ($L_{\parallel} [\hbox{\AA}] =
4.137 \times 10^{5} / d [\hbox{\AA}] B^{*}_{\parallel} [\mathrm{T}]$.) For
$B^{*}_{\parallel} = 0.8\mathrm{T}$ and $d = 150 \hbox{\AA}$, $L_{\parallel} =
2700 \hbox{\AA} $ which is approximately twenty times the spacing between
electrons in a given layer and thirty times larger than the quantized cyclotron
orbit radius $\ell \equiv (\hbar c / e B_{\perp})^{1/2}$ within an individual
layer. Significant drops in the excitation gap are already seen at fields of
0.1T implying enormous phase coherent correlation lengths must exist. Again this
shows the highly-collective long-range nature of the ordering in this system.

In the next subsection we shall briefly outline a detailed model which explains
all these observed effects.

\subsection{Interlayer Phase Coherence}
\label{sec:coherence}

The essential physics of spontaneous inter-layer phase coherence can be examined
from a microscopic point of view \cite{ahmz1,gapless,harfok,usPRL,Ilong} or a
macroscopic Chern-Simons field theory point of view
\cite{wenandzee,ezawa,usPRL,Ilong}, but it is perhaps most easily visualized in
the simple variational wave function which places the spins purely in the XY
plane \cite{Ilong}
\begin{equation}
|\psi\rangle = \prod_{k} \left\{c^{\dagger}_{k\uparrow} +
c^{\dagger}_{k\downarrow} e^{i\varphi}\right\} |0\rangle.
\label{eq:variational}
\end{equation} 
Note for example, that if $\varphi=0$ then we have precisely the non-interacting
single Slater determinant ground state in which electrons are in the symmetric
state which, as discussed previously in the analysis of the effective Lagrangian
in eq.~(\ref{eq:easyplane}), minimizes the tunneling energy. This means that the
system has a definite total number of particles ($\nu=1$ exactly) but an
indefinite number of particles in each layer. In the absence of inter-layer
tunneling, the particle number in each layer is a good quantum number. Hence
this wave function represents a state of spontaneously broken symmetry
\cite{wenandzee,ezawa,Ilong} in the same sense that the BCS state for a
superconductor has indefinite (total) particle number but a definite phase
relationship between states of different particle number.

In the absence of tunneling ($t=0$) the energy can not depend on the phase angle
$\varphi$ and the system exhibits a global $U(1)$ symmetry associated with
conservation of particle number in each layer \cite{wenandzee}. One can imagine
allowing $\varphi$ to vary slowly with position to produce excited states. Given
the $U(1)$ symmetry, the effective Hartree-Fock energy functional for these
states is restricted to have the leading form
\begin{equation}
H = \frac{1}{2}\rho_{s}\int d^{2}r |\nabla\varphi|^{2} + \ldots\,\,.
\label{eq:xymod}
\end{equation}
The origin of the finite `spin stiffness' $\rho_{s}$ is the loss of exchange
energy which occurs when $\varphi$ varies with position. Imagine that two
particles approach each other. They are in a linear superposition of states in
each of the layers (even though there is no tunneling!). If they are
characterized by the same phase $\varphi$, then the wave function is symmetric
under pseudospin exchange and so the spatial wave function is antisymmetric and
must vanish as the particles approach each other. This lowers the Coulomb
energy. If a phase gradient exists then there is a larger amplitude for the
particles to be near each other and hence the energy is higher. This loss of
exchange energy is the source of the finite spin stiffness and is what causes
the system to spontaneously `magnetize'.

We see immediately that the $U(1)$ symmetry leads to eq.~(\ref{eq:xymod}) which
defines an effective XY model which will contain vortex excitations which
interact logarithmically. \cite{goldenfeld,boulevard}
In a superconducting film the vortices interact
logarithmically because of the kinetic energy cost of the supercurrents
circulating around the vortex centers. Here the same logarithm appears, but
it is due to the potential energy cost (loss of exchange) associated with the
phase gradients (circulating pseudo-spin currents).

Hartree-Fock estimates \cite{Ilong} indicate that the spin stiffness $\rho_{s}$
and hence the Kosterlitz-Thouless (KT) critical temperature are on the scale of
0.5~K in typical samples. Vortices in the $\varphi$ field are reminiscent of
Laughlin's fractionally charged quasiparticles but in this case carry charges
$\pm\frac{1}{2} e$ and can be left- or right-handed for a total of four
`flavors' \cite{usPRL,Ilong}. It is also possible to show \cite{Ilong,II} that
the presence of spontaneous magnetization due to the finite spin stiffness means
that the charge excitation gap is finite (even though the tunnel splitting is
zero). Thus the QHE survives \cite{Ilong} the limit $\Delta_{\mathrm{SAS}}
\longrightarrow 0$.

Since the `charge' conjugate to the phase $\varphi$ is the $z$ component of the
pseudo spin $S^{z}$, the pseudospin `supercurrent'
\begin{equation}
\vec{J} =\rho_{s} \mathbf{\vec{\nabla}}\varphi
\end{equation}
represents oppositely directed charge currents in each layer. Below the KT
transition temperature, such current flow will be dissipationless (in linear
response) just as in an ordinary superfluid. Likewise there will be a linearly
dispersing collective Goldstone mode as in a superfluid
\cite{ahmz1,wenandzee,ezawa,usPRL,Ilong} rather than a mode with quadratic
dispersion as in the SU(2) symmetric ferromagnet. (This is somewhat akin to the
difference between an ideal bose gas and a repulsively interacting bose gas.)

If found, this Kosterlitz-Thouless transition would be the first example of a
finite-temperature phase transition in a QHE system. The transition itself has
not yet been observed due to the tunneling amplitude $t$ being significant in
samples having the layers close enough together to have strong correlations. As
we have seen above however, significant effects which imply the existence of
long-range XY order correlations have been found. Whether or not an appropriate
sample can be constructed to observe the phase transition is an open question at
this point.

\boxedtext{\begin{exercise}
Following the method used to derive eq.~(\ref{eq:1105228}), show that the
collective mode for the Lagrangian in eq.~(\ref{eq:easyplane}) has linear rather
than quadratic dispersion due to the presence of the $\beta$ term. (Assume
$\Delta=t=0$.) Hint: Consider small fluctuations of the magnetization away from
$\vec{m} = (1,0,0)$ and choose an appropriate gauge for $\cal A$ for this
circumstance.

Present a qualitative argument that layer imbalance caused by $\Delta$ does not
fundamentally change any of the results described in this section but rather
simply renormalizes quantities like the collective mode velocity. That is,
explain why the $\nu=1$ QHE state is robust against charge imbalance. (This is
an important signature of the underlying physics. Certain other interlayer
correlated states (such as the one at total filling $\nu=1/2$) are quite
sensitive to charge imbalance \cite{GMbook}.)
\label{ex:981201}
\end{exercise}}

\subsection{Interlayer Tunneling and Tilted Field Effects}

As mentioned earlier, a finite tunneling amplitude $t$ between the layers breaks
the $U(1)$ symmetry
\begin{equation}
H_{\mathrm{eff}} = \int d^{2}r \left[\frac{1}{2}\rho_{s}
\vert\nabla\varphi\vert^{2} - nt \cos{\varphi}\right]
\label{eq:H_eff}
\end{equation}
by giving a preference to symmetric tunneling states. This can be seen from the
tunneling Hamiltonian
\begin{equation}
H_{\mathrm{T}} = - t \int d^{2}r \left\{\psi_{\uparrow}^{\dagger} (\mathbf{r})
\psi_{\downarrow} (\mathbf{r}) + \psi_{\downarrow}^{\dagger} (\mathbf{r})
\psi_{\uparrow} (\mathbf{r})\right\}
\label{eq:tunnel}
\end{equation}
which can be written in the spin representation as
\begin{equation}
H_{\mathrm{T}} = - 2t \int d^{2}r S^{x}(\mathbf{r}).
\end{equation}
(Recall that the eigenstates of $S^{x}$ are symmetric and antisymmetric
combinations of up and down.) 

As the separation $d$ increases, a critical point $d^{*}$ is reached at which
the magnetization vanishes and the ordered phase is destroyed by quantum
fluctuations \cite{usPRL,Ilong}. This is illustrated in
fig.~(\ref{fig:qhe-noqhe}). For \textit{finite} tunneling $t$, we will see below
that the collective mode becomes massive and quantum fluctuations will be less
severe. Hence the phase boundary in fig.~(\ref{fig:qhe-noqhe}) curves upward
with increasing $\Delta_{\mathrm{SAS}}$.

\begin{figure}
\centerline{\epsfysize=10cm
 \epsffile{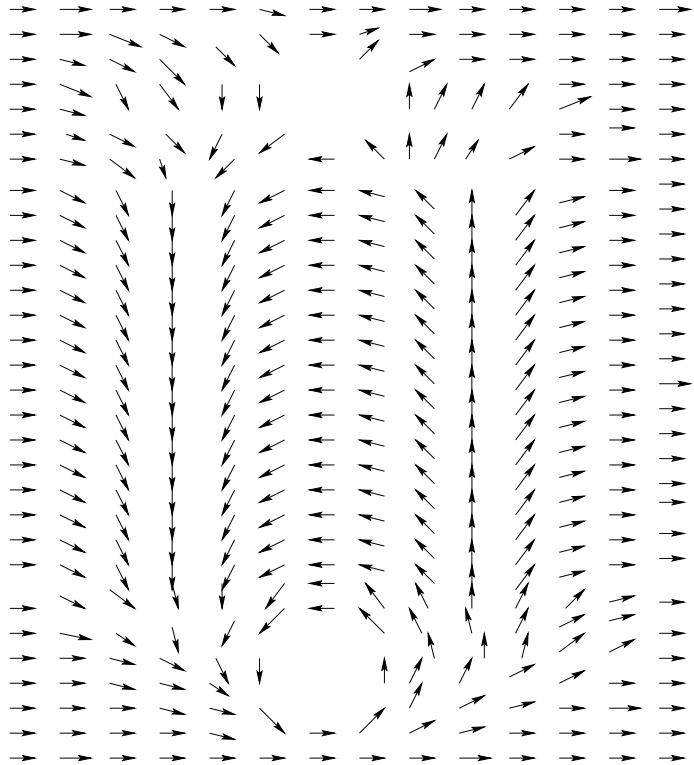}}
\caption[]{Meron pair connected by a domain wall. Each meron carries a charge
$e/2$ which tries to repel the other one.}
\label{fig:meron_string}
\end{figure}

The introduction of finite tunneling amplitude destroys the U(1) symmetry and
makes the simple vortex-pair configuration extremely expensive. To lower the
energy the system distorts the spin deviations into a domain wall or `string'
connecting the vortex cores as shown in fig.~(\ref{fig:meron_string}).
The spins are oriented in the $\hat{x}$ direction everywhere except in the
central domain wall region where they tumble rapidly through $2\pi$. The domain
wall has a fixed energy per unit length and so the vortices are now confined by
a linear `string tension' rather than logarithmically. We can estimate the
string tension by examining the energy of a domain wall of infinite length. The
optimal form for a domain wall lying along the $y$ axis is given by
\begin{equation}
\varphi(\vec{r}) = 2 \arcsin{[\tanh{(\lambda x)}]},
\label{eq:soliton}
\end{equation}
where the characteristic width of the string is
\begin{equation}
\lambda^{-1} = \left[\frac{2\pi\ell^{2}\rho_{s}}{t}\right]^\frac{1}{2}.
\end{equation}
The resulting string tension is
\begin{equation}
T_{0} = 8 \left[\frac{t\rho_{s}}{2\pi\ell^{2}}\right]^\frac{1}{2}.
\label{eq:tension_{0}}
\end{equation}
Provided the string is long enough ($R\lambda \gg 1$), the total energy of a
segment of length $R$ will be well-approximated by the expression
\begin{equation}
E_{\mathrm{pair}}' = 2E_{\mathrm{mc}}' + \frac{e^{2}}{4R} + T_{0}R.
\label{string_pair}
\end{equation}
This is minimized at $R^{*} = \sqrt{e^{2}/4T_{0}}$. The linear confinement
brings the charged vortices closer together and rapidly increases the Coulomb
energy. In the limit of very large tunneling, the meron pair shrinks and the
single-particle excitation (hole or extra spin-reversed electron) limit must be
recovered.

The presence of parallel field $B_{\parallel}$ field can be conveniently
described with the gauge choice
\begin{equation}
\vec A_{\parallel} = xB_{\parallel} \hat{z}
\end{equation}
where $\hat{z}$ is the growth direction. In this gauge the tunneling amplitude
transforms to
\begin{equation}
t \rightarrow t\; e^{iQx}
\end{equation}
and the energy becomes
\begin{equation}
H = \int d^{2}r \left[\frac{1}{2} \rho_{s} \vert\vec{\nabla}\varphi\vert^{2} -
\frac{t}{2\pi\ell^{2}}\cos{(\varphi - Qx)}\right]
\end{equation}
where $Q = 2\pi /L_{\parallel}$ and $L_{\parallel}$ is the length associated
with one quantum of flux for the loops shown in fig.~\ref{fig:figloop}. This is
the so-called Pokrovsky-Talopov model which exhibits a
commensurate-incommensurate phase transition. At low $B_{\parallel}$, $Q$ is
small and the low energy state has $\varphi \approx Qx$; i.e. the local spin
orientation `tumbles'. In contrast, at large $B_{\parallel}$ the gradient cost
is too large and we have $\varphi \approx \hbox{constant}$. It is possible to
show \cite{Ilong,II} that this phase transition semiquantitatively explains the
rapid drop and subsequent leveling off of the activation energy vs.
$B_{\parallel}$ seen in fig.~(\ref{fig:arrhenius}).

\boxedtext{\begin{exercise}
Derive eq.~(\ref{eq:soliton}) for the form of the `soliton' that minimizes the energy cost
for the Hamiltonian in eq.~(\ref{eq:H_eff}).
\label{ex:string}
\end{exercise}}

\section{Acknowledgments}

Much of my work on the quantum Hall effect has been in collaboration with Allan
MacDonald. The more recent work on quantum Hall ferromagnets has also been done
in collaboration with M.\ Abolfath, L.\ Belkhir, L.\ Brey, R.\ C\^{o}t\'{e}, H.
Fertig, P.\ Henelius, K.\ Moon, H.\ Mori, J.\ J.\ Palacios, A.\ Sandvik, H.
Stoof, C.\ Timm, K.\ Yang, D.\ Yoshioka, S.\ C.\ Zhang, and L.\ Zheng. It is a
pleasure to acknowledge many useful conversations with S.\ Das Sarma, M.\ P.\ A.
Fisher, N.\ Read, and S.\ Sachdev.

It is a pleasure to thank Ms.~Daphne Klemme for her expert typesetting
of my scribbled notes and Jairo Sinova for numerous helpful comments on
the manuscript.

This work was supported by NSF DMR-9714055.

\appendix
\renewcommand{\theequation}{\Alph{chapter}.\arabic{equation}}

\newcounter{asaveeqn}
\renewcommand{\alpheqn}{\setcounter{asaveeqn}{\value{equation}}%
\stepcounter{asaveeqn}\setcounter{equation}{0}%
\renewcommand{\theequation}{\mbox{\Alph{chapter}.\arabic{asaveeqn}\alph{equation}}}}
\renewcommand{\reseteqn}{\setcounter{equation}{\value{asaveeqn}}%
\renewcommand{\theequation}{\Alph{chapter}.\arabic{equation}}}

\chapter{Lowest Landau Level Projection}
\label{app:projection}

A convenient formulation of quantum mechanics within the subspace of the lowest
Landau level (LLL) was developed by Girvin and Jach \cite{girvinjach}, and was exploited
by Girvin, MacDonald and Platzman in the magneto-roton theory of collective
excitations of the incompressible states responsible for the fractional quantum
Hall effect \cite{GMP}. Here we briefly review this formalism. See also
Ref.~\cite{stonebook}.

We first consider the one-body case and choose the symmetric gauge. The
single-particle eigenfunctions of kinetic energy and angular momentum in the LLL
are given in Eq.~(\ref{eq:symmgauge})
\begin{equation}
\phi_m(z)=\frac{1}{(2\pi 2^m m!)^{1/2}}\> z^m\>
 \exp{\biggl( -\frac{\vert z\vert^2}{4}\biggr)} ,
\label{eq3.10}
\end{equation}
where $m$ is a non-negative integer, and $z = (x + iy)/\ell$. From
(\ref{eq3.10}) it is clear that any wave function in the LLL can be written in
the form
\begin{equation}
\psi (z)=f(z)\> e^{-\frac{\vert z\vert^2}{4}}
\label{eq3.20}
\end{equation}
where $f(z)$ is an analytic function of $z$, so the subspace in the LLL is
isomorphic to the Hilbert space of analytic functions
\cite{girvinjach,Bargman,stonebook}. Following Bargman \cite{girvinjach,Bargman}, we define
the inner product of two analytic functions as
\begin{equation}
(f, g)=\int d\mu (z)\> f^\ast (z)\> g(z),
\label{eq3.30}
\end{equation}
where
\begin{equation}
d\mu (z)\equiv (2\pi )^{-1}\> dxdy\> e^{-\frac{\vert z\vert^2}{2}} .
\label{eq3.40}
\end{equation}

Now we can define bosonic ladder operators that connect $\phi_m$ to $\phi_{m\pm
1}$ (and which act on the polynomial part of $\phi_m$ only):
\alpheqn{
\begin{eqnarray}
a^\dagger &=& {z\over \sqrt{2}} ,\label{eq3.50a}\\
a &=& \sqrt{2}\> \frac{\partial}{\partial z} ,\label{eq3.50b}
\end{eqnarray}}
\reseteqn

\noindent so that
\alpheqn{
\begin{eqnarray}
a^\dagger\> \varphi_m &=& \sqrt{m+1}\> \varphi_{m+1} ,\label{eq3.60a}\\
a\> \varphi_m &=& \sqrt{m}\> \varphi_{m-1} ,\label{eq3.60b}\\
(f, a^\dagger\; g) &=& (a\; f, g) , \label{eq3.60c}\\
(f, a\; g) &=& (a^\dagger\; f, g) .\label{eq3.60d}
\end{eqnarray}}
\reseteqn

\noindent All operators that have non-zero matrix elements only within the LLL
can be expressed in terms of $a$ and $a^\dagger$. It is essential to notice that
the adjoint of $a^\dagger$ is not $z^\ast/\sqrt{2}$ but $a\equiv
\sqrt{2}\partial/\partial z$, because $z^\ast$ connects states in the LLL to
higher Landau levels. Actually $a$ is the projection of $z^\ast/\sqrt{2}$ onto
the LLL as seen clearly in the following expression:
\[
(f, \frac{z^\ast}{\sqrt{2}}\; g)=(\frac{z}{\sqrt{2}}\; f, g)=(a^\dagger\; f,
g)=(f, a\; g).
\]
So we find
\begin{equation}
\overline{z^\ast}=2\frac{\partial}{\partial z}, 
\label{eq3.70}
\end{equation}
where the overbar indicates projection onto the LLL. Since $\overline{z^\ast}$
and $z$ do not commute, we need to be very careful to properly order the operators
before projection.  A little thought shows that in order to project an operator which is a
combination of $z^\ast$ and $z$, we must first normal order all the $z^\ast$'s to the
left of the $z$'s, and then replace $z^\ast$ by $\overline{z^\ast}$. With this 
rule in mind and (\ref{eq3.70}), we can easily project onto the LLL any operator that
involves space coordinates only.

\begin{figure}
\centerline{\epsfxsize=10cm
 \epsffile{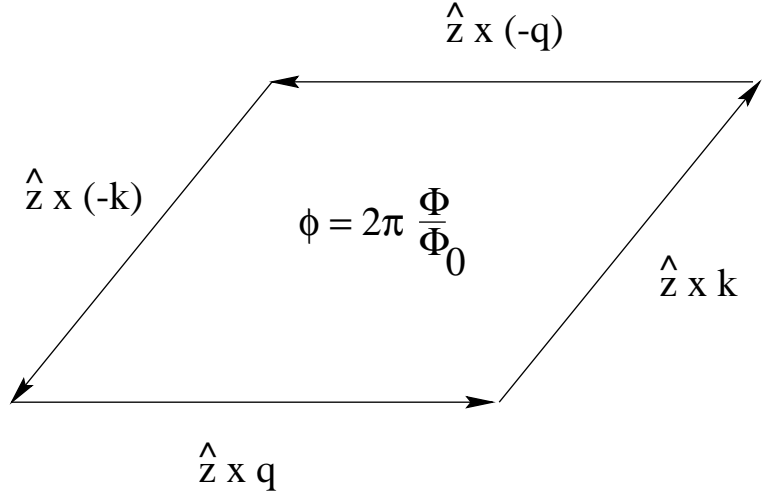}}
\caption[]{Illustration of magnetic translations and phase factors. When an
electron travels around a parallelogram (generated by
$\tau_{q}\tau_{k}\tau_{-q}\tau_{-k}$) it picks up a phase $\phi = 2\pi
\frac{\Phi}{\Phi_{0}} = q\wedge k$, where $\Phi$ is the flux enclosed in the
parallelogram and $\Phi_{0}$ is the flux quantum.}
\label{fig:magtrans}
\end{figure}

For example, the one-body density operator in momentum space is
\[
\rho_{\mathbf{q}} = \frac{1}{\sqrt{A}}\> e^{-i\mathbf{q}\cdot \mathbf{r}} =
\frac{1}{\sqrt{A}}\> e^{-\frac{i}{2}(q^\ast z + qz^\ast)} = \frac{1}{\sqrt{A}}\>
e^{-\frac{i}{2}qz^\ast}\; e^{-\frac{i}{2}q^\ast z} ,
\]
where $A$ is the area of the system, and $q=q_x + iq_y$. Hence
\begin{equation}
\overline{\rho_q} = \frac{1}{\sqrt{A}}\> e^{-iq\frac{\partial}{\partial z}}\;
e^{-\frac{i}{2}q^\ast z} = \frac{1}{\sqrt{A}}\> e^{-\frac{\vert q\vert^2}{4}}\;
\tau_q ,
\label{eq3.80}
\end{equation}
where
\begin{equation}
\tau_q = e^{-iq\frac{\partial}{\partial z} - \frac{i}{2}q^\ast z}
\label{eq3.90}
\end{equation}
is a unitary operator satisfying the closed Lie algebra
\alpheqn{
\begin{eqnarray}
\tau_q\tau_k &=& \tau_{q+k}\> e^{\frac{i}{2}q\wedge k} ,\label{eq3.100a}\\
{}[\tau_q, \tau_k] &=& 2i\; \tau_{q+k}\> \sin{\frac{q\wedge k}{2}},
\label{eq3.100b}
\end{eqnarray}}
\reseteqn

\noindent where $q\wedge k \equiv \ell^2(\mathbf{q}\times \mathbf{k}) \cdot
\hat{\mathbf{z}}$. We also have $\tau_q\tau_k\> \tau_{-q}\tau_{-k} = e^{iq\wedge
k}$. This is a familiar feature of the group of translations in a magnetic
field, because $q\wedge k$ is exactly the phase generated by the flux in the
parallelogram generated by $\mathbf{q}\ell^2$ and $\mathbf{k}\ell^2$. Hence the
$\tau$'s form a representation of the magnetic translation group [see
Fig.~(\ref{fig:magtrans})].
In fact $\tau_{q}$ translates the particle a distance $\ell^{2}\hat{\mathbf{z}}
\times \mathbf{q}$. This means that different wave vector components of the
charge density do not commute. It is from here that non-trivial dynamics arises
even though the kinetic energy is totally quenched in the LLL subspace.

This formalism is readily generalized to the case of many particles with spin,
as we will show next. In a system with area $A$ and $N$ particles the projected
charge and spin density operators are
\alpheqn{
\begin{eqnarray}
\overline{\rho _{q}} &=& \frac{1}{\sqrt{A}}\> \sum_{i=1}^N
\overline{e^{-i\mathbf{q} \cdot \mathbf{r}_i}} = \frac{1}{\sqrt{A}}\>
\sum_{i=1}^N e^{-\frac{\vert q\vert^2}{4}}\> \tau_{q}(i) \label{eq3.110a}\\
\overline{S_{q}^\mu} &=& \frac{1}{\sqrt{A}}\> \sum_{i=1}^N
\overline{e^{-i\mathbf{q}\cdot \mathbf{r}_i}}\> S_i^\mu = \frac{1}{\sqrt{A}}\>
\sum_{i=1}^N e^{-\frac{\vert q\vert^2}{4}}\> \tau_{q}(i)\> S_i^\mu ,
\label{eq3.110b}
\end{eqnarray}}
\reseteqn

\noindent where $\tau_{q}(i)$ is the magnetic translation operator for the $i$th
particle and $S_i^\mu$ is the $\mu$th component of the spin operator for the
$i$th particle. We immediately find that unlike the unprojected operators, the
projected spin and charge density operators do not commute:
\begin{equation}
[\bar{\rho}_{k}, \bar{S}_{q}^\mu] = \frac{2i}{\sqrt{A}}\> e^{\frac{\vert k +
q\vert^2 - \vert k\vert^2 - \vert q\vert^2}{4}}\> \overline{S_{k + q}^\mu}\>
\sin{\biggl(\frac{k\wedge q}{2}\biggr)} \neq 0.
\label{eq3.120}
\end{equation}
This implies that within the LLL, the dynamics of spin and charge are entangled,
i.e., when you rotate spin, charge gets moved. As a consequence of that, spin
textures carry charge as discussed in the text.

\chapter{Berry's Phase and Adiabatic Transport}
\label{app:BerryPhase}

Consider a quantum system with a Hamiltonian $H_{\vec{R}}$ which depends on a
set of externally controlled parameters represented by the vector $\vec{R}$.
Assume that for some domain of $\vec{R}$ there is always a finite excitation gap
separating the ground state energy from the rest of the spectrum of
$H_{\vec{R}}$. Consider now the situation where the parameters $\vec{R}(t)$ are
slowly varied around a closed loop in parameter space in a time interval $T$
\begin{equation}
\vec{R}(0) = \vec{R}(T).
\end{equation}
If the circuit is transversed sufficiently slowly so that $h/T \ll
\Delta_{\mathrm{min}}$ where $\Delta_{\mathrm{min}}$ is the minimum excitation
gap along the circuit, then the state will evolve \textit{adiabatically}. That
is, the state will always be the local ground state $\Psi_{\vec{R}(t)}^{(0)}$ of
the instantaneous Hamiltonian $H_{\vec{R}(t)}$. Given the complete set of energy
eigenstates for a given $\vec{R}$
\begin{equation}
H_{\vec{R}} \Psi_{\vec{R}}^{(j)} = \epsilon_{\vec{R}}^{(j)}
\Psi_{\vec{R}}^{(j)},
\end{equation}
the solution of the time-dependent Schr\"{o}dinger equation
\begin{equation}
i\hbar \frac{\partial\psi(\vec{r},t)}{\partial t} = H_{\vec{R}(t)}
\psi(\vec{r},t)
\label{eq:berry3}
\end{equation}
is
\begin{eqnarray}
\psi(\vec{r},t) &=& \Psi_{\vec{R}(t)}^{(0)}(\vec{r}\,)\; e^{i\gamma(t)}\; 
e^{-\frac{i}{\hbar}\int_{0}^{t}dt'\; \epsilon_{\vec{R}(t')}^{(0)}}\nonumber\\
&&+\sum_{j\neq 0} a_{j}(t)\; \Psi_{\vec{R}(t)}^{(j)}.
\end{eqnarray}
The adiabatic approximation consists of neglecting the admixture of excited
states represented by the second term. In the limit of extremely slow variation
of $\vec{R}(t)$, this becomes exact as long as the excitation gap remains
finite. The only unknown at this point is the Berry Phase \cite{Berry}
$\gamma(t)$ which can be found by requiring that $\psi(\vec{r},t)$ satisfy the
time-dependent Schr\"{o}dinger equation. The LHS of eq.~(\ref{eq:berry3}) is
\begin{eqnarray}
i\hbar \frac{\partial\psi(\vec{r},t)}{\partial t} &=&
\left[-\hbar\dot{\gamma}(t) + \epsilon_{\vec{R}(t)}^{(0)}\right]\;
\psi(\vec{r},t)\nonumber\\
&&+i\hbar\dot{R}^{\mu}\; \left[\frac{\partial}{\partial R^{\mu}}\;
\Psi_{\vec{R}(t)}^{(0)}(\vec{r}\,)\right]\; e^{i\gamma(t)}\; 
e^{-\frac{i}{\hbar}\int_{0}^{t}dt'\; \epsilon_{\vec{R}(t')}^{(0)}}
\label{eq:berry5}
\end{eqnarray}
if we neglect the $a_{j}(t)$ for $j > 0$. The RHS of eq.~(\ref{eq:berry3}) is
\begin{equation}
H_{\vec{R}(t)}\; \psi(\vec{r},t) = \epsilon_{\vec{R}(t)}^{(0)}\; \psi(\vec{r},t)
\label{eq:berry6}
\end{equation}
within the same approximation. Now using the completeness relation
\begin{equation}
\left|\frac{\partial}{\partial R^{\mu}}\; \Psi_{\vec{R}}^{(0)}\right\rangle =
\sum_{j=0}^{\infty} \left|\Psi_{\vec{R}}^{(j)}\right\rangle\;
\left\langle\Psi_{\vec{R}}^{(j)} \left|\frac{\partial}{\partial
R^{\mu}}\right.\; \Psi_{\vec{R}}^{(0)}\right\rangle.
\end{equation}
In the adiabatic limit we can neglect the excited state contributions so
eq.~(\ref{eq:berry5}) becomes
\begin{equation}
i\hbar \frac{\partial\psi}{\partial t} = \left[-\hbar\dot{\gamma}(t) +
i\hbar\dot{R}^{\mu}\; \left\langle\Psi_{\vec{R}}^{(0)}
\left|\frac{\partial}{\partial R^{\mu}}\right.\;
\Psi_{\vec{R}(t)}^{(0)}\right\rangle + \epsilon_{\vec{R}(t)}^{(0)}\right]\; \psi .
\end{equation}
This matches eq.~(\ref{eq:berry6}) provided
\begin{equation}
\dot{\gamma}(t) = i\dot{R}^{\mu}(t)\; \left\langle\Psi_{\vec{R}(t)}^{(0)}
\left|\frac{\partial}{\partial R^{\mu}}\right.\;
\Psi_{\vec{R}(t)}^{(0)}\right\rangle.
\end{equation}
The constraint $\left\langle\Psi_{\vec{R}}^{(0)}
\left|\Psi_{\vec{R}}^{(0)}\right.\right\rangle = 1$ guarantees that
$\dot{\gamma}$ is purely real.

Notice that there is a kind of gauge freedom here. For each $\vec{R}$ we have a
different set of basis states and we are free to choose their phases
independently. We can think of this as a gauge choice in the \textit{parameter}
space. Hence $\dot{\gamma}$ and $\gamma$ are `gauge dependent' quantities. It is
often possible to choose a gauge in which $\dot{\gamma}$ vanishes. The key
insight of Berry \cite{Berry} however was that this is not always the case. For
some problems involving a closed-circuit $\Gamma$ in parameter space the
\textit{gauge invariant} phase
\begin{equation}
\gamma_{\mathrm{Berry}} \equiv \int_{0}^{T} dt\; \dot{\gamma} = i \oint_{\Gamma}
dR^{\mu}\; \left\langle\Psi_{\vec{R}}^{(0)} \left|\frac{\partial}{\partial
R^{\mu}}\right.\; \Psi_{\vec{R}}^{(0)}\right\rangle 
\end{equation}
is non-zero. This is a gauge invariant quantity because the system returns to
its starting point in parameter space and the arbitrary phase choice drops out
of the answer. This is precisely analogous to the result in electrodynamics that
the line integral of the vector potential around a closed loop is gauge
invariant. In fact it is useful to define the `Berry connection' $\mathcal{A}$
on the parameter space by
\begin{equation}
\mathcal{A}^{\mu}(\vec{R}\,) = i\; \left\langle\Psi_{\vec{R}}^{(0)}
\left|\frac{\partial}{\partial R^{\mu}}\right.\;
\Psi_{\vec{R}}^{(0)}\right\rangle
\label{eq:berry11}
\end{equation}
which gives the suggestive formula
\begin{equation}
\gamma_{\mathrm{Berry}} = \oint_{\Gamma} d\vec{R} \cdot \mathcal{A}(\vec{r}\,).
\end{equation}
Notice that the Berry's phase is a purely geometric object independent of the
particular velocity $\dot{R}^{\mu}(t)$ and dependent solely on the path taken in
parameter space. It is often easiest to evaluate this expression using Stokes
theorem since the curl of $\mathcal{A}$ is a gauge invariant quantity.

As a simple example \cite{Berry} let us consider the Aharonov-Bohm effect where
$\mathcal{A}$ will turn out to literally be the electromagnetic vector
potential. Let there be an infinitely long solenoid running along the $z$ axis.
Consider a particle with charge $q$ trapped inside a box by a potential $V$
\begin{equation}
H = \frac{1}{2m}\; \left(\vec{p} - \frac{q}{c} \vec{A}\right)^{2} +
V\left(\vec{r} - \vec{R}(t)\right).
\label{eq:berry13}
\end{equation}
The position of the box is moved along a closed path $\vec{R}(t)$ which
encircles the solenoid but keeps the particle outside the region of magnetic
flux. Let $\chi^{(0)}\left(\vec{r} - \vec{R}(t)\right)$ be the adiabatic wave
function in the absence of the vector potential. Because the particle only sees
the vector potential in a region where it has no curl, the exact wave function
in the presence of $\vec{A}$ is readily constructed
\begin{equation}
\Psi_{\vec{R}(t)}^{(0)}(\vec{r}\,) = e^{\frac{i}{\hbar} \frac{q}{c}
\int_{\vec{R}(t)}^{\vec{r}} d\vec{r}' \cdot \vec{A}(\vec{r}')}\;
\chi^{(0)}\left(\vec{r} - \vec{R}(t)\right)
\label{eq:berry14}
\end{equation}
where the precise choice of integration path is immaterial since it is interior
to the box where $\vec{A}$ has no curl. It is straightforward to verify that
$\Psi_{\vec{R}(t)}^{(0)}$ exactly solves the Schr\"{o}dinger equation for the
Hamiltonian in eq.~(\ref{eq:berry13}) in the adiabatic limit.

The arbitrary decision to start the line integral in eq.~(\ref{eq:berry14}) at
$\vec{R}$ constitutes a gauge choice in parameter space for the Berry
connection. Using eq.~(\ref{eq:berry11}) the Berry connection is easily found to
be 
\begin{equation}
\mathcal{A}^{\mu}(\vec{R}\,) = +\frac{q}{\hbar c}\; A^{\mu}(\vec{R}\,)
\end{equation}
and the Berry phase for the circuit around the flux tube is simply the 
Aharonov-Bohm phase
\begin{equation}
\gamma_{\mathrm{Berry}} = \oint dR^{\mu}\; \mathcal{A}^{\mu} = 2\pi
\frac{\Phi}{\Phi_{0}}
\end{equation}
where $\Phi$ is the flux in the solenoid and $\Phi_{0} \equiv hc/q$ is the flux
quantum.

As a second example \cite{Berry} let us consider a quantum spin with Hamiltonian
\begin{equation}
H = -\vec{\Delta}(t) \cdot \vec{S}.
\label{eq:berry17}
\end{equation}
The gap to the first excited state is $\hbar |\vec{\Delta}|$ and so the circuit
in parameter space must avoid the origin $\vec{\Delta} = \vec{0}$ where the
spectrum has a degeneracy. Clearly the adiabatic ground state has
\begin{equation}
\left\langle\Psi_{\vec{\Delta}}^{(0)} \left|\vec{S}\right|
\Psi_{\vec{\Delta}}^{(0)}\right\rangle = \hbar S
\frac{\vec{\Delta}}{|\vec{\Delta}|}.
\end{equation}
If the orientation of $\vec{\Delta}$ is defined by polar angle $\theta$ and
azimuthal angle $\varphi$, the same must be true for $\langle\vec{S}\rangle$. An
appropriate set of states obeying this for the case $S = \frac{1}{2}$ is
\begin{equation}
|\psi_{\theta,\varphi}\rangle = \left(\begin{array}{c}
\cos{\frac{\theta}{2}}\\
\sin{\frac{\theta}{2}}\; e^{i\varphi}\end{array}\right)
\end{equation}
since these obey
\begin{equation}
\left\langle\psi_{\theta,\varphi} \left|S^{z}\right|
\psi_{\theta,\varphi}\right\rangle = \hbar S \left(\cos^{2}{\frac{\theta}{2}} -
\sin^{2}{\frac{\theta}{2}}\right) = \hbar S \cos{\theta}
\end{equation}
and
\begin{equation}
\left\langle\psi_{\theta,\varphi} \left|S^{x} + iS^{y}\right|
\psi_{\theta,\varphi}\right\rangle = \left\langle\psi_{\theta,\varphi}
\left|S^{+}\right| \psi_{\theta,\varphi}\right\rangle = \hbar S \sin{\theta}\;
e^{i\varphi}.
\end{equation}
Consider the Berry's phase for the case where $\vec{\Delta}$ rotates slowly
about the $z$ axis at constant $\theta$
\begin{eqnarray}
\gamma_{\mathrm{Berry}} &=& i \int_{0}^{2\pi} d\varphi\;
\left\langle\psi_{\theta,\varphi} \left|\frac{\partial}{\partial\varphi}\right.
\psi_{\theta,\varphi}\right\rangle\nonumber\\
&=& i \int_{0}^{2\pi} d\varphi\; \left(\cos{\frac{\theta}{2}}\;
\sin{\frac{\theta}{2}}\; e^{-i\varphi}\right)\; \left(\begin{array}{c}
0\\
i \sin{\frac{\theta}{2}}\; e^{i\varphi}\end{array}\right)\nonumber\\
&=& -S \int_{0}^{2\pi} d\varphi\; (1 - \cos{\theta})\nonumber\\
&=& -S \int_{0}^{2\pi} d\varphi\; \int_{\cos{\theta}}^{1} d\cos{\theta'} =
-S\Omega
\label{eq:berry22}
\end{eqnarray}
where $\Omega$ is the solid angle subtended by the path as viewed from the
origin of the parameter space. This is precisely the Aharonov-Bohm phase one
expects for a charge $-S$ particle traveling on the surface of a unit sphere
surrounding a magnetic monopole. It turns out that it is the degeneracy in the
spectrum at the origin which produces the monopole \cite{Berry}.

Notice that there is a singularity in the connection at the `south pole' $\theta
= \pi$. This can be viewed as the Dirac string (solenoid containing one quantum
of flux) that is attached to the monopole. If we had chosen the basis
\begin{equation}
e^{-i\varphi}\; |\psi_{\theta,\varphi}\rangle
\end{equation}
the singularity would have been at the north pole. The reader is directed to
Berry's original paper \cite{Berry} for further details.

In order to correctly reproduce the Berry phase in a path integral for the spin
whose Hamiltonian is given by eq.~(\ref{eq:berry17}), the Lagrangian must be 
\begin{equation}
\mathcal{L} = \hbar S\; \left\{-\dot{m}^{\mu}\mathcal{A}^{\mu} +
\Delta^{\mu}m^{\mu} + \lambda(m^{\mu}m^{\mu} - 1)\right\}
\end{equation}
where $\vec{m}$ is the spin coordinate on a unit sphere, $\lambda$ enforces the
length constraint, and
\begin{equation}
\vec{\nabla}_{m} \times \vec{\mathcal{A}} = \vec{m}
\end{equation}
is the monopole vector potential. As discussed in the text in
section~\ref{sec:qhf}, this Lagrangian correctly reproduces the spin precession
equations of motion.

\backmatter

\end{document}